\documentclass[aps,twocolumn,amsmath,amssymb,secnumarabic,eqsecnum,prb]{revtex4-1}
\usepackage{graphicx}
\usepackage[colorlinks=false]{hyperref}

\newcommand {\be}{\begin{equation}}
\newcommand {\ee} {\end{equation}}
\newcommand {\bea}{\begin{eqnarray}}
\newcommand {\eea} {\end{eqnarray}}

\newcommand{\bk}{{\bf p}}

\newcommand{\half}{{\textstyle{\frac{1}{2}}}}

\begin{document}
\title{Kubo formulas for viscosity: Hall viscosity, Ward identities, and the relation with conductivity}
\author{Barry Bradlyn}
\email{barry.bradlyn@yale.edu}
\author{Moshe Goldstein}
\author{N.\ Read}
\affiliation{Department of Physics, Yale
University, P.O. Box 208120, New Haven, CT 06520-8120, USA}
\date{November 13, 2012}
\begin{abstract}
Motivated by recent work on Hall viscosity, we derive from first principles the Kubo formulas for
the stress-stress response function at zero wavevector that can be used to define the full complex
frequency-dependent viscosity tensor, both with and without a uniform magnetic field. The formulas in
the existing literature are frequently incomplete, incorrect, or lack a derivation; in particular, Hall
viscosity is overlooked. Our approach begins from the response to a uniform external strain field, which
is an active time-dependent coordinate transformation in $d$ space dimensions. These transformations form
the group GL$(d,\mathbb{R})$ of invertible matrices, and the infinitesimal generators are called strain
generators. These enable us to express the Kubo formula in different ways, related by Ward identities;
some of these make contact with the adiabatic transport approach. The importance of retaining contact
terms, analogous to the diamagnetic term in the familiar Kubo formula for conductivity, is emphasized. For
Galilean-invariant systems, we derive a relation between the stress response tensor and the conductivity
tensor that is valid at all frequencies and in both the presence and absence of a magnetic field. In the
presence of a magnetic field and at low frequency, this yields a relation between the Hall viscosity, the
$q^2$ part of the Hall conductivity, the inverse compressibility (suitably defined), and the diverging
part of the shear viscosity (if any); this relation generalizes a result found recently by others. We
show that the
correct value of the Hall viscosity at zero frequency can be obtained (at least in the absence of
low-frequency bulk and shear viscosity) by assuming that there is an orbital spin per particle that
couples to a perturbing electromagnetic field as a magnetization per particle. We study several examples
as checks on our formulation. We also present formulas for the stress response that directly generalize
the Berry (adiabatic) curvature expressions for zero-frequency Hall conductivity or viscosity to the full
tensors at all frequencies.
\end{abstract}
\maketitle
\section{Introduction}\label{intro}

There has been great interest recently in the viscosity of quantum fluids, coming from various
directions. These directions include a conjectured lower bound on the ratio of the shear viscosity
to entropy density of a fluid, from the AdS/CFT correspondence \cite{Kovtun2005}; the properties of
an interacting gas of fermions with interactions described by $s$-wave scattering at or near the
unitarity limit \cite{Cao2011}; and the so-called Hall viscosity, an antisymmetric part of the
viscosity tensor analogous to Hall conductivity, which has been calculated for several gapped
topological phases \cite{Avron1995,Read2009,Read2010}.

Viscosity, whether in a solid or fluid, is essentially the set of transport coefficients describing
the relaxation of a deviation of the momentum density from its value in (possibly only local)
equilibrium. Hence it is necessary that momentum be conserved in order even to consider viscosity.
If the momentum density at $\mathbf{x}$ at time $t$ is $\mathbf{g}(\mathbf{x},t)$, the continuity
equation for momentum is
\begin{equation}
\frac {\partial g_\nu(\mathbf{x},t)}{\partial t}+\partial_\mu \tau _{\mu\nu}(\mathbf{x},t)=0,
\label{cont_eq_mom}
\end{equation}
(Greek indices $\mu$, $\nu$, \ldots, $=1$, \ldots, $d$ refer to space components, $d$ is the dimension
of space, and repeated Greek indices are summed). The
stress tensor operator $\tau_{\mu\nu}(\mathbf{x},t)$ plays a central role in calculating viscosity.
The viscosity tensor in a fluid can be defined as the expectation of the stress due to a
time-varying ``strain''. Unlike in a solid, in a fluid with no external fields present, an intrinsic
local static strain tensor is not defined in full
in any completely natural way, but we can define its trace as ${\rm tr}\, u= \delta(L^d)/L^d$, where $L^d$
is the volume, for a uniform dilation, or locally using the average particle number density $\overline{n}$
as ${\rm tr}\, u= -\delta\overline{n}/\overline{n}$. The local expected stress does not respond to
a change in the shape of a box confining the fluid (which can be considered as an attempt to impose a
static strain $u_{\alpha\beta}$), except that there is a response of the pressure to a
change in volume; the pressure is the expectation of the trace of the stress tensor, divided
by the dimension of space. However, time-dependent strain has an analog, which is the matrix of
gradients of the velocity field $\mathbf{v}$, the average velocity of the fluid:
\begin{equation}
\frac{\partial u_{\alpha\beta}}{\partial t}=\frac{\partial v_\beta}{\partial x_\alpha}. \nonumber
\end{equation}
The change in the average stress tensor $\langle\tau_{\mu\nu}\rangle$ from its equilibrium value can
be formally expanded in time derivatives, as
\begin{equation}
\delta\left\langle\tau_{\mu\nu}\right\rangle=-\lambda_{\mu\nu\alpha\beta} u_{\alpha\beta}
-\eta_{\mu\nu\alpha\beta}
\frac{\partial u_{\alpha\beta}}{\partial t}+\dots, \label{stresslinresponse}
\end{equation}
where $\lambda_{\mu\nu\alpha\beta}$ is the tensor of elastic moduli and $\eta_{\mu\nu\alpha\beta}$
is  the viscosity tensor. Here we may view this as holding between local quantities at the same position
in space from a long-wavelength point of view, or to zeroth order in spatial derivatives. Then in a fluid,
\be
\lambda_{\mu\nu\alpha\beta}=\kappa^{-1}\delta_{\mu\nu}\delta_{\alpha\beta},
\ee
that is, only the trace of $\tau$ responds, and only to ${\rm tr}\, u$, corresponding to a change
in volume or local density of the fluid. The coefficient $\kappa^{-1}$ is the
inverse compressibility, $\kappa^{-1}=-L^d(\partial
P/\partial (L^d))_N$, the derivative of the pressure with respect to volume $L^d$ of the fluid, taken
at fixed particle number $N$.
So far, we did not use rotational invariance (isotropy) of the fluid; if we do, then the stress tensor
is symmetric, and only the more familiar symmetrized rate of strain
\be
\frac{1}{2}\left(\frac{\partial v_\alpha}{\partial x_\beta}+\frac{\partial v_\beta}{\partial
x_\alpha}\right)
\ee
enters, so that $\eta_{\mu\nu\alpha\beta}$ is also symmetric under the exchange
$\alpha\leftrightarrow\beta$. Another part of the expected stress is the momentum flux
$m \overline{n}v_\mu v_\nu$ (where $m$ is the particle mass), which is very important in fluid mechanics
(e.g.\ in obtaining the Navier-Stokes equations),
but will not be encountered in this paper as we consider only linear response to strains.

Without assuming rotation invariance, we can further distinguish some parts of the viscosity tensor.
$\eta$ can be divided into a symmetric and antisymmetric part under exchange of the first and the last
pair of indices \cite{Avron1995}:
\begin{align}
\eta_{\mu\nu\alpha\beta}&=\eta^S_{\mu\nu\alpha\beta}+\eta^A_{\mu\nu\alpha\beta},\nonumber \\
\eta^S_{\mu\nu\alpha\beta}&=\eta^S_{\alpha\beta\mu\nu}, \nonumber \\
\eta^A_{\mu\nu\alpha\beta}&=-\eta^A_{\alpha\beta\mu\nu}.
\end{align}
For the zero-frequency parts that we consider at the moment, only the symmetric part of
$\eta^S$ contributes to dissipation. For a rotationally-invariant $d$-dimensional system it has only
two independent components,
\begin{equation} \label{eqn:etas}
\eta^S_{\mu\nu\alpha\beta} =
\zeta \delta_{\mu \nu} \delta_{\alpha \beta} +
\eta^\text{sh}
\left(\delta_{\mu \alpha} \delta_{\nu \beta} + \delta_{\mu \beta} \delta_{\nu \alpha} -
\frac{2}{d} \delta_{\mu \nu} \delta_{\alpha \beta} \right),
\end{equation}
with $\zeta$ the bulk viscosity and $\eta^\text{sh}$ the shear viscosity.
$\eta^A$ can only be nonzero when time-reversal symmetry is broken, and in rotationally-invariant two
dimensional system has only one independent component,
\begin{equation} \label{eqn:etah1}
\eta^A_{\mu\nu\alpha\beta}=\eta^H\left(\delta_{\nu\alpha}\epsilon_{\mu\beta}
-\delta_{\mu\beta}\epsilon_{\alpha\nu}\right),
\end{equation}
with $\eta^H$ the so-called Hall viscosity\cite{Avron1995,Read2009}. For gapped quantum systems at zero
temperature, it has been shown using the adiabatic approach to quantum transport that
\begin{equation} \label{eqn:etah2}
\eta^H=\half\hbar\overline{n}\overline{s},
\end{equation}
where $\overline{n}$ is the expected particle number density in the ground state, and $\overline{s}$
is minus the average orbital spin per particle\cite{Read2009,Read2010}; see also Refs.\
\onlinecite{Haldane2009,Nicolis2011}. $\hbar$ is Planck's constant,
which we usually set to $1$. We should point out that in Ref.\ \onlinecite{Read2010}, there
is an unfortunate sign mistake in the definition of the stress tensor in eq.\ (2.18) of that
reference, which propagated through the paper, though the adiabatic curvature results are correct.
This means that all viscosities calculated there should have the opposite sign; above we have also
reversed the sign in the definition of the scalar $\eta^H$, so as to retain the memorable formula, eq.\
(\ref{eqn:etah2}).

The purpose of this paper is to develop a variety of Kubo formulas from which the viscosity tensor,
which in general is frequency dependent and complex, can be calculated in quantum fluids at zero or
non-zero temperatures. A motivation for doing so is to define the Hall viscosity from a Kubo formula,
and thus make contact with traditional approaches. In addition, we consider the relation of the
viscosity tensor to the conductivity tensor in Galilean-invariant systems, in which the latter
tensor is the momentum-momentum density response; the relation comes from the continuity equation,
eq.\ (\ref{cont_eq_mom}), and (along with some other relations in this paper) it can be called a
Ward identity. This has been studied previously, or else frequently is used as the {\em definition}
of the viscosity for calculation purposes; our goal is to do it allowing for the possibility of a
Hall viscosity. Moreover, we will do the same in the presence of a magnetic field. In this case,
total (kinetic) momentum is not time-independent, but precesses at the cyclotron frequency. We
will show that we can nonetheless derive parallel formulas in this case. In particular, we
recover a formula of Hoyos and Son \cite{Hoyos2012} that relates the Hall viscosity in a quantum
Hall state
to the order $q^2$ part of the Hall conductivity at wavevector $\mathbf{q}$ and the inverse
compressibility $\kappa^{-1}$ above (suitably defined in the presence of the magnetic field),
and generalize it further. These results bring us closer to finding experimental techniques
with which the Hall viscosity can be measured. We also study several example systems, to
validate our approach to Kubo formulas for the viscosity tensor. Finally, we show, in a
partially heuristic way, that the Hall viscosity can be rederived from macroscopic
electrodynamics by using the relation with conductivity, and assuming that at low frequencies
the system behaves as if there is an orbital spin $-\overline{s}$ per particle that acts as a
contribution to  magnetization density.

The basic strategy of our work is to define the viscosity from the response of the stress to an
external field. This external field enters (in some gauge choices) as a spatial metric, or if we
do not assume rotational invariance, as what we will call a {\em strain}; it is generally assumed
to be constant in space. (This strain field is external, and can have arbitrary time dependence.
This does
not contradict our statement that a static strain is not fully defined in a fluid, as that was
for an intrinsic strain, which would be a property defined given any state of the fluid in no
external field. But where the intrinsic strain is defined, i.e.\ for its static trace and for
its time derivative, the stress responses to either of the two strains should agree.) It is well
known (especially in gravitation theory and high-energy physics) that the stress tensor is the
change in the Lagrangian or Hamiltonian with respect to the metric. We build on this to consider
the underlying response function, that gives the viscosity, as the next order response of this
stress to the time-derivative of the strain field. This approach has the advantage of making contact
with previous work in which the Hall viscosity of some systems was calculated
\cite{Avron1995,Read2009,Read2010} using the adiabatic transport technique. It closely parallels the
case of
conductivity, in which one considers the response of the current, which is the change in Lagrangian
or Hamiltonian with the vector potential, to an electric field, viewed as the time-derivative of
the perturbing vector potential, setting that vector potential to zero at the end. Our approach
uses the standard Hamiltonian (canonical) operator formalism, plus linear response. We do not
make any assumptions that hydrodynamic behavior or local equilibrium holds, or use constitutive
relations, beyond motivating the names for some parts of our expressions, such as compressibility
and viscosity. We focus on quantum systems, but a similar discussion could be given in the
classical setting.

Early work on response-function type formulas for transport coefficients was done by Green, Kubo,
Mori, Kadanoff and Martin, and Luttinger, starting in the 1950s
\cite{Green1952,Kubo1957,Mori1958,Kadanoff1963,Luttinger1964}. Green and Mori initially
used somewhat phenomenological methods. ``Mechanical'' formulations as the response to an external
field, similar to ours, were used by Kubo\cite{Kubo1957} for the electrical conductivity case, and
by Luttinger\cite{Luttinger1964} for thermal transport and viscosity, and for the most part the
final formulas agree with those of Green and Mori. Kubo and Luttinger began by deriving the retarded
response function to an externally applied field, and then transformed to expressions that are not
as familiar today, involving an integration over an imaginary time variable, as well as one over real
time. By contrast, most references from the last forty years follow Kadanoff and Martin
\cite{Kadanoff1963} in using only retarded response functions to obtain ``Kubo-type'' formulas, now
usually called simply Kubo formulas. (We too use only this formalism.)
For the particular case of viscosity, many authors including Luttinger \cite{Luttinger1964} base the
definition on the continuity equation, eq.\ (\ref{cont_eq_mom}), and his ``mechanical'' formulation
is a study of the response of the momentum density to an electric field (which couples to the
number current, equal to the momentum density times the mass in a Galilean-invariant system of
particles that all have the same mass and unit charge). This leads him to a correct formula as a
stress-stress response function which, however, he writes in terms of an additional integral
over an imaginary-time interval. In many other papers, including the interesting recent Ref.\
\onlinecite{Taylor2010}, the starting formula is the momentum-momentum response, which has to
be expanded to order $q^2$ as the wavevector $\mathbf{q}$ tends to zero, to obtain the viscosity
tensor. (We note that this approach cannot be employed to define viscosity of a finite system, which was
defined using a strain in the adiabatic approach.) In some of these papers the stress-stress form is
not used at all. But in most papers that do give a stress-stress form for the response function that
yields the viscosity, the formulas frequently are incomplete, lack a derivation, contain pitfalls for
the unwary, or are simply incorrect. In particular, we are not aware of any derivations from an
external strain field in the many-body literature. Hence, and because so much time has elapsed and
notations have changed since the 1960s, we feel justified in revisiting these formal matters here.

In order to explain the issues that need to be corrected in the stress-stress form, we will use the
more familiar case of
conductivity as an analogy. For the (complex) conductivity tensor at zero wavevector, it was shown
by Kubo\cite{Kubo1957} that the approach outlined above gives rise to the Kubo formula
\begin{align}
  \lefteqn{\sigma_{\mu\nu}(\omega) =
  \frac{i \overline{n}}{m \omega^+}\delta_{\mu\nu}}\quad&\nonumber\\
 &{}+ \frac{1}{\omega^+}
  \int_0^\infty dt\, e^{i \omega^+ t}
  \int d^dx\,
  \left\langle \left[ j_{\mu}(\mathbf{x},t), j_{\nu}(\mathbf{0},0) \right] \right\rangle_0.
  \label{kubo_right}
\end{align}
Here $j_\nu(\mathbf{x},t)$ is the current density operator (equal to $g_\nu/m$ in a Galilean-invariant
system), $\omega^+=\omega+i\epsilon$, the limit $\epsilon\to 0^+$ is implicit, and the expectation
is taken in the unperturbed ground state or statistical ensemble, denoted $\langle\ldots\rangle_0$.
We set the charge of the particles to $1$ throughout this paper. This expression gives the response
of the expectation of the current density to a uniform external electric field. (When multiplied by $m^2$,
it is also the $\mathbf{q}\to0$ limit of the momentum-momentum response mentioned above; this fact will
not play any role just now). The first term is the so-called diamagnetic
current, which (in the approach we are using at the moment) arises because the current density in the
presence of a perturbing vector potential $A_\nu$ is actually
$j_\nu(\mathbf{x},t)-A_\nu(\mathbf{x},t)n(\mathbf{x},t)/m$ (because the Hamiltonian is quadratic in
$A_\nu$), where $n(\mathbf{x},t)$ is the number density operator.
The second term is the current-current retarded response function (though we usually reserve such
terminology for the complete expression). We note that textbook derivations of linear response
usually assume that the perturbing field appears only linearly in the Hamiltonian (except possibly
when considering conductivity), and then no such ``contact'' terms (i.e.\ terms like the diamagnetic
term in the conductivity) appear.

It would not be wise to drop the diamagnetic current term from the conductivity. Recall
the Sokhotski-Plemelj formula (for $\omega$ real),
\be
\frac{1}{\omega+i\epsilon}={\rm PP}\,\frac{1}{\omega}-i\pi \delta(\omega),
\ee
where $\rm PP$ denotes the principal part (both terms become meaningful once substituted into an integral).
Even if one wants to find only the real part of the conductivity, the diamagnetic term contributes a
$\delta$-function at zero frequency. In some cases, such as for fermions without impurity scattering
(disorder), both in Fermi liquids and paired superfluids, the time-integral term vanishes and the
diamagnetic term is the full response. At zero frequency, one would say that the conductivity is infinite,
which is correct.

For the real part, some authors instead use the formula
\be
{\rm Re}\,\sigma_{\mu\nu}(\omega) =
\frac{{\rm Re}\,
  \int_0^\infty dt\, e^{i \omega^+ t}
  \int d^dx\,
  \left\langle \left[ j_{\mu}(\mathbf{x},t), j_{\nu}(\mathbf{0},0) \right] \right\rangle_0}{\omega},
  \label{kubo_wrong}
 \ee
in which the $\pi\delta(\omega)$ coming from use of the Sokhotski-Plemelj formulas has been dropped,
both in the
diamagnetic term, and in the time-integral term (the latter would contain the imaginary part of the time
integral, in place of the real part here). This is correct when these $\delta$-functions cancel, which
does happen for the case with impurity scattering. In the simple Drude approximation, the complex
conductivity tensor is
\be
\sigma_{\mu\nu}(\omega)=\frac{i\overline{n}}{m(\omega+i/\tau_{\rm imp})},
\ee
where $1/\tau_{\rm imp}$ is the impurity scattering rate; there is no $\delta(\omega)$ piece in the real
part. In that case the real part of the time integral, which is kept, gives the correct broadened Drude
peak, in the simplest approximation. But in the limit as the impurity scattering rate goes to zero, the
correct result at zero frequency should be diverging, while use of the above formula gives a result
that increases as the rate gets smaller, but jumps to zero when there is strictly no impurity scattering.
One would like to criticize this behavior for not being continuous, however if the value at zero
scattering was infinity it would not be continuous either. Looking at the full function of $\omega$,
not only at $\omega=0$, we would like to say that the correct result approaches a $\delta$-function
 continuously as the scattering rate goes to zero. This makes sense only if we interpret ${\rm
 Re}\,\sigma_{\mu\nu}(\omega)$ as a
distribution for any scattering rate; then it is correct to say that as the rate goes to zero, it
approaches a $\delta$-function continuously (in the space of distributions). But this continuity
with the limit of zero scattering is lost if the above form ${\rm Re}\,\int_0^\infty dt\ldots/\omega$
is used.

The use of eq.\ (\ref{kubo_wrong}) also leads to difficulties with the Kramers-Kronig (KK) relations,
that do not occur with the correct form, eq.\ (\ref{kubo_right}), even if the diamagnetic term is
dropped (because that complex term obeys the KK relations). The difficulty can be remedied only by
reinstating the $\pi\delta(\omega)$ times the imaginary part of the $\omega=0$ time-integral. However,
the diamagnetic term cannot be recovered in this way, unless one has for example an argument that the
real part of $\sigma$ contains no $\delta$-function at zero.

The issues in the literature on Kubo stress-stress formulas for viscosity are very similar to these.
Let us now give
one of our forms (slightly simplified) for the response function from which we obtain the viscosity:
\begin{eqnarray}
\lefteqn{\chi_{\mu\nu\alpha\beta}(\omega)=\frac{1}{i\omega^+}\left\{
\left< \left.\frac{\partial \tau_{\mu\nu}(\mathbf{0})}{\partial \lambda_{\alpha\beta}}\right|_{\lambda=0}
\right>_0\right.}&&\nonumber\\
&&{}\left.+i\int_0^\infty{dt
\int d^dx\,e^{i\omega^+t}}
\left<\left[\tau_{\mu\nu}(\mathbf{x},t),\tau_{\alpha\beta}(\mathbf{0},0)\right]\right>_0\right\}.
\end{eqnarray}
The time-integral term is what would be expected for the stress-stress response. The first term, which
is frequency independent except for the $1/\omega^+$ factor, is what we call (following long-time usage
in the high-energy literature) a contact term (this has no connection at all with the so-called 
``contact'' in the theory of interacting Fermi gases at the unitarity limit
\cite{Tan2008_1,Tan2008_2,Tan2008_3}). Without giving the full details here, $\lambda_{\alpha\beta}$
is the external strain, and varying the Hamiltonian with respect to $\lambda_{\mu\nu}$ gives (minus)
the stress tensor $\tau_{\mu\nu}(\mathbf{x})$. The stress still depends on $\lambda$, like the
current above. Thus the response of the expectation of the stress to the strain contains the contact
term, which is one more derivative with respect to $\lambda_{\alpha\beta}$; it is directly analogous
to the diamagnetic conductivity. By contrast, stress-stress response formulas in the literature usually
omit this term, an exception being the early work of Luttinger \cite{Luttinger1964}, whose expression
is equivalent to this, but is written in a way that may now seem obscure (the relation can be found
in Kubo\cite{Kubo1957}). We will now give reasons why the contact term has to be retained.

We pointed out earlier that in a homogeneous fluid, the expectation of the stress is
\be
\langle\tau_{\mu\nu}(\mathbf{x},t)\rangle_0=P\delta_{\mu\nu},
\ee
and that a static strain affects this at first order, the coefficient being the inverse compressibility.
It follows that there will be a part of the response function,
\be
\chi_{\mu\nu\alpha\beta}(\omega)\sim
\frac{i\kappa^{-1}}{\omega^+}\delta_{\mu\nu}\delta_{\alpha\beta}+\ldots
\ee
as $\omega\to0$, and any remaining $1/\omega^+$ term has vanishing trace on the indices $\mu\nu$ and
on $\alpha\beta$
(this result is valid at non-zero as well as at zero temperature). The coefficient $\kappa^{-1}$ is
equal to the zero frequency limit of the response function $-i\omega^+\chi_{\mu\mu\alpha\alpha}/d^2$,
and there is a contribution from the contact term as well as from the time-integral term (we discuss
this more formally in Sec.\ \ref{nondiv} below). Thus dropping the contact term spoils this relation.
It is a feature of our approach that this term can be easily identified as the inverse compressibility,
because it is the response to a static external strain.

We then define the complex viscosity tensor to be
\be
\eta_{\mu\nu\alpha\beta}(\omega)=\chi_{\mu\nu\alpha\beta}(\omega)-
\frac{i\kappa^{-1}}{\omega^+}\delta_{\mu\nu}\delta_{\alpha\beta}.
\ee
(We note that Luttinger \cite{Luttinger1964} recognized the need to subtract such a term to obtain
the viscosity tensor, however he derives it using hydrodynamic arguments and local equilibrium.)
It follows from the preceding remarks that the bulk viscosity cannot diverge as $\omega\to0$ (see
the definitions above, for rotationally-invariant systems). The shear viscosity has a contribution
from the contact term, and so the real part has a $\delta(\omega)$ contribution. This may sometimes
be cancelled by another from the time-integral term. In particular, this occurs in a gapped system at
zero temperature, which should have no dissipative viscosity at $\omega=0$. It also occurs in an
interacting Fermi liquid at positive temperature $T$, which has a finite shear viscosity at $\omega=0$
that tends to infinity as the temperature goes to zero. Thus, similarly to the case of the conductivity
of a Fermi gas with impurities, the $\delta$-function terms must cancel. Either of these cancelations
is spoiled if the contact term is dropped.

Some authors, who consider fluids at positive temperatures, use an expression for the real part of
the viscosity,
\begin{eqnarray}
\lefteqn{{\rm Re}\,\eta_{\mu\nu\alpha\beta}(\omega)=}\quad&&\nonumber\\
&&{}\frac{{\rm Re}\,\int_0^\infty{dt
\int d^dx\,e^{i\omega^+t}}
\left<\left[\tau_{\mu\nu}(\mathbf{0},t),\tau_{\alpha\beta}(\mathbf{x},0)\right]\right>_0}{\omega},\quad
\end{eqnarray}
where $\omega$ in the denominator is real ($i\epsilon$ is dropped).
In particular, Kadanoff and Martin give a related form, which can be obtained from this using
the fluctuation-dissipation theorem and letting $\omega\to0$. In this form, $\delta$-functions
$\delta(\omega)$ are all dropped, which removes the $\kappa^{-1}$ term, and gives the viscosity
correctly only if the remaining $\delta(\omega)$'s do cancel in the real part of our full expression
above. As with the conductivity of a Fermi gas with impurities, use of this form in an interacting
Fermi liquid as $T\to0$ gives discontinuous behavior of the shear viscosity, which ought to be
continuous when viewed as a distribution. It also gives zero for a non-interacting Fermi gas, instead
of infinity. While one may say that in either of these two limits, hydrodynamics is not well defined,
it is preferable to have continuous behavior of our theories, and these are reasons to retain the
contact term.

A few authors go a step further than this, asserting that the complex viscosity is given by
\begin{eqnarray}
\lefteqn{\eta_{\mu\nu\alpha\beta}(\omega)\stackrel{?}{=}}\quad&&\nonumber\\
&&{}\frac{\int_0^\infty{dt
\int d^dx\,e^{i\omega^+t}}
\left<\left[\tau_{\mu\nu}(\mathbf{0},t),\tau_{\alpha\beta}(\mathbf{x},0)\right]\right>_0}{\omega},\quad
\end{eqnarray}
with $1/\omega$, not the more correct $1/\omega^+$.
This is incorrect, as the right hand side usually has a divergence as $\omega\to0$ in the trace part,
which is a contribution to, but not equal to, $\kappa^{-1}$. Further, if the $\delta(\omega)$'s are
to cancel in the real part of the correct expression, it must also have a $1/\omega$ divergence in
the imaginary part of the shear viscosity, that is cancelled by that due to the contact term. [The
recent paper by Taylor and Randeria \cite{Taylor2010} correctly states that the contact term (in
one particular form)
contributes to the imaginary part of the viscosity.] These latter forms also lead to difficulties
with the KK relations, as in the conductivity case. There is sufficient information in these forms to
recover a version (given by the last expression, but with $\omega$ replaced by $\omega^+$ in the
denominator) that satisfies the KK relations, but the contact term cannot be fully recovered in this way.

A further reason to retain the contact term is that then the Ward identity relation with the conductivity
tensor has a simple form,
\be
  \sigma_{\nu\beta}(\mathbf{q},\omega) =  \frac{i \overline{n}}{m \omega^+}\delta_{\nu\beta}+\frac{q_\mu
  q_\alpha }{m^2\omega^{+2}} \chi_{\mu\nu\alpha\beta}(\omega)+\mathcal{O}(q^4),
\ee
in zero magnetic field in a Galilean-invariant system. We see that there is a term $ i\kappa^{-1}q_\nu
q_\beta/(m^2\omega^{+3})$ on the right hand side, which is sometimes wrongly omitted (e.g.\ in Ref.\
\onlinecite{Taylor2010}).

Now we will describe some of the highlights of our work in the present paper. We begin in Sec.\ \ref{sec2}
with some technical background used to set up the stress response expressions. As mentioned above, we work
systematically with the response to an external ``strain'' field. This can be introduced in two ways in
particular. Again, an analogy with conductivity may be helpful here. The conductivity is the response of
the number current density to an external electric field, and the field is taken to be uniform in space,
and have frequency $\omega$. The field can be represented in two ways in particular, which are simply two
gauge choices. One is as a scalar potential,
which depends linearly on position, because the field is uniform. The other is as a vector potential which
is constant in space, and whose time derivative gives the electric field (and so is non-zero even when
$\omega=0$). Both formulations appear in the literature; the latter is very commonplace, while the first
has the drawback that the linearly-varying scalar potential is not compatible with periodic boundary
conditions, and the choice of suitable boundary conditions for a finite-size calculation becomes more
problematic, though this is frequently ignored. On the other hand, in the first formulation the external
electric field appears only linearly in the Hamiltonian, so there is no contact term in the response
function, whereas in the formulation using the vector potential, the external field appears quadratically,
and so there is a contact term---the diamagnetic current term, as discussed above. Last, we note that the
transformation from one formulation to the other is a unitary transformation, simply implementing the
change of gauge in the quantum theory.

Similar alternatives appear in the case of stress response. It is known that stress is the derivative of
the Hamiltonian with strain, so we may begin with a Hamiltonian with a general, spatially-uniform strain
that depends on time (it appears nonlinearly in the Hamiltonian). A time-independent strain can be
eliminated from the Hamiltonian by a coordinate transformation, which for spatially-constant strain is a
linear transformation. The two coordinate systems used to describe our system, which are related by this
transformation, are referred to as the $\mathbf{x}$ and $\mathbf{X}$ variables, respectively. These
coordinate transformations take the place of U(1) gauge transformations; the relation can be understood
if we realize that the conserved quantity corresponding here to particle number in the conductivity case
is the total momentum of the particles, and so in the generators of infinitesimal transformations, the
$x$-dependence must be multiplied by the momentum of the particle on which it acts:
\be
J_{\alpha\beta}=-\half\sum_i \{x^i_\alpha,p^i_\beta\}
\ee
for zero magnetic field, which leads to
coordinate transformations (for the conductivity problem, the corresponding operators are $\sum_i
x^i_\alpha$). These generators will be called {\em strain generators}; they play a central
role in our approach. In fact, the time derivative of a strain generator gives the corresponding component
of the integrated stress tensor. This follows either from expanding the (Fourier-transformed) continuity
equation to first order in wavevector, or by seeing that the time derivative is the commutator with the
Hamiltonian, which thus produces the change in the Hamiltonian with a uniform strain, which we know is the
(integrated) stress.

As the strain generators generate linear transformations of the coordinates (and inverse ones of the
momenta, as the transformations are canonical), they obey the relations of the Lie algebra
$\mathfrak{gl}(d,\mathbb{R})$ of the group of invertible matrices GL$(d,\mathbb{R})$ which describes
(invertible) linear transformations. We further generalize this to the case with a magnetic field
(in $d=2$ dimensions). Previous work \cite{Read2009,Read2010} considered mainly transformations in the
subgroup SL$(d,\mathbb{R})$, consisting of shears and rotations but not dilations, because of the apparent
difficulty of including dilations in the presence of a magnetic field. Here we overcome this difficulty
using a technical trick: when the system is dilated, for consistency the magnetic field must be rescaled
as an inverse length-squared. To do this, we promote the magnetic field strength to be a quantum variable
$\mathcal{B}$, for which the canonically conjugate momentum $\mathcal{P}$ does not appear in the
Hamiltonian. Then operator expressions for the strain generators can be written down as before, and the
integrated stress tensor is again the commutator of these with the Hamiltonian. Finally, the states with
which we work are assumed to have small width in $B$, so that the results correspond to a fixed magnetic
field as in conventional approaches.

With this formalism in hand, we are ready to tackle stress response in Sec. \ref{sec3}. One place to start
is in the $\mathbf{X}$ variables, in which the time-derivative of the strain (but not the strain itself)
appears linearly in the Hamiltonian, multiplied by the strain generator, similar to the conductivity case
when using the scalar potential to represent the external field; there is then no contact term in the
response. This ``stress-strain'' form can then be integrated by parts on time, to obtain the
``stress-stress'' form which is similar to what was already discussed above, and includes a contact
term. (It can also be obtained as the response to a strain by working directly in the $\mathbf{x}$
variables, similar to the vector-potential formulation in the conductivity case.) A different integration
by parts produces instead the ``strain-strain'' form, which is a strain-strain response function, and
is useful in making contact with the adiabatic approach. These three forms have direct parallels with
Kubo's Theorem 2 (there is a typo in the second line of the stated theorem in Ref.\ \onlinecite{Kubo1957};
the dot should be over the second occurrence of $\phi_{BA}$, not the first).
There are several variations on these results, including formulas for the response in the local stress
$\tau_{\mu\nu}$ instead of in the integrated stress, and the use of periodic boundary conditions in some
of these formulations, or of a confining potential that is handled similarly to the magnetic field. We
argue that the inverse compressibility must be subtracted from the foregoing response to obtain the
viscosity tensor as described already above; in a magnetic field, this becomes the inverse ``internal''
compressibility, the partial derivative with respect to area taken with the flux through the system
fixed, not the
field strength. We argue that the bulk viscosity is never diverging at zero frequency (unlike the shear
viscosity, which may), and discuss the form of standard sum rules and positivity constraints on the
spectral density. We emphasize that these results are for all components of the viscosity tensor,
including Hall viscosity.

Sec.\ \ref{sec4} is devoted to the relation of the stress response, and hence viscosity, to the
conductivity itself. The relation is between the stress response at zero wavevector, and the second
derivative of the conductivity with respect to wavevector, and holds for all frequencies. This has
been discussed previously in
zero magnetic field, and is sometimes used to define the viscosity. Our derivation includes a uniform
magnetic field (with zero field as a special case). This relation may then be studied at low frequencies,
where if the bulk and shear viscosities are zero, there is a relation between the Hall viscosity, the
second derivative of the Hall conductivity, and the internal compressibility, which agrees with recent
work by Hoyos and Son \cite{Hoyos2012}. We find that the relation still holds if these dissipative
viscosity coefficients are non-zero but finite, and can be generalized to allow for diverging shear
viscosity at zero frequency also.

In Sec.\ \ref{sec5}, we study several simple examples as checks on our formulation. These include the
free (non-interacting) Fermi gas in zero magnetic field, which possesses a diverging shear viscosity.
Other examples are a non-interacting Fermi gas in non-zero magnetic field, for which we can find the
full frequency-dependent response function, and fractional quantum Hall states, for which we recover
the Hall viscosity in agreement with previous work. Finally, we consider paired states of fermions for
pairing functions with non-zero angular momentum, for which we recover the Hall viscosity in agreement
with Refs.\ \onlinecite{Read2009,Read2010}, for a mean-field model in which the Hamiltonian is
quadratic in field operators. In each case, we can verify the relation with conductivity, using results
of some previous authors (who did not consider viscosity).

Finally, in Sec.\ \ref{electro}, we show that using macroscopic electrodynamics, together with the
assumption (in the spirit of an effective theory) that an external perturbing magnetic field couples to
the orbital spin $-\overline{s}$ of each particle as a magnetization, we again recover the Hall viscosity
result, eq.\ (\ref{eqn:etah2}).

In Appendix \ref{app_stress}, we show how a standard expression for the stress tensor in
a rotationally-invariant interacting particle system can be obtained by varying the spatial metric in
the Hamiltonian.
In Appendix \ref{app}, we describe a formalism for including a confining potential to make a
finite-volume system in an infinite space, and show that it works in the example of free fermions in
a harmonic potential. In Appendix \ref{appC}, we show that
the three forms of Kubo formula for stress response can be written down in periodic boundary conditions,
using derivatives of the ground state instead of strain generators. This makes contact with the adiabatic
approach to quantum transport, in which Hall conductivity and Hall viscosity are written as Berry
curvatures (and then as a Chern number in the case of Hall conductivity) and generalizes it to
non-zero frequency, and to gapless systems. In Appendix \ref{appB}, we show that some assumptions
of time-translation invariance that were used at some points in the derivations are justified.


\section{Stress tensor and strain generators} \label{sec2}

In this section, we provide background needed for the viscosity linear response calculation. We discuss
Hamiltonians with a strain that could be static, and others (related by a coordinate transformation,
implemented by a unitary transformation) that contain only the time-derivative of the strain, times a
strain generator. The basic relation giving the integral of the stress tensor as the time derivative of
the strain generator is obtained. We consider separately the two cases of zero magnetic field (in Sec.\
\ref{sec21}), which is mostly straightforward, and non-zero magnetic field (in Sec.\ \ref{sec22}), which
is less so.

\subsection{Zero magnetic field}\label{sec21}

We begin by considering a Hamiltonian $H_0$ for a system of interacting particles in infinite
$d$-dimensional space:
\begin{equation}
H_0^{(\mathbf{x})}=\frac{1}{2m}\sum_ip^i_\mu p^i_\mu+\frac{1}{2}\sum_{i\neq j}V(\mathbf{x}^i-\mathbf{x}^j),
\end{equation}
where $\mathbf{x}^i$ are the coordinates of the $i$th particle ($i=1$, \ldots, $N$), and $\mathbf{p}^i$
are their momenta, with
\be
[x_\mu^j,p_\nu^k]=i\delta_{\mu\nu}\delta_{jk}.
\ee
(Here and in what follows, we use Roman letters $i$, $j$, \ldots, for particle indices.) We could be
more general by introducing an anisotropic mass in the kinetic term; this modification is simple to make
and will not be done explicitly. In general we do not assume the interaction potential $V$ is rotationally
invariant. Then we introduce a spatially-uniform strain $\Lambda$, that is a linear transformation in
the coordinates (with matrix $\Lambda_{\mu\nu}$), and the corresponding inverse transformation of the
conjugate momenta\cite{Read2010}. We then allow $\Lambda$ to depend on time $t$. This transformation
is viewed actively, as changing the Hamiltonian of the system, relative to the same variables
$\mathbf{x}^i$, $\mathbf{p}^i$. The Hamiltonian for this system is then
\begin{equation}
H_\Lambda(t)=\frac{g^{\mu\nu}(t)}{2m}\sum_i{p^i_\mu p^i_\nu}+\frac{1}{2}\sum_{i\neq
j}{V\left(\Lambda^T(t)(\mathbf{x}^i-\mathbf{x}^j)\right)}, \label{hammetric}
\end{equation}
where
\begin{align}
g_{\mu\nu}(t)&=\Lambda_{\mu\alpha}(t)\Lambda_{\nu\alpha}(t), \nonumber \\
g^{\mu\nu}(t)&=\Lambda^{-1}_{\alpha\mu}(t)\Lambda^{-1}_{\alpha\nu}(t).
\end{align}
We also define the matrix $\lambda_{\mu\nu}$ by
\be
\Lambda=e^\lambda
\ee
as matrices. Here $g_{\mu\nu}$ is the metric, and $g^{\mu\nu}$ is the inverse metric. If the Hamiltonian
is rotationally invariant, then the strain enters only through the metric $g_{\mu\nu}$ and its inverse.
We shall work primarily in units where $\hbar=e=1$, unless otherwise stated.

The strain transformation is canonical (and in fact does not mix coordinates and momenta), so it can
be implemented by a unitary transformation $S(t)$ for each $t$:
\begin{equation}
H_\Lambda(t)=S(t) H_0^{(\mathbf{x})} S^{-1}(t). \label{straintrans}
\end{equation}
We can parametrize the strain transformation $S(t)$ in terms of generators $J_{\mu\nu}$ such that
\begin{equation}
S(t)=\exp(-i\lambda_{\mu\nu}(t)J_{\mu\nu}). \label{finitestrain}
\end{equation}
Inserting this into Eq.~(\ref{straintrans}) we find that the self-adjoint strain generators $J$ must obey
\begin{align}
i\left[J_{\mu\nu},p^j_{\alpha}\right]&=\delta_{\alpha\mu}p^j_\nu, \nonumber \\
i\left[J_{\mu\nu},x^j_{\alpha}\right]&=-\delta_{\alpha\nu}x^j_\mu. \label{vartrans}
\end{align}
It follows that the strain generators must also satisfy
\begin{equation}
i\left[J_{\mu\nu},J_{\alpha\beta}\right]=\delta_{\mu\beta}J_{\alpha\nu}-\delta_{\nu\alpha}J_{\mu\beta},
\label{straincomm}
\end{equation}
which are the commutation relations of the Lie algebra $\mathfrak{gl}(d,\mathbb{R})$, associated with
the group GL$(d,\mathbb{R})$ of linear coordinate transformations. In particular, the antisymmetric part
of $J_{\mu\nu}$ is simply $-1/2$ times the angular momentum operator; for example in three dimensions,
\be
L_\sigma = -\epsilon_{\sigma \mu \nu} J_{\mu \nu}.
\ee
We can satisfy Eqs.~(\ref{vartrans}) and (\ref{straincomm}) by choosing \cite{Read2010}
\begin{align}
J_{\mu\nu}&=-\frac{1}{2}\sum_i{\left\{x^i_\mu,p^i_\nu\right\}}. \label{jnofield}
\end{align}

With $H_\Lambda(t)$ as the starting point, we now use a similar strain transformation but with a
different point of view. We make a time-dependent change of variables from $\mathbf{x}^i$, $\mathbf{p}^i$
(which we term the $\mathbf{x}$ variables) to $\mathbf{X}^i=\Lambda^T(t) \mathbf{x}^i$ and
$\mathbf{P}^i=\Lambda^{-1}(t)\mathbf{p}^i$ (which we term the $\mathbf{X}$ variables), which is again a
canonical transformation implemented by the same $S$ is defined as above: $\Lambda^T(t)\mathbf{x}^i=S(t)
\mathbf{x}^i S^{-1}(t)$, and similarly for $\mathbf{p}^i$.
In the Heisenberg picture, the time-dependence of operators is determined by the Heisenberg equation of
motion, for example for an operator $A$ in the $\mathbf{x}$ variables
\begin{equation}
i\frac{ dA}{dt}=[A,H_\Lambda(t)] +i\frac{\partial A}{\partial t}
\end{equation}
where $\partial A/\partial t$ means the derivative acting on the explicit time dependence of $A$. This
equation of motion requires a choice of the canonical variables used (here $\mathbf{x}^i$,
$\mathbf{p}^i$), which are viewed as having no explicit time dependence. But then due to the
time-dependence of the change of variables, when the $\mathbf{X}$ variables are viewed as having no
explicit time dependence, the resulting equations of motion will not agree with simple changing
variable unless an additional term is included in the Hamiltonian. For an operator $B$, using the
$\mathbf{X}$ variables, one has
\begin{equation}
i\frac{ dB}{dt}=[B,H] +i\frac{\partial B}{\partial t},
\end{equation}
where
\begin{align}
H&=H_0^{(\mathbf{X})}+H_1,\\
H_1&= -i\left(\frac{\partial S}{\partial t}\right)_{\mathbf{x},\mathbf{p}}S^{-1},
\end{align}
in which $H_0^{(\mathbf{X})}=H_\Lambda(t)$ is viewed as a function of $\mathbf{X}^i$ and $\mathbf{P}^i$,
and so is the same functional form as $H_0^{(\mathbf{x})}$ above, but with $\mathbf{x}^i$ and
$\mathbf{p}^i$ replaced by $\mathbf{X}^i$ and $\mathbf{P}^i$.
In the partial (explicit) time derivative of $S$, $\mathbf{x}^i$ and $\mathbf{p}^i$ are to be viewed
as time-independent. However, one can show that
\be
-i\left(\frac{\partial S}{\partial
t}\right)_{\mathbf{x},\mathbf{p}}S^{-1}(t)=-iS^{-1}(t)\left(\frac{\partial
S}{\partial t}\right)_{\mathbf{X},\mathbf{P}}.
\ee
$S$ can be expressed in terms of strain generators for the $\mathbf{X}$ variables,
\be
J_{\mu\nu}^{(\mathbf{X})}=-\half\sum_i\{X^i_\mu,P^i_\nu\},
\ee
which are related to $J_{\mu\nu}$ by a linear transformation. The change of canonical variables is
similar to that which occurs when passing to a rotating frame in mechanics, which in fact is a particular
case of the above derivation.

As we will be interested only in linear response, we can expand the perturbation $-iS^{-1}\frac{\partial
S}{\partial t}$ to first order in $\lambda$ to find that in the $\mathbf{X}$ variables the term $H_1$ is
\be
H_1 =-\frac{\partial\lambda_{\mu\nu}}{\partial t}J_{\mu\nu}.
\ee
(This form of a perturbation to a rate of strain is fairly common in the literature, see e.g.\ Ref.\
\onlinecite{Taylor2010}.)
At the same time, the distinction of $\mathbf{x}$ and $\mathbf{X}$ can be dropped, and we usually use
$\mathbf{x}$ variables to lighten notation.
We have thus mapped the Hamiltonian for a system with a time varying metric to that of a system in a fixed
Euclidean metric, with a perturbation generating time-dependent deformations. The viscosity can now be
computed as the response of an appropriately defined stress tensor to this strain perturbation. This is
analogous to the computation of the conductivity $\sigma_{\mu\nu}$, where one can consider either the
response of the current to a time-varying vector potential, or --- after an appropriate gauge
transformation --- the response of the current to a scalar potential.

To define the stress tensor, let us start with the continuity equation for momentum density
$\mathbf{g}(\mathbf{x})$, defined in zero magnetic field and in the absence of strain (and so using
$\mathbf{x}$ variables and Hamiltonian $H_0^{(\mathbf{x})}$) by
\begin{equation}
  \mathbf{g}(\mathbf{x})=\frac{1}{2}\sum_i\left\{\mathbf{p}^i,\delta(\mathbf{x}-\mathbf{x}^i)\right\}.
\end{equation}
In the absence of other external forces, the continuity equation reads
\begin{equation}
\frac {\partial g_\nu(\mathbf{x},t)}{\partial t}+\partial_\mu \tau _{\mu\nu}^{(0)}(\mathbf{x},t)=0,
\label{contnofield}
\end{equation}
where $\tau_{\mu\nu}^{(0)}(\mathbf{x})$ is the stress tensor operator in the absence of strain, and
$\partial_\mu=\partial/\partial x_\mu$.
A standard expression for $\tau_{\mu\nu}^{(0)}$ for the Hamiltonian $H_0^{(\mathbf{x})}$ is the 
Irving-Kirkwood form \cite{Irving1950} (also used in Refs.\
\onlinecite{Forster1975,Cooper1997}); we derive it within our framework in Appendix \ref{app_stress}.
We note that the continuity equation does not completely determine $\tau_{\mu\nu}^{(0)}$, because any
operator with vanishing divergence (such as a curl of something) could be added to $\tau$ without
violating the equation.
This should not affect physical results, and in particular will not be an issue when the stress tensor
is integrated over all space, as it often will be when the viscosity is calculated.

 Because we are interested in the long-wavelength behavior of the stress tensor, we will examine
 Eq.~(\ref{contnofield}) in Fourier space. To leading order in wavevector $\mathbf{q}$, we have
\begin{equation}
\frac{\partial}{\partial
t}\left(\sum_j\left(p_\nu^j-\frac{iq_\mu}{2}\left\{x_\mu^j,p_\nu^j\right\}\right)\right)+iq_\mu \tau
_{\mu\nu}^{(0)}(\mathbf{q}=0)=0. \nonumber
\end{equation}
We define $T_{\mu\nu}^{(0)}=\tau_{\mu\nu}^{(0)}(\mathbf{q}=0)$ or equivalently $T_{\mu\nu}^{(0)}=\int d^dx
\, \tau_{\mu\nu}^{(0)}(\mathbf{x})$, the integrated stress tensor.
Then, because $H_0^{(\mathbf{x})}$ is translationally invariant,
\begin{equation}
T_{\mu\nu}^{(0)}=-\frac{\partial J_{\mu\nu}}{\partial t}=-i\left[H_0,J_{\mu\nu}\right]. \label{momflux}
\end{equation}
This relation between the stress and the strain generators is a central result of this section. (As an
example, for non-interacting particles, $T_{\mu\nu}^{(0)}=\sum_i p^i_\mu p^i_\nu/m$, the momentum flux.)
Since the antisymmetric part of $J_{\mu\nu}$ is the angular momentum, it is clear that the integrated
stress tensor $T_{\mu\nu}^{(0)}$ is symmetric (that is, the antisymmetric part is zero) when the
Hamiltonian is rotationally invariant. We can also view the result in another way: referring to eq.\
(\ref{straintrans}), we see that
\be
T_{\mu\nu}^{(0)}=-\left.\frac{\partial H_\Lambda}{\partial
\lambda_{\mu\nu}}\right|_{\Lambda=I}.\label{stress_str_rel}
\ee
This is consistent with the idea that the stress tensor can be obtained by varying the Hamiltonian with
respect to the metric, but here is slightly more general as we do not assume rotational invariance. In
the rotationally-invariant case, we can also write
\be
T_{\mu\nu}^{(0)}=-2\frac{\partial H_\Lambda}{\partial g_{\mu\nu}},
\label{stress_met_rel}
\ee
which again is clearly symmetric, because $g_{\mu\nu}$ is so
(we note that the corresponding expression in Ref.\ \onlinecite{Read2010} has the wrong sign).

Strictly speaking, because the Hamiltonian $H_0$ contains no confining potential, for repulsive or for
no interactions, there is no normalizable ground state. We will nonetheless go ahead somewhat informally
(as in many papers in the literature), and evaluate the response in a normalizable state in which
the particles occupy a region (or ``box'') of volume $L^d$, and we will treat the state as if it were
an eigenstate. Such expectation values will be written as $\langle\ldots \rangle_0$. (A similar approach
can be used for non-zero temperature, and most of the following is also valid for that case.) In a large
system (i.e.\ as $L$ and $N$ become large with the density remaining fixed---we refer to this as the
thermodynamic limit) this procedure may possibly be justified over intermediate time scales. In any case,
our results will be seen to make sense. More formally, and completely generally, we can add to
$H_\Lambda(t)$ an explicit ``confining'' potential term $U$,
\be
U=\sum_i u(\mathbf{x}^i).
\ee
The single-particle potential $u(\mathbf{x})$ has no $\Lambda$ dependence, and can be used to represent
a box (say, by using a hard-wall potential) the shape of which is fixed in $\mathbf{x}$ variables even
as $\Lambda$ varies. Then in the $\mathbf{X}$ variables, the potential becomes $U=\sum_i
u(\Lambda^{T-1}\mathbf{X})$, so the shape of the box varies with $\Lambda$ (these conventions agree
with Ref.\ \onlinecite{Read2010}, except that there a periodic boundary condition was used instead, and
the shape was chosen to be a square in the $\mathbf{x}$ variables). The potential modifies the
continuity equation for momentum density by the inclusion of an external force due to the potential,
while the stress tensor remains unchanged. We can also extend our formalism to include the potential,
in such a way that an analog of eq.\ (\ref{momflux}) holds, with $H_0+U$ in place of $H_0$, by modifying
the strain generators. In this way, we can justify all the later results that we present here less
formally by ignoring the potential that confines the system to a finite volume. This is sketched
in Appendix~\ref{app}.

Returning to the original approach without a confining potential, at zeroth order in strain, we can
consider the expectation of the stress. Then use of eq.~(\ref{stress_str_rel}) and the
Hellmann-Feynman theorem gives
\begin{equation}
\left<T_{\mu\nu}^{(0)}\right>_0=-\left(\left.\frac{\partial E(\lambda)}{\partial
\lambda_{\mu\nu}}\right|_{\Lambda=I}\right)_N,
\end{equation}
where $E(\lambda)=\langle H_\Lambda\rangle_0$ is the energy eigenvalue of the $\lambda$-dependent
ground state of $H_\Lambda$ (for time-independent $\Lambda$), and we choose to emphasize that
particle number $N$ is held fixed. By evaluating $E$ in the $\mathbf{X}$ variables, in which a
time-independent $\Lambda$ enters only through the shape and size of the box, we realize that when
the state is a homogeneous fluid, the ground state energy depends on the volume of the box but not on
its shape (in leading order in the thermodynamic limit). As $-\partial E/\partial (L^d)=P$, the pressure,
we have therefore
\be
\langle T_{\mu\nu}^{(0)}\rangle_0=PL^d\delta_{\mu\nu},\label{Tmunuexp}
\ee
(to leading order) as would be expected from the standard result,
$\langle\tau_{\mu\nu}^{(0)}(\mathbf{x})\rangle_0=P\delta_{\mu\nu}$. For non-zero temperature, we obtain
the same result, as the derivative of the expectation of energy is taken with the probabilities held
fixed, in order to use the Hellmann-Feynman theorem under the thermal average. This implies that the
derivative is taken with the entropy fixed, in agreement with the thermodynamic definition of pressure.
(A similar method is used to find an expression for the pressure in Ref.\ \onlinecite{Taylor2010}.)
There are certain subtleties about the argument for this result (for either zero or non-zero
temperature), which we will briefly discuss later in Sec.\ \ref{nondiv}, however the result is still
valid. We emphasize that we did not use rotational invariance to obtain it.

We also point out that if we use eq.\ (\ref{momflux}) and then argue that the ground state is an
eigenstate of $H_0$, we will conclude that all components of $\langle T_{\mu\nu}^{(0)}\rangle_0$
vanish. This is incorrect in general because in the absence of the potential $U$, the normalizable
state that we use is not an eigenstate of $H_0$ for repulsive or no interactions, while if the potential
$U$ is included, there is a normalizable ground (energy eigen-) state of $H_0+U$, but again the
argument is blocked.

In order to derive linear response of the stress to a time-dependent strain, we will need the stress to
next order in the strain. For this it is convenient to notice that $H_\Lambda$ is translationally
invariant, and so the sum of $\mathbf{p}^i$ is conserved. This means that the density of $\mathbf{p}$
momentum obeys a continuity equation, even in the presence of the time-dependent strain $\Lambda(t)$:
\begin{equation}
\frac {\partial g_\nu(\mathbf{x},t)}{\partial t}+\partial_\mu \tau _{\mu\nu}(\mathbf{x},t)=0.
\end{equation}
Here $\tau_{\mu\nu}$ depends on $\Lambda$; to illustrate the form of this, note that the momentum flux
term part of $\tau_{\mu\nu}$ integrated over space is
\be
T_{\mu\nu}=\sum_i \Lambda^{-1}_{\alpha\mu}\Lambda^{-1}_{\alpha\beta}\frac{p^i_\beta p^i_\nu}{m} +\ldots
\ee
where the omitted terms refer to interactions. Following the same derivation as before, we see that
\be
T_{\mu\nu}=-i[H_\Lambda,J_{\mu\nu}].
\ee
Then expanding to order $\lambda$, we have
\be
T_{\mu\nu}=-i[H_0^{(\mathbf{x})},J_{\mu\nu}]+
\lambda_{\alpha\beta}[[H_0^{(\mathbf{x})},J_{\alpha\beta}],J_{\mu\nu}]+O(\lambda^2).
\ee
However, when we calculate response in the $\mathbf{X}$ variables, we will use the stress tensor in
those variables, which is defined by transformation of tensors, so
\be
T_{\mu\nu}^{(\mathbf{X})}=\Lambda_{\alpha\mu}\Lambda^{-1}_{\nu\beta}T_{\alpha\beta}.\label{tmunuX}
\ee
In terms of the $\mathbf{X}$ variables, this has the form $T_{\mu\nu}^{(\mathbf{X})}=\sum_i P^i_\mu
P^i_\nu/m+\ldots$ (where the omitted terms are from interactions), and coincides with $T_{\mu\nu}^{(0)}$
with $\mathbf{X}^i$ and $\mathbf{P}^i$ in place of $\mathbf{x}^i$ and $\mathbf{p}^i$.
In the $\mathbf{x}$ variables, it can be expanded as
\be
T_{\mu\nu}^{(\mathbf{X})}=-i[H_0^{(\mathbf{x})},J_{\mu\nu}]+
\lambda_{\alpha\beta}[[H_0^{(\mathbf{x})},J_{\mu\nu}],J_{\alpha\beta}]
+O(\lambda^2),\label{tmunuXexp}
\ee
in which the order in the double commutator is reversed, compared with $T_{\mu\nu}$. It is this
integrated stress $T_{\mu\nu}^{(\mathbf{X})}$ that we believe constitutes a natural starting point for
the calculation of viscosity. However, we will see that in a homogeneous fluid, the difference in results
from using either $T_{\mu\nu}$ or $T_{\mu\nu}^{(\mathbf{X})}$ is negligible in the thermodynamic limit.


\subsection{Nonzero magnetic field in two dimensions}\label{sec22}

We now turn to the problem of generalizing the preceding set-up in the presence of an external magnetic
field $B$; we concentrate on two dimensions. Most of the work is in finding the strain generators for this
case; we will do so by considering separately the cases of pure shear deformations ($\det \Lambda(t)=1$)
and pure dilations ($\Lambda_{\mu\nu}(t)\propto\delta_{\mu\nu}$). The unstrained Hamiltonian can be taken
to be
\be
H_0^{(\mathbf{x})}=\frac{1}{2m}\sum_{i}\pi^i_\mu \pi^i_\mu+\frac{1}{2}\sum_{i\neq
j}V(\mathbf{x}^i-\mathbf{x}^j),
\ee
where
\begin{align}
\hbox{\boldmath$\pi$}^i&=\mathbf{p}^i-\mathbf{A}(\mathbf{x}^i), \nonumber\\
\left[x_\mu^j,\pi_\nu^k\right]&=i\delta_{jk}\delta_{\mu\nu}, \nonumber\\
\left[\pi_\mu^j,\pi_\nu^k\right]&=iB\delta_{jk}\epsilon_{\mu\nu}.
\end{align}
 We usually assume that the interaction $V$ is independent of the magnetic field. Following the procedure
 of Section~\ref{sec21} above, we seek strain generators $J_{\mu\nu}$ satisfying
\begin{align}
i\left[J_{\mu\nu},x^j_\alpha\right]&= -\delta_{\alpha\nu}x^j_\mu,  \nonumber \\
i\left[J_{\mu\nu},\pi^j_\alpha\right]&=\delta_{\mu\alpha}\pi^j_\nu. \label{eqprime2}
\end{align}

First, let us consider the case of pure shear deformations. The Hamiltonian for the system in the
presence of the time-varying strain with $\det \Lambda=1$ is given by
\begin{equation}
H_\Lambda(t)=\frac{1}{2m}\sum_{i}g^{\mu\nu}(t)\pi^i_\mu \pi^i_\nu+\frac{1}{2}\sum_{i\neq
j}V\left(\Lambda^T(t)(\mathbf{x}^i-\mathbf{x}^j)\right). \label{maghamt}
\end{equation}
The condition $\det \Lambda=1$ implies that $\mathrm{tr}\lambda=0$.
If we attempt naively to generalize the generators from zero magnetic field by taking
\begin{equation}
\tilde{J}^{\mathrm{sh}}_{\mu\nu}=-\sum_i\frac{1}{2}\left\{x^i_\mu,\pi^i_\nu\right\},
\end{equation}
we find that
\begin{align}
i\left[\tilde{J}^{\mathrm{sh}}_{\mu\nu},x^j_\alpha\right]=-\delta_{\alpha\nu}x^j_\mu,\nonumber \\
i\left[\tilde{J}^{\mathrm{sh}}_{\mu\nu},\pi^j_\alpha\right]=\delta_{\mu\alpha}\pi^j_\nu+B\epsilon_{\nu\alpha}
x^j_\mu.\nonumber
\end{align}
If we define
\begin{equation}
\tilde{S}=\exp\left(-i\mathrm{tr}(\lambda^T\tilde{J}^{\mathrm{sh}})\right), \nonumber
\end{equation}
then these relations imply that the coordinates $\mathbf{x}^i$ transform correctly, while the momenta
$\hbox{\boldmath$\pi$}^i$ transform as
\begin{align}
\tilde{S}^{\dag}\pi^i_\mu\tilde{S}=\pi^i_\mu+\lambda_{\mu\nu}\pi_\nu+B\epsilon_{\alpha\mu}\lambda_{\nu\alpha}
x^i_\nu+\mathcal{O}(\lambda^2). \nonumber
\end{align}
But note that if $\lambda_{\mu\nu}$ is traceless, then the curl of the extra term containing $B$ is $0$,
and so it is just a $\lambda$ dependent gauge transformation. Thus, we see that $\tilde{J}$ generates
the desired strain transformation along with the gauge transformation
\begin{equation}
A_\mu\rightarrow A_\mu-B\lambda_{\alpha \nu}\epsilon_{\nu\mu}x_{\alpha}. \nonumber
\end{equation}
An integration of this gauge term shows that, since $\lambda$ is traceless, the gauge transformation
is given by
\begin{align}
\phi&=-\frac{B}{2}\lambda_{\mu\nu}\epsilon_{\nu\alpha}x_\mu x_\alpha, \nonumber \\
A_\mu&\rightarrow A_\mu+\partial_\mu\phi. \nonumber
\end{align}
In order to remove this unwanted gauge transformation, as well as to make this $J_{\mu\nu}$ traceless,
we define
\begin{align}
J^{\mathrm{sh}}_{\mu\nu}&=\tilde{J}^{\mathrm{sh}}_{\mu\nu}-\frac{1}{2}\mathrm{tr}(\tilde{J}^{\mathrm{sh}})
\delta_{\mu\nu}+\frac{B}{2}\sum_i\epsilon_{\nu\alpha}x^i_\mu x^i_{\alpha}\label{eqmagstraintless}\\
&=\sum_i\left(-\frac{1}{2}\left\{x^i_\mu,\pi^i_\nu\right\}+\frac{1}{4}\delta_{\mu\nu}\left\{x^i_\alpha,
\pi^i_\alpha\right\}+\frac{B}{2}\epsilon_{\nu\alpha}x^i_\mu x^i_{\alpha}\right).\nonumber
\end{align}
A short calculation shows that $J^{\mathrm{sh}}_{\mu\nu}$ defined in this way reproduces the traceless
part of the transformations~(\ref{eqprime2}), and therefore it is the desired traceless strain operator.

Next, we consider the case of a pure dilation. We will soon see that we must define strains of the system
so that they rescale the magnetic field in such a way that the magnetic flux $\Phi=L^2B$ through the
system stays fixed while its shape is strained. With a fixed particle number $N$, and defining the
filling factor for the region of area $L^2$ occupied by the particles as $\nu=2\pi N/(BL^2)$ (as usual),
this has the effect that we consider deformations at fixed filling factor $\nu$. For a dilation, we have
\begin{equation}
\lambda_{\mu\nu}=\frac{1}{2}\mathrm{tr}(\lambda)\delta_{\mu\nu}. \nonumber
\end{equation}
In accordance with Eq.~(\ref{eqprime2}), we seek a dilation generator $K$ satisfying
\begin{align}
i\left[K,x^j_\mu\right]&=-x_\mu^j, \nonumber \\
i\left[K,\pi^j_\mu\right]&=\pi_\mu^j. \label{dilationcomm}
\end{align}
Writing $\pi_\mu^i=p^i_\mu-A_\mu(\mathbf{x}^i)$, we see that these imply that
\begin{align}
i\left[K,x^j_\mu\right]&=-x_\mu^j, \nonumber \\
i\left[K,p^j_\mu\right]&=p_\mu^j, \nonumber \\
i\left[K,A_\mu(\mathbf{x}^j)\right]&=A_\mu(\mathbf{x}^j). \label{magdilcomms}
\end{align}
At first glance, it appears that we are in a quandary --- the first and third of these equations are
inconsistent, unless the magnetic field strength $B$ also transforms under the action of $K$. To
accomplish this, we promote the field strength $B$ to a dynamical variable, represented by an operator
$\mathcal{B}$, whose eigenvalues are values $B$. This enlarges the Hilbert space of the system to include
states with different magnetic fields; we will continue to consider only $B>0$. At this point, we are
naturally motivated to introduce the momentum $\mathcal{P}$ conjugate to $\mathcal{B}$, such that
\begin{equation}
\left[\mathcal{B},\mathcal{P}\right]=i.
\end{equation}
$\mathcal{B}$ and $\mathcal{P}$ commute with $\mathbf{x}^i$ and $\mathbf{p}^i$ for all $i$.
Note that the vector potential is now a function of $\mathcal{B}$, but is independent of $\mathcal{P}$.
We then have
\be
\pi^i_\mu=p^i_\mu-A_\mu(\mathbf{x}^i,\mathcal{B}),
\ee
and
\be
\left[\pi_\mu^j,\pi_\nu^k\right]=i\mathcal{B}\delta_{jk}\epsilon_{\mu\nu}.
\ee
For consistency with eq.\ (\ref{dilationcomm}), we must also have
\be
i[K,\mathcal{B}]=2\mathcal{B}.
\label{K_B_comm}
\ee
The Hamiltonian~(\ref{maghamt}) now becomes
\begin{align}
H_\Lambda(t)=&\frac{1}{2m}\sum_{i}g^{\mu\nu}(t)(p^i_\mu-A_\mu(\mathbf{x}^i,\mathcal{B}))
(p^i_\nu-A_\nu(\mathbf{x}^i,\mathcal{B}))\nonumber\\
&+\frac{1}{2}\sum_{i\neq j}V\left(\Lambda^T(t)(\mathbf{x}^i-\mathbf{x}^j)\right), \label{maghamfieldop}
\end{align}
where we have expressed $\hbox{\boldmath$\pi$}$ in terms of the canonical momentum and vector potential
in order to make explicit the dependence of the Hamiltonian on $\mathcal{B}$. (Again, an anisotropic
mass can be introduced in the kinetic terms if desired.) Since $H$ is independent of $\mathcal{P}$,
states with given eigenvalues $B$ of $\mathcal{B}$ retain those values for all times. Eigenstates of
$\mathcal{B}$ are not normalizable, however, instead we can use normalized packets with very small width
in $B$ to calculate expectations in linear response.

A gauge-invariant dilation generator satisfying the commutation relations~(\ref{dilationcomm})
and (\ref{K_B_comm}) is then given by
\be
K=-\frac{1}{2}\sum_{i}\left\{x_\mu^i,\pi_\mu^i\right\}+ \left
\{\mathcal{B},\Xi(\{\mathbf{x}^i\},\mathcal{B})\right\}, \label{dilationgen}
\ee
where we have introduced the ``kinetic momentum''
$\Xi(\{\mathbf{x}^i\},\mathcal{B})=\mathcal{P}-\sum_i\mathcal{A}(\mathbf{x}^i,B)$ conjugate to
$\mathcal{B}$, and $\{\mathbf{x}^i\}=\{\mathbf{x}^i:i=1,\ldots,N\}$ is the set of all $\mathbf{x}^i$'s.
Under a gauge transformation generated by some scalar function $\phi(\mathbf{x},B)$ we have
\begin{align}
A_\mu(\mathbf{x},B)&\rightarrow A_\mu+\partial_\mu\phi(\mathbf{x},B), \nonumber \\
\mathcal{A}(\mathbf{x},B)&\rightarrow\mathcal{A}+ \partial_B\phi(\mathbf{x},B). \label{gaugeconds}
\end{align}
To complete our definition of $K$, we specify that if the gauge choice is the symmetric gauge for all $B$,
that is $A_\mu(\mathbf{x})=-\half B\epsilon_{\mu\nu}x_\nu$ (which is preserved by $\mathbf{x}$-independent
gauge transformations), then $\mathcal{A}$ is a function of $B$ only, independent of $\mathbf{x}$. Then
\be
[\pi_\mu^j,\{\mathcal{B},\Xi(\{\mathbf{x}^k\},\mathcal{B})\}]=i
\epsilon_{\mu\nu}\mathcal{B}x_\nu^j,\label{pi_Xi_comm}
\ee
and this result of course is gauge covariant.

Putting it all together, we have thus shown that
\begin{align}
 J_{\mu\nu}&=J^{\mathrm{sh}}_{\mu\nu}+\half K\delta_{\mu\nu} \nonumber \\
 &=\sum_i\left[\vphantom{\frac{1}{N}}\half\left(-\left\{x_\mu^i,\pi_\nu^i\right\}
 +\mathcal{B}\epsilon_{\nu\alpha}x_\mu^ix_\alpha^i\right)\right]\nonumber\\
 &\hbox{}\quad+\frac{1}{2}\delta_{\mu\nu}\left\{\mathcal{B},\Xi(\{\mathbf{x}^i\},\mathcal{B})\right\}
 \label{magjcompact}
 \end{align}
gives the strain generators for two dimensional systems in a magnetic field, which satisfy eqs.\
(\ref{eqprime2}), (\ref{K_B_comm}). These generators also satisfy the $\mathfrak{gl}(2,\mathbb{R})$
commutation relations~(\ref{straincomm}), and the antisymmetric part is $-1/2$ times the gauge-invariant
rotation generator (``angular momentum''). We note also that we consider only the range $B>0$. It can then
be shown that the operator $\left\{\mathcal{B},\Xi\right\}$ is a bona-fide self-adjoint operator, which is
\emph{not} the case for the operator $\Xi$ alone, on this range.

As in Section \ref{sec21}, we can now apply the canonical transformation
$S=\exp(-i\lambda_{\mu\nu}J_{\mu\nu})$ to transform to the Hamiltonian in $\mathbf{X}$ variables, in
which the corresponding kinetic momenta are $\mathbf{\Pi}^i$ (the variables $\mathcal{B}$ and $\mathcal{P}$
in $\mathbf{X}$ variables should be distinguished from those in the $\mathbf{x}$ variables also, but we
will not introduce additional notation for this; at this point, it should be clear from the context). We
find, as in the zero magnetic field case, that the system with time-varying strain is equivalent to one
with Hamiltonian
 \begin{align}
 H&=\frac{1}{2m}\sum_i\Pi_\mu^i\Pi_\mu^i+\sum_{i\neq
 j}V(\mathbf{X}^i-\mathbf{X}^j)-\frac{\partial\lambda_{\mu\nu}}{\partial t}J_{\mu\nu}\nonumber \\
 &=H_0^{(\mathbf{X})}+H_1,\label{magtransformedham}
 \end{align}
 to first order in $\lambda$.

As above, we can obtain an expression for the stress tensor in the presence of a magnetic field by
 considering the continuity equation for the kinetic momentum density (with time-dependence obtained from
 $H_0^{(\mathbf{x})}$, which is now generalized to include $\mathcal{B}$), which reads
\begin{equation}
\frac{\partial g_\nu(\mathbf{x})}{\partial t}+\partial_\mu \tau
_{\mu\nu}^{(0)}(\mathbf{x})=\frac{\mathcal{B}}{m}\epsilon_{\nu\alpha}g_\alpha(\mathbf{x})
\label{fieldcont},
\end{equation}
with the kinetic momentum density given by
\begin{equation}
\mathbf{g}(\mathbf{x})=\frac{1}{2}\sum_i\left\{\hbox{\boldmath$\pi$}^i,\delta(\mathbf{x}-\mathbf{x}^i)\right\}.
\end{equation}
As in the previous section, we Fourier transform this equation, and to first order in wavevector $
\mathbf{q}$ we find (again, $T_{\mu\nu}^{(0)}=\int d^2x\, \tau_{\mu\nu}^{(0)}(\mathbf{x}$))
\begin{align}
\frac{\partial}{\partial
t}&\left(\sum_j\left(\pi_\nu^j-\frac{iq_\mu}{2}\left\{x_\mu^j,\pi_\nu^j\right\}\right)\right)+iq_\mu T
_{\mu\nu}^{(0)}=\nonumber\\
&=\frac{\mathcal{B}}{m}\epsilon_{\nu\alpha}\sum_j\left(\pi_\mu^j-\frac{iq_\mu}{2}\left\{x_\mu^j,
\pi_\alpha^j\right\}\right).
\end{align}
Now, the first term on either side cancels since only the Lorentz force breaks conservation of total
kinetic momentum. Thus
\begin{equation}
T _{\mu\nu}^{(0)}=\frac{1}{2}\sum_i\left(\frac{\partial}{\partial
t}\left\{x_\mu^i,\pi_\nu^i\right\}-\frac{\mathcal{B}}{m}\epsilon_{\nu\alpha}\left\{x_\mu^i,
\pi_\alpha^i\right\}\right).
\end{equation}
We see that the term in the time derivative matches the first term in Eq.~(\ref{magjcompact}) for the
strain generator. Then we have
\begin{align}
\frac{\partial}{\partial
t}&\left(J_{\mu\nu}+\frac{1}{2}\sum_i\left\{x_\mu^i,\pi_\nu^i\right\}\right)=\nonumber \\
&=\frac{\mathcal{B}}{2m}\sum_i\left(\epsilon_{\nu\alpha}\left(\pi_\mu^ix_\alpha^i+x_\mu^i\pi_\alpha^i\right)+
\delta_{\mu\nu}\epsilon_{\beta\alpha}x_\beta^i\pi_\alpha^i\right) \nonumber \\
&=\frac{\mathcal{B}}{2m}\epsilon_{\nu\alpha}\sum_i\left\{x_\mu^i,\pi_\alpha^i\right\}.
\end{align}
Thus finally
\begin{equation}
T _{\mu\nu}^{(0)}=-\frac{\partial J_{\mu\nu}}{\partial t}=-i\left[H_0^{(\mathbf{x})},J_{\mu\nu}\right],
\label{momflux2}
\end{equation}
just as in Eq.~(\ref{momflux}) above. The definitions for $T_{\mu\nu}$ and for $T_{\mu\nu}^{(\mathbf{X})}$
and their expansions to order $\Lambda$ in $\mathbf{x}$ variables are the same as in zero magnetic field.

Again, at zeroth order in the strain, we can express the expectation of the stress in terms of
thermodynamic properties, if we assume the state is a normalizable ground (and eigen-) state of $H_0$
for each value of $\mathcal{B}$, and as a function of $\mathcal{B}$ is concentrated near a value $B$. In
the case of a two-dimensional system with a magnetic field, the field itself provides confinement,
so normalizable eigenstates exist for given $B$, and we may consider a disk of fluid that covers an
area $L^2$.
Then we expect that the average total stress $\langle T_{\mu\nu}^{(0)}\rangle_0 $ can be decomposed into a
pressure term and a magnetization term\cite{Cooper1997} as
\begin{equation}
\langle T_{\mu\nu}^{(0)}\rangle_0=\delta_{\mu\nu}(PL^2-MB),
\end{equation}
where $P=-(\partial E/\partial (L^2))_{N,B}$ is the pressure, and $M=-(\partial E/\partial B)_{N,L^2}$
is the total magnetization. We define the internal pressure\cite{Cooper1997} as
\begin{equation}
P_{\mathrm{int}}=P-\frac{MB}{L^2}
\ee
or as
\be
P_{\mathrm{int}}=-\left(\frac{\partial E}{\partial (L^2)}\right)_{\nu,N},
\end{equation}
where the partial derivative is at fixed $N$ and fixed filling factor, as we specified before. As pointed
out by Cooper et.\ al.\cite{Cooper1997}, in a homogeneous fluid, the usual pressure $P$ is the change
in energy under a change in the size of the box (at fixed $B$), and so includes a contribution from
the Lorentz force acting on the boundary current that is related to the magnetization. This part is
removed by defining the internal pressure, or equivalently by taking the derivative with the flux
through the system held fixed.

In our framework, derivatives with respect to strain are taken in exactly that way. If we calculate the
expectation of the stress using eq.\ (\ref{momflux2}), then we must be careful to recall that the state is
a wavepacket in $B$, and so not an energy eigenstate of $H_0$, because the energy for given $B$ generally
depends on $B$. Now when calculating the expectation of an operator that may depend on $B$, but not on
$\mathcal{P}$ (i.e.\ does not contain $\mathcal{P}$), we can take the expectation in the $H_0$
eigenstate for each $B$, and then average the result over $B$ using the dependence of the wavefunction
on $B$. For the traceless (or shear) part of $J_{\mu\nu}$, $\mathcal{P}$ does not appear, and so the fact
that the state is an eigenstate of $H_0$ for each $B$ can be used to conclude that the expectation of
$T_{\mu\nu}^{(0)}$ is zero. This cannot be done for the trace, so the result can be non-zero. On the other
hand, the stress, including its trace, is itself independent of $\mathcal{P}$, and so the result is the
average over a small range of $B$ of the result for each $B$, and the latter can be related to the
derivative of energy with the strain (holding $\nu$ and $N$ fixed) using the Hellmann-Feynman theorem just
as in the zero magnetic field case. (This is done in $\mathbf{x}$ variables, using $H_\Lambda$, and we
emphasize that then $\mathbf{x}^i$ and $\mathcal{B}$ are viewed as fixed when taking derivatives with
respect to strain $\lambda$.) Then by averaging over a sufficiently small range of $B$, the result for
the trace is just the internal pressure. Then the full result is
\begin{equation}
\left<T _{\mu\nu}^{(0)}\right>_0=\delta_{\mu\nu}P_{\mathrm{int}}(L)L^2,
\label{exptmunu}
\end{equation}
and this result can be considered as exact, rather than just as the order $L^2$ part as the size goes
to infinity, if $P_{\mathrm{int}}(L)$ is defined in this way at finite $L$, but tends to the
thermodynamic $P_{\mathrm{int}}$ as $L\to\infty$. For non-zero temperature, we should include all
energy eigenstates, weighted by their Gibbs weight, but then a similar difficulty as in the zero
magnetic field case in infinite size reappears: most states are spread over arbitrarily large volumes.
We may deal with this in the same way as in Sec.\ \ref{sec21}, or by including a confining potential as
in Appendix \ref{app}. The result takes the same form as in eq.\ (\ref{exptmunu}).


\section{Kubo formulas for viscosity}\label{sec3}
In this section, to calculate the viscosity, we first compute the retarded response function
$X_{\mu\nu\alpha\beta}$ of the integrated stress tensor $T _{\mu\nu}$ to the perturbation $H_1$ in Sec.\
\ref{sec31}. Initially, we consider separately the zero magnetic field case in any dimension and the case
of nonzero magnetic field in two dimensions.  In Sec.\ \ref{seclocal}, we give the response in the form
of an intensive response function $\chi$, which can be defined in periodic boundary conditions also.
Then in Sec.\ \ref{viscfromX} we relate $X $ and $\chi$ to the viscosity tensor. We show in Sec.\
\ref{nondiv} that a leading part of $\chi$ at low frequency is the inverse compressibility, and finally
in Sec.\ \ref{sumrules} discuss sum rules and positivity properties of $X$ and $\chi$.

\subsection{The response function from strain generators}
\label{sec31}

We work in the $\mathbf{X}$ variables, and calculate the response of the integrated stress
$T_{\mu\nu}^{(\mathbf{X})}$ to the perturbation $H_1$, using linear response theory. Dropping the
superscript $(\mathbf{X})$, the change in the stress tensor to first order is given by
\begin{equation}
\left<T _{\mu\nu}\right>(t)-\left<T _{\mu\nu}\right>_0=-\int_{-\infty}^{t}dt'\,X
_{\mu\nu\alpha\beta}(t-t')\frac{\partial\lambda_{\alpha\beta}(t')}{\partial t'}, \label{stresschange}
\end{equation}
where the retarded response function $X $ is given by
\begin{equation}
X _{\mu\nu\alpha\beta}(t)=-\lim_{\epsilon\rightarrow 0^+}i\Theta(t)\left<\left[T
_{\mu\nu}(t),J_{\alpha\beta}(0)\right]\right>_0e^{-\epsilon t}. \label{chitime}
\end{equation}
These expressions are to be evaluated with $\lambda=0$, so we may now drop the distinction between
$\mathbf{X}$ and $\mathbf{x}$, and corresponding superscripts. The time evolution here is taken with
respect to $H_0$, and the expectation is taken in the unperturbed ground state of $H_0$ (in zero
magnetic field, it is again subject to the same caveats as in section~\ref{sec21}, which are addressed
further in Appendix~\ref{app}). The exponential damping ensures that the system was unperturbed infinitely
far in the past. Fourier transforming Eq.~(\ref{chitime}), we find that in the frequency domain
\begin{equation}
X_{\mu\nu\alpha\beta}(\omega)=-\lim_{\epsilon\rightarrow0^+}i\int_0^{\infty}
{dt\,e^{i\omega^+t}\left<\left[T _{\mu\nu}(t),J_{\alpha\beta}(0)\right]\right>_0}, \label{chifreq1}
\end{equation}
where $\omega^+=\omega+i\epsilon$ (the $\epsilon\to0^+$ limit will be left implicit from here on). We
call this the stress-strain form of the response function. It corresponds directly to the response of
the stress to an applied spatially-uniform rate of strain $\partial\lambda_{\alpha\beta}/\partial t$.
No rotational invariance has been assumed in the derivation.

Using the relation~(\ref{momflux}), we can express Eq.~(\ref{chifreq1}) in two additional  equivalent
forms. The second form of the Kubo formula is the stress-stress form
\begin{align}
X _{\mu\nu\alpha\beta}(\omega)=&\frac{1}{\omega^+}\left(\left<\left[T
_{\mu\nu}(0),J_{\alpha\beta}(0)\right]\right>_0\vphantom{\int}\right. \nonumber \\
&\left.+\int_0^{\infty}{dt\,e^{i\omega^+t}\left<\left[T _{\mu\nu}(t),T _{\alpha\beta}(0)\right]\right>_0}
\right). \label{stressstress}
\end{align}
(In obtaining this, we had to use time-translation invariance of the correlation function in eq.\
(\ref{chifreq1}) to shift the time dependence onto the operator $J_{\alpha\beta}$, and then back after
using the identity (\ref{momflux}) and integrating by parts on $t$.)
The time-integral term is what one might have expected for the response function, as the deformation
of shape couples directly to the stress tensor through the metric. In the additional contact term
(the equal-time commutator term), the coefficient of $1/\omega^+$ is purely imaginary, because it is
the expectation of a commutator of self-adjoint operators. The complete expression is directly analogous
to the standard Kubo formula for conductivity in terms of the current-current response, and the contact
term in eq.~(\ref{stressstress}) is analogous to the diamagnetic conductivity. The latter will be
discussed further in Sec.\ \ref{seclocal} below.

Lastly, we have the strain-strain form of the response function
\begin{align}
X _{\mu\nu\alpha\beta}(\omega)=&-i\left<\left[J_{\mu\nu}(0),J_{\alpha\beta}(0)\right]\right>_0 \nonumber \\
&+\omega^+\int_0^{\infty}{dt\,e^{i\omega^+t}\left<\left[J_{\mu\nu}(t),J_{\alpha\beta}(0)\right]\right>_0}.
\label{strainstrain}
\end{align}
(In this case, the identity (\ref{momflux}) and integration by parts was used on the operator $T_{\mu\nu}$
in eq.\ (\ref{chifreq1}), and time-translation invariance was not invoked.)
This form of the response function is closely connected with the adiabatic formalism for
viscosity\cite{Avron1995,Read2009,Read2010}. For systems with non-degenerate ground states and an energy
gap, the contact term (the equal-time strain-strain commutator term) is the adiabatic curvature associated
with deformation of the metric, and gives the full response as $\omega\to0$. Note this part is real and is
manifestly antisymmetric under exchanging the pair $\mu\nu$ with $\alpha\beta$. It gives the simplest way
to see that the Hall viscosity is connected with the orbital spin density in these cases \cite{Read2010}.
See also Appendix \ref{appC}.

A similar analysis applies in the magnetic field case. Because the equations of motion
Eqs.~(\ref{momflux}) and~(\ref{momflux2}) are functionally identical, the same three forms of Kubo formula
Eqs.~(\ref{chitime}-\ref{strainstrain}) for the response of $T $ to $H_1$ hold even in the presence of a
magnetic field, provided one uses the appropriate strain generator as given in Eq.~(\ref{magjcompact}).

The different forms of the Kubo formula can be viewed as related by use of Ward identities. Generally,
Ward identities are the consequences of symmetries or conservation laws for response or correlation
functions. In our case, the key relation, Eq.~(\ref{momflux}) or~(\ref{momflux2}), was obtained by
expanding the Fourier-transformed continuity equation to first order in wavevector.

There are some technical points about the derivation to discuss. These center around the assumption
of time-translation invariance, which is usually assumed to hold for response functions such as
eq.\ (\ref{chitime}), on the basis that the unperturbed ground state is an eigenstate of the
unperturbed Hamiltonian (or similarly at non-zero temperature, because the thermal weighting
is stationary). As we have mentioned already, for the systems we consider, in zero magnetic field
a normalizable ground state is generally not available unless a confining potential is included; this
is because the system is in infinite volume, so that the strain generators can be defined. In
Appendix \ref{appB}, we show that the preceding Kubo formulas are certainly valid as written in
the presence of non-zero magnetic field, using the formalism of Sec.\ \ref{sec22}, and also more
generally using the formalism of Appendix \ref{app}.

For rotationally-invariant systems, in which $H_0$ commutes with angular momentum (the antisymmetric part
of $J_{\mu\nu}$), the symmetries of $X$ can be read off from the formulas. First, $T_{\mu\nu}$ is
symmetric, and therefore $X_{\mu\nu\alpha\beta}$ is symmetric under $\mu\leftrightarrow\nu$. Next, in
the stress-strain and stress-stress forms the part of $X$ antisymmetric under
$\alpha\leftrightarrow\beta$ also vanishes, using the assumption that the ground state is an
$H_0$ eigenstate (more detailed or careful arguments can be given along lines discussed at the end of
Sec.\ \ref{sumrules} and in Appendix \ref{appB} below). This holds
without assuming the ground state is an angular momentum eigenstate.
Finally, for the strain-strain form, at first sight it may be less obvious that the part of $X$ that
is antisymmetric under $\mu\leftrightarrow\nu$ vanishes. However, for these components one can see that
the time-integral term cancels the contact term (and similarly for the part antisymmetric
under $\alpha\leftrightarrow\beta$), again without assuming the ground state is an angular
momentum eigenstate.


\subsection{Intensive form of the response and periodic boundary conditions}\label{seclocal}

The stress-stress form, eq.\ (\ref{stressstress}), of the response $X_{\mu\nu\alpha\beta}(\omega)$, which
is an extensive quantity, can also be understood in another way, as the response to a change in
$\lambda$ using the Hamiltonian $H_\Lambda$ in the $\mathbf{x}$ variables. In the rotationally-invariant
case, this is the same as the response to a change in metric, and this point of view may be familiar to
some readers. Using this approach, we can also obtain a Kubo formula in terms of intensive quantities
only, which is compatible  with periodic boundary conditions, and will be useful later.

To rederive eq.\ (\ref{stressstress}), we work in $\mathbf{x}$ variables, and recall that the
integrated stress tensor we are using is $T_{\mu\nu}^{(\mathbf{X})}$, eq.\ (\ref{tmunuX}), and its
expansion to order $\lambda$ in the $\mathbf{x}$ variables was given in eq.\ (\ref{tmunuXexp}) (and the
same forms are valid with a magnetic field). Then the linear response of $T_{\mu\nu}^{(\mathbf{X})}$ to
the strain, divided by $-i\omega^+$ so that this is actually the response to $d\lambda_{\alpha\beta}/dt$,
and finally dropping the distinction between $\mathbf{x}$ and $\mathbf{X}$ variables (as we
require only the linear response), gives exactly eq.\ (\ref{stressstress}). The contact term came
from the expansion of the stress to order $\lambda$ in eq.\ (\ref{tmunuXexp}), just like the
familiar diamagnetic term in the conductivity response comes from expanding the current operator to
order $A_\mu$ ($A_\mu$ being the perturbing vector potential, the response to which is conductivity).

A similar approach works for the linear response of the local stress tensor
$\tau_{\mu\nu}^{(\mathbf{X})}(\mathbf{X})=(\det \Lambda)^{-1}\Lambda_{\alpha\mu}
\Lambda_{\nu\beta}^{-1}\tau_{\alpha\beta}(\mathbf{x})$ to the uniform strain [the $(\det \Lambda)^{-1}$
factor is present because this transforms as a density, or formally because the $\delta$-function should
be written in $\mathbf{X}$ space rather than $\mathbf{x}$ space]; the result is the same as eq.\
(\ref{stressstress}), but with $\tau_{\mu\nu}(\mathbf{0})$ in place of $T_{\mu\nu}$. Here we have used the
same boundary conditions as for the previous derivation, that is, a system in an infinite volume. But a
similar derivation also works for {\em periodic} boundary conditions. These are defined in $\mathbf{x}$
variables as periodic boundary conditions on a box (or ``unit cell'') of fixed shape and size (independent
of $\Lambda$), say a cube. The Hamiltonian is $H_\Lambda$ as before, except for the different boundary
conditions, and uninteresting changes to the interaction potential to ensure it is periodic. When a
magnetic field is present (in two dimensions), the flux through the unit cell is fixed independent of
$\Lambda$ also, and must be an integer number of flux quanta (the flux quantum is $2\pi$ in our units);
use of the operator $\mathcal{B}$ is not required here. In $\mathbf{X}$ variables, the box has periodic
boundary conditions described by $\Lambda$, as $\mathbf{X}=\Lambda^T\mathbf{x}$, while the Hamiltonian
$H_0^{(\mathbf{X})}$ is independent of $\Lambda$, except possibly through the interaction potential,
the boundary conditions, and also the magnetic field strength varies with $\Lambda$ so that the flux
through the unit cell stays fixed. We note that with these boundary conditions, translational invariance
holds strictly. Eq.\ (\ref{tmunuXexp}) does not hold due to the non-existence of $J_{\mu\nu}$ in a
finite-size system with these boundary conditions, but there is still an expansion of
$\tau_{\mu\nu}^{(\mathbf{X})}(\mathbf{X})$ to order $\lambda$ in the $\mathbf{x}$ variables:
\be
 \tau_{\mu\nu}^{(\mathbf{X})}(\mathbf{X})=\tau_{\mu\nu}^{(0)}(\mathbf{x})+\left.\frac{\partial
 \tau_{\mu\nu}^{(\mathbf{X})}(\mathbf{X})}{\partial
 \lambda_{\alpha\beta}}\right|_{\lambda=0}\lambda_{\alpha\beta}+O(\lambda^2).
 \ee
Another benefit of these boundary conditions is that normalizable energy eigenstates always exist.
Then the Kubo formalism, working in $\mathbf{x}$ variables, leads to the result for linear response of
the stress at $\mathbf{x}=\mathbf{0}$ in the ground state (or at nonzero temperature) to a uniform rate
of strain, which we call $\chi$:
\begin{eqnarray}
\lefteqn{\chi_{\mu\nu\alpha\beta}(\omega)=\frac{1}{i\omega^+}\left\{
\left< \left.\frac{\partial \tau_{\mu\nu}^{(\mathbf{X})}(\mathbf{0})}{\partial
\lambda_{\alpha\beta}}\right|_{\lambda=0} \right>_0\right.}&&\nonumber\\
&&{}\left.+i\int_0^\infty{dt
\int d^dx\,e^{i\omega^+t}}
\left<\left[\tau_{\mu\nu}^{(0)}(\mathbf{0},t),\tau_{\alpha\beta}^{(0)}(\mathbf{x},0)\right]\right>_0
\right\};
\label{chi1}
\end{eqnarray}
the integral over $\mathbf{x}$ is restricted to the box. (There is of course also a similar formula for
$X$ with these boundary conditions; one expects the leading, extensive part of $X$ to be independent of
the choice of boundary conditions.) Now we can take the infinite size limit
(with particle density held fixed, as always). In the limit,
because $\tau_{\mu\nu}^{(\mathbf{X})}(\mathbf{0})$ is a local operator, its expansion to
order $\lambda_{\alpha\beta}$ in terms of the commutator with $J_{\alpha\beta}$ is again valid. Then
finally we have [dropping the superscript $(0)$]
\begin{eqnarray}
\lefteqn{\chi_{\mu\nu\alpha\beta}(\omega)=\frac{1}{\omega^+}\left\{\vphantom{\int_0^\infty}
\left< [\tau_{\mu\nu}(\mathbf{0}),J_{\alpha\beta}]\right>_0\right.}&&\nonumber\\
&&{}\left.+\int_0^\infty{dt
\int d^dx\,e^{i\omega^+t}}
\left<\left[\tau_{\mu\nu}(\mathbf{0},t),\tau_{\alpha\beta}(\mathbf{x},0)\right]\right>_0\right\}.
\label{chi2}
\end{eqnarray}
This agrees with the argument sketched just above, which began with $N$ particles in infinite space.

We can also transform this stress-stress expression into a stress-strain form. As we have already passed
to the infinite system, we can use the relation eq.\ (\ref{momflux}) and integration by parts again,
to obtain
\be
\chi_{\mu\nu\alpha\beta}(\omega)=-i\int_0^\infty
dt\,e^{i\omega^+t}\langle[\tau_{\mu\nu}(\mathbf{0},t),J_{\alpha\beta}
(0)]\rangle_0.
\label{chi3}
\ee
Note that a strain-strain form for the response is not available in terms of intensive quantities.


\subsection{Viscosity from the response function}\label{viscfromX}

It is natural to ask how the extensive and intensive forms are related. $\chi_{\mu\nu\alpha\beta}(\omega)$
is not simply $X_{\mu\nu\alpha\beta}(\omega)/L^d$, because of the following simple point. Recall that $X$
is the response of $\langle T_{\mu\nu}^{(\mathbf{X})}\rangle$ to strain, while $\chi$ is the response of
$\langle\tau_{\mu\nu}^{(\mathbf{X})}(\mathbf{0})\rangle$ instead. Assuming a homogeneous fluid state,
these should be related by
\be
\langle T_{\mu\nu}^{(\mathbf{X})}\rangle = L^d\langle\tau_{\mu\nu}^{(\mathbf{X})}(\mathbf{0})\rangle,
\ee
where $L^d$ means the volume of the system, which is the $\Lambda$-dependent volume of the box,
$L^d=L^d|_{\Lambda=I}\det \Lambda$, calculated either in the $\mathbf{X}$ variables as the volume of the
$\Lambda$-dependent box using the fixed, standard Euclidean metric, or in $\mathbf{x}$ variables for a
fixed box but with the $\Lambda$-dependent metric. (Actually, without rotation invariance, these metrics
might not be uniquely defined by our models, however the formula is still correct, because the important
point is how the volume varies with $\Lambda$, which is always through $\det \Lambda$ only.) Then for the
response at first order, we obtain
\be
X_{\mu\nu\alpha\beta}(\omega)/L^d=\chi_{\mu\nu\alpha\beta}(\omega)
-\frac{i}{\omega^+}\delta_{\mu\nu}\delta_{\alpha\beta}P,
\label{Xchirel}
\ee
where in the last term we use the result at zeroth order in the strain, that in the thermodynamic limit
the expectation of the local stress is
\be
\langle\tau_{\mu\nu}^{(0)}(\mathbf{0})\rangle=\delta_{\mu\nu}P,
\label{tauexp}
\ee
and this becomes $\delta_{\mu\nu}P_{\mathrm{int}}$ in the presence of a magnetic field in two dimensions.

The response function $\chi$ is almost exactly what we need to obtain the viscosity, which is supposed
to be
the local stress response to a uniform $\partial \lambda_{\mu\nu}/\partial t$, and so is an intensive
quantity. However, the expectation value
of the stress that we just discussed will respond to {\em static} (time-independent) strains---see
for example
the hydrodynamic forms (response local in space and time) in eq.\ (\ref{stresslinresponse}). As $\chi$
is the response to the time-derivative of the strain, these elastic
moduli at zero frequency will show up as the coefficients of singularities $\sim i/\omega^+$ in $\chi$.
If non-zero, they might be confused with viscosity coefficients that happen to diverge at $\omega=0$.

To deal with this and obtain expressions for the viscosity tensor at all frequencies, we will remove
from $\chi$ these static, and thus equilibrium, elastic coefficients, by subtracting
the zero-frequency value of the response to strain, that is the elastic moduli divided by $i\omega^+$,
and call the remainder the viscosity tensor.
For the homogeneous fluids we consider, the expectation of the stress
obeys eq.~(\ref{tauexp}), which has an obvious generalization in the presence of a static strain, and is
affected only by a dilation. Then we arrive at our definition for the viscosity tensor
at frequency $\omega$ in zero magnetic field,
\begin{align}\label{viscnofield1}
\eta_{\mu\nu\alpha\beta}(\omega)&=\chi_{\mu\nu\alpha\beta}(\omega)+
\frac{i}{\omega^+}\delta_{\mu\nu}\delta_{\alpha\beta}L^d\left(\frac{\partial P}{\partial (L^d)}\right)_{N},
\end{align}
or, in terms of the inverse compressibility $\kappa^{-1}=-L^d\left(\partial{P}/\partial{(L^d)}\right)_N$,
\begin{equation}
\eta_{\mu\nu\alpha\beta}(\omega)=\chi_{\mu\nu\alpha\beta}(\omega)
-\frac{i\kappa^{-1}}{\omega^+}\delta_{\mu\nu}\delta_{\alpha\beta}. \label{viscnofield}
\end{equation}
This, along with Eqs.~(\ref{chifreq1}-\ref{strainstrain}) and (\ref{Xchirel}), or Eqs.\ (\ref{chi2})
and (\ref{chi3}), give the Kubo formulas for viscosity in the absence of magnetic field.

The case of two dimensions with a magnetic field is similar, but the dilations involve a rescaling
of magnetic field, as we have seen. Then we arrive at
\begin{align}
\eta_{\mu\nu\alpha\beta}(\omega)&=\chi_{\mu\nu\alpha\beta}(\omega)
+\frac{i}{\omega^+}\delta_{\mu\nu}\delta_{\alpha\beta}L^2\left(\frac{\partial
P_{\mathrm{int}}}{\partial(L^2)}\right)_{\nu,N}\nonumber\\
&=\chi_{\mu\nu\alpha\beta}(\omega)
-\frac{i\kappa^{-1}_\text{int}}{\omega^+}\delta_{\mu\nu}\delta_{\alpha\beta}, \label{viscfield}
\end{align}
where $\kappa^{-1}_{\mathrm{int}}=-L^2\left(\partial{P_{\mathrm{int}}}/\partial{(L^2)}\right)_{\nu,N}$
is the ``inverse internal compressibility'', with the second partial derivative taken with $BL^2$ (or
the filling factor) held fixed.
If we write the energy of the system as $E(N,L^2,B) = L^2 \varepsilon(\nu,B)$,
where $\varepsilon(\nu, B)$ is the energy density as a function of Landau level filling factor
$\nu=\overline{n}\phi_0/B$, where $\overline{n}=N/L^2$ is the particle density and $\phi_0 = h c/e$ is
the flux quantum ($\phi_0=2\pi$ in our units), then
\begin{align}
P_{\mathrm{int}}(\nu,B)&=B\left(\frac{\partial\varepsilon(\nu,B)}{\partial
B}\right)_\nu-\varepsilon(\nu,B),\label{pint}\\
  \kappa^{-1}_\text{int}(\nu,B) &= B^2\left( \frac{\partial^2 \varepsilon(\nu, B)}{\partial
  B^2}\right)_\nu.\label{eqn:kint_hoyosson}
\end{align}

In the fractional quantum Hall effect, one encounters incompressible fluids, which have vanishing
compressibility at zero temperature. This refers to the usual compressibility, which can be related to
the change in density with chemical potential {\em with the magnetic field fixed}, and so differs from
the internal compressibility considered here, which is well-defined and usually non-zero and finite.
(Later, we will see that the latter can be extracted from the $q^2$ part of the conductivity.)

Further, there are similar results at non-zero temperature. For the (internal) compressibility, there is
the question of what is held fixed in taking the partial derivative of the pressure: the temperature or
the entropy. We will discuss this point in Sec.\ \ref{nondiv}, and argue that it is the entropy, as
mentioned in Ref.\ \onlinecite{Taylor2010}. After this point, our discussion applies to both zero and
non-zero temperature, unless otherwise noted.

\subsection{Non-divergence of bulk viscosity}\label{nondiv}

Next we comment on the subtraction of terms containing the thermodynamic inverse (internal)
compressibility in the expressions for the viscosity in terms of the response function,
Eqs.~(\ref{viscnofield}) and (\ref{viscfield}), in either the extensive or intensive forms. We will be
led to the striking conclusion that the bulk viscosity is finite, not infinite. Consider first the case
of zero magnetic field in $d$ dimensions, and the extensive response $X$ in finite size. We can show in
general that the leading contribution to the diagonal or trace part $X_{\mathrm{d}} \equiv
X_{\mu\mu\nu\nu}/d^2$ of the response function as $\omega\rightarrow0$ is given by the derivative of
$PL^d$ with $\lambda$. In the stress-stress form, $X_{\mathrm{d}}$ is
\begin{align}
X_{\mathrm{d}}(\omega)&=\frac{1}{d\omega^+}\left<\left[T (0),K(0)\right]\right>_0\nonumber \\
&+\frac{1}{\omega^+}\int_0^{\infty}{dte^{i\omega^+t}\left<\left[T (t),T(0)\right]\right>_0},
\end{align}
where we have defined $T \equiv T_{\mu\mu}/d$
with $d$ the dimension of space. Asymptotically as $\omega\rightarrow 0$,
\begin{align}
\omega^+X_{\mathrm{d}}&\sim\frac{1}{d}\left<\left[T (0),K(0)\right]\right>_0\nonumber \\
&+\lim_{\omega\rightarrow 0}\int_0^{\infty}dt\,e^{i\omega^+t}\left<\left[T (t),T(0)\right]\right>_0.
\end{align}
This is the response of the system to a static dilation at zero wavevector. By inserting a complete set
of energy eigenstates $\left|e\right>$ of $H_0$ into the commutator in the second line, we obtain
\begin{equation}
\omega^+X_{\mathrm{d}}\sim\frac{1}{d}\left<\left[T (0),K(0)\right]\right>_0+2i\sum_{e:E_e\neq
E_0}\frac{\left|\left<0\right|T\left|e\right>\right|^2}{E_0-E_e}\label{time_ind_pert}
\end{equation}
(the restricted sum over states is really a principal part, arising from careful use of $\epsilon\to 0$).
As explained already in Sec.\ \ref{seclocal}, working in the $\mathbf{x}$ variables the first term on
right-hand side is the expectation of the change in the operator $T^{(\mathbf{X})}$ under a dilation. Then
time-independent perturbation theory using the $\mathbf{x}$ variables shows that the right-hand side is
the change in the expectation of $T^{(\mathbf{X})}$ due to the dilation, the second term being due to
the change of the ground state. Thus
\begin{align}
\omega^+X_{\mathrm{d}}\sim-i\left(\frac{\partial\left<T\right>}{\partial
(L^d)}\right)_{N}&=-i\left(\frac{\partial(PL^d)}{\partial (L^d)}\right)_N\nonumber\\
&=i(\kappa^{-1}-P)L^d,
\label{traceresidue}
\end{align}
showing that the response to a dilation at leading order as $\omega\to0$ is given by the difference between
the inverse compressibility and the pressure  --- it is an elastic response. This is
precisely the term we subtract from $X$ to get the viscosity in Eq.~(\ref{viscnofield1}). We conclude
that there can be no divergent bulk viscosity at zero frequency. The
argument goes through similarly in the presence of a magnetic field in $d=2$, with $P$ and
$\kappa^{-1}$ replaced by $P_{\mathrm{int}}$ and $\kappa^{-1}_{\mathrm{int}}$ respectively.
It also goes through similarly for the intensive formulation of viscosity in terms of
$\chi$, eq.\ (\ref{chi1}).

This argument may give rise to some unease. The shear modulus of a fluid
is of course zero, while the preceding argument may lead us to expect that it is given
by a similar time-independent perturbation theory expression, which (as we will see) is non-zero
in some cases. This must be related to the fact that the conventional way to obtain the elastic
moduli (including the inverse compressibility or bulk modulus), which are susceptibilities, from
the stress-stress response function, would be by taking the thermodynamic limit of the (intensive)
response first, then taking $\omega\to0$ before $\mathbf{q}\to0$. But we have $\mathbf{q}=0$, then
took $\omega\to0$, and finally the thermodynamic limit.

One can find susceptibilities directly at $\mathbf{q}=0$, provided one is careful with the order in
which one takes the derivative with respect to strain and the thermodynamic limit. The
usual ``thermodynamic'' formulas involve finite size, but the size is treated as large and
discreteness effects are ignored when differentiating the ground state energy, which corresponds to
taking the limit first. (Here and in the remainder of this section we concentrate on the case of
zero temperature.) The perturbation
theory formula for the derivative (in finite size), as in the right-hand side of
eq.~(\ref{time_ind_pert}), takes the derivative of the expectation in a state that varies continuously
with strain. If the ground state energy level does not cross others as the strain is varied, the result
is most likely independent of the order of the limit and the derivative. We expect this is the case for
pure dilations, leading to the pressure and inverse compressibility. But perturbation theory can break
down if energy levels cross, especially if they do so on a set in $\lambda$ space that becomes dense as
the size goes to infinity. Then if by ``ground state'' we mean the lowest energy state for given
strain (as in the thermodynamic formulas), this state vector changes discontinuously with strain,
and this effect cannot be picked up in perturbation theory. Taking the derivative after the limit will
give a different result from
taking it before, at whatever strain it is taken. This is the case for shear strains, in some
gapless systems.

As an example, consider the free Fermi gas. Using periodic boundary conditions on a rhomboid-shaped box, we
can examine the expectation of $H_0$ and of $T_{\mu\nu}$, and their variation with shape to first
order. In $\mathbf{X}$ variables, $\Lambda$ describes the shape of the box, and the metric in $H_0$ and
in $T_{\mu\nu}$ is
the standard one. Then as is well-known,
the many-particle ground state for a given box is constructed by occupying all single-particle
plane-wave states with wavevectors $\mathbf{k}$ inside the Fermi sphere; the radius of the sphere is
chosen to obtain the correct particle number $N$. Under shear and dilation, the $\mathbf{k}$ points
move around, and every so often points enter or leave the Fermi sphere. Under a pure dilation, the
volume of the Fermi sphere changes so that exactly the ``same'' set of $\mathbf{k}$ points (up to
a rescaling) is always occupied, and no
levels cross. But during a nonzero shear (or shear and dilation together), $\mathbf{k}$ points
do enter or leave the Fermi sphere, and so the ground state energy levels cross. In the limit, the
ground state energy density depends on the particle density in the system, but not on its shape.
Hence derivatives of this energy density with respect to shear vanish, as expected. Derivatives with
respect to dilations give the pressure and inverse compressibility, and because no levels cross as the
scale factor $\det\Lambda$ is varied, the same result would be obtained by differentiating the energy of
the finite volume system, and taking the limit afterwards. But for derivatives with respect to shear,
in finite size the second derivative is nonzero, though it becomes undefined on a set of measure zero
at which ground state levels cross.
Through the above formulas, this leads to an infinite shear viscosity, while the bulk viscosity is zero
(we give further details for the free Fermi gas in Section~\ref{freefermi} below).
We expect, though we do not have a rigorous proof, similar behavior for an interacting Fermi gas in a
Fermi-liquid phase, so that the pressure and compressibility can be obtained by using dilations
either before or after the thermodynamic limit. This means the $i/\omega^+$ terms in $X_{\mathrm{d}}/L^d$
or the corresponding part of $\chi$ are canceled by the subtraction, and the bulk viscosity cannot
be infinite. But the shear viscosity will be infinite at zero temperature, in agreement with
calculations of its temperature dependence \cite{Abrikosov1959}.

For the first derivative of ground state energy with respect to shear, that is for the traceless part
of the expectation of the stress, the situation is slightly different from that for the second
derivative. Taking the limit first, the ground state energy density will be independent of shear. If,
as an interval of a path in $\lambda$ space is traversed (with ${\rm tr}\,\lambda$ fixed), there are
no level crossings, then the limit of the first derivative along the path will also be zero. But suppose
there are level crossings on this path, and the spacing of them goes to zero in the limit, so that the set
of positions of level crossings is dense (as for the free Fermi gas). As the derivative taken after the
limit is zero, the lowest energy levels as a function of position on the path must be close to a sequence
of overlapping parabolas, with the minimum of each at the same energy. We have seen that the second
derivative is of order one, and so the first derivative on any of these curves, anywhere within the
interval in which it is the lowest, will tend to zero in the limit, because the distance in $\lambda$
from the minimum of that curve goes to zero, as the level crossings become dense. So (except on a set of
measure zero in $\lambda$ on which the first derivative is not defined) the traceless local (or intensive)
stress in a fluid state does go to zero in the thermodynamic limit, as claimed earlier.

At non-zero temperature, one can make different but related arguments. In particular, the trace part
$\omega^+X_\mathrm{d}$ is now identified as $-i(P-\kappa_{S}^{-1})L^d$, or similarly with a magnetic field
in two dimensions, where the inverse isentropic compressibility, $\kappa_{S}^{-1}$, is
$\kappa_{S}^{-1}=-L^d(\partial P/\partial (L^d))_{N,S}$ at fixed entropy $S$ (and also fixed $\nu$ for
the isentropic internal compressibility). This is because, as for the pressure, the response function can
be identified as the partial derivative taken under the thermal average, with the probabilities and hence
the entropy, held fixed. These compressibilities must be non-negative for stability. The appearance of
the isentropic compressibility in this limit of the response is frequently obtained from hydrodynamic
considerations, rather than directly from the stress-stress response, as here.

In the preceding arguments, we have taken the frequency to zero in the response function before the
thermodynamic limit. However, we will be using the intensive functions $X/L^d$, $\chi$ and $\eta$ in the
thermodynamic limit, and the behavior of these as the frequency tends to zero subsequently. The different
order of limits does not appear to be a problem. For example, in the zero-temperature case, we can use the
intensive form in $\mathbf{x}$ variables to study the response of the trace of $\tau$ to a low-frequency
dilation. The leading part comes from the ground state adiabatically following the dilation, and gives the
inverse compressibility. The real part of the bulk viscosity involves transitions to excited states of the
unperturbed system, for which the available phase space is usually small (as in a Fermi liquid, for
example) or zero. Hence we expect that at zero temperature, in general the real part of the bulk viscosity
actually goes to zero at zero frequency.

\subsection{Spectral density, sum rules, and positivity}\label{sumrules}

We can derive a spectral density for the viscosity tensor, and a sum rule for it, by following a
standard method, starting from the convenient stress-strain formulation, in either extensive or
intensive forms. For the extensive form, using Eqs.~(\ref{chifreq1}), we define the spectral
density function by \cite{Forster1975}
\be
X''_{\mu\nu\alpha\beta}(\omega)=-\half i\int_{-\infty}^\infty dt\, e^{i\omega t}
\langle [T_{\mu\nu}(t),J_{\alpha\beta}(0)]\rangle.\label{specdens}
\ee
In many cases of linear response theory, such a function would be the imaginary part of the
corresponding retarded response function, although for transport functions, such as conductivity as well
as viscosity, the division by $-i\omega$ (in the current-current, respectively stress-stress, forms)
means that the spectral density is proportional to the real part of the conductivity, if certain
symmetries such as time-reversal and reflection symmetry are unbroken. But this is not generally the
case: $X''$ is not in general real \cite{Forster1975}, but does consist (in a finite size system) of a
sum of $\delta$-functions in $\omega$, with tensor-valued coefficients.

There is a spectral representation,
\be
X_{\mu\nu\alpha\beta}(\omega)=\frac{i}{\pi}\int_{-\infty}^\infty
d\omega'\,\frac{X''_{\mu\nu\alpha\beta}(\omega')}{\omega^+-\omega'}.
\ee
This shows that if $X''$ is real, it is the real part of $X$, as expected. $X_{\mu\nu\alpha\beta}(\omega)$
is analytic in the upper-half complex $\omega$ plane, and there are also
corresponding Kramers-Kronig relations between $X''$ and the complementary
part $X'_{\mu\nu\alpha\beta}(\omega)=[X_{\mu\nu\alpha\beta}(\omega)-X''_{\mu\nu\alpha\beta}(\omega)]/i$
(for real $\omega$),
which is the imaginary part of $X$ when $X''$ is real.
Finally, the definition leads immediately to a sum rule for the total spectral density,
\be
\int_{-\infty}^\infty \frac{d\omega}{\pi}\, X''_{\mu\nu\alpha\beta}(\omega)=-i\langle
[T_{\mu\nu}(0),J_{\alpha\beta}(0)]\rangle.
\ee
The ``sum'' on the right-hand side is real, because the expectation of a commutator of
self-adjoint operators is imaginary. The sum rule can also be viewed as describing the $\omega\to\infty$
limit of $X_{\mu\nu\alpha\beta}(\omega)$, using the spectral representation on the one hand,
and integration by parts from the stress-strain definition of $X$, together with the Riemann-Lebesgue
lemma, on the other; this explains its relation with the
contact term in the stress-stress form. There are related results for $\chi_{\mu\nu\alpha\beta}$ and
for $\eta_{\mu\nu\alpha\beta}$. (Similar sum rules were also discussed in Ref.\ \onlinecite{Taylor2010},
but only in the absence of Hall viscosity.) One would expect that the right-hand side is symmetric
under $\mu\nu\leftrightarrow\alpha\beta$, which can be shown in certain limits, as we discuss below.
That is, the Hall viscosity cancels from the sum rule. This would then be similar to the case
of conductivity, in which the Hall conductivity cancels from the sum rule.

For further arguments, the stress-stress form of the spectral density is most convenient. The
only complication here is the factor $1/\omega^+$ in the formula for $X$, which causes the appearance
of $\delta(\omega)$ terms in the spectral density, in addition to $\delta$-functions that come from the
stress-stress time-integral term. The former correspond to the terms discussed in Sec.\ \ref{nondiv}. We
can obtain the spectral density by multiplying $X''$ in eq.\ (\ref{specdens}) by $i\omega$, which means
differentiating with respect to $t$ under the integral, and then using relation (\ref{momflux}) once again
(after shifting the $t$-dependence onto $J_{\alpha\beta}$). This has the effect of removing any
$\delta(\omega)$ terms from $X''$, one of which we know is the inverse compressibility term. Reinstating
these terms, one has
\begin{align}
X''_{\mu\nu\alpha\beta}(\omega)&=\pi C_{\mu\nu\alpha\beta}\delta(\omega)\nonumber\\
&\quad{}+ \frac{1}{2\omega}\int_{-\infty}^\infty dt\, e^{i\omega t}\langle
[T_{\mu\nu}(t),T_{\alpha\beta}(0)]\rangle
\end{align}
(in finite size, the time-integral expression should not produce any $\delta$-function at $\omega=0$),
where the constant tensor $C_{\mu\nu\alpha\beta}$ is
\begin{align}
C_{\mu\nu\alpha\beta}=&-i\left<\left[T _{\mu\nu}(0),J_{\alpha\beta}(0)\right]\right>_0\vphantom{\int}
\nonumber \\
&\quad{}-i\int_0^{\infty}{dt\,e^{-\epsilon t}\left<\left[T _{\mu\nu}(t),T
_{\alpha\beta}(0)\right]\right>_0},
\end{align}
and is real.
Then, using arguments presented in e.g.\ Ref.\ \onlinecite{Forster1975}, one can almost conclude
that $X''(\omega)$, viewed as a matrix with rows and
columns indexed by the pairs $\mu\nu$ and $\alpha\beta$, respectively, should be Hermitian, and
also positive semidefinite, for all $\omega$, and that at zero frequency it (i.e.\
$C_{\mu\nu\alpha\beta}$) should be real. We say ``almost'' because Forster's
discussion \cite{Forster1975} does not include the contact terms in our stress-stress form,
which contribute to $X''$ only at zero frequency. Hermiticity holds at non-zero frequencies, and
the time-integral term in $C_{\mu\nu\alpha\beta}$ is real and symmetric, so to
obtain symmetry of the matrix at zero frequency, we would need the contact term
coefficient $-i\langle[T_{\mu\nu},J_{\alpha\beta}]\rangle$
to be symmetric under the exchange $\mu\nu\leftrightarrow\alpha\beta$. This is not yet obvious in
general, and we discuss it further below.
For positivity, the energy absorption argument \cite{Forster1975} does allow for contact terms, and
those that occur in the positivity statement at zero frequency are automatically symmetrized, as they
arise from the second derivative of the Hamiltonian with respect to the perturbing field, in our case
of $H_\Lambda$ with respect to $\lambda$. The part of
$C_{\mu\nu\alpha\beta}\delta(\omega)$ that vanishes on taking the trace on $\mu\nu$ and on
$\alpha\beta$ would either be zero, or would represent an infinite shear
viscosity, as discussed in Sec.\ \ref{nondiv}.
At zero frequency, the infinite shear viscosity (if any) must be positive. For arbitrary frequencies,
in simple cases such as with time-reversal and reflection symmetries, in which case the $X''$ matrix is
symmetric, these conditions imply that the real parts of the shear and bulk viscosities must be
positive (like the real part of the symmetric conductivity tensor). In general, at non-zero frequency,
the condition that the Hermitian matrix $X''$ be positive semidefinite involves
the imaginary part of the Hall viscosity, and not only the shear and bulk viscosities (just like the case
of conductivity, in which the imaginary part of the Hall conductivity enters); note that the discussion
in Sec.\ \ref{intro} was for frequency-independent or zero-frequency viscosity coefficients only.
At zero frequency, one statement of positivity is for the intensive function $\chi''(\omega)$, which
differs from $X''$ by the pressure term in the trace part; it implies that the inverse (internal)
compressibility should be non-negative, in agreement with a consequence of thermodynamic stability.
On the other hand, positivity of the trace part of $X''$ implies that $\kappa_\mathrm{int}^{-1}-P\geq0$
as well. This can also be viewed as a consequence of thermodynamic stability, by using ${\rm tr}\,\lambda$
as a ``generalized coordinate'', rather than the volume $L^d$.

 If the zero-temperature real part of the shear viscosity is a $\delta$-function (that results in part
 from the contact term), then at non-zero temperature in an interacting system in the thermodynamic
 limit, the shear viscosity at zero frequency will generally be neither zero nor infinity, and as a
 function of frequency the $\delta$-function becomes broadened. In this case the $\delta$-function
 is cancelled by a contribution from the time-integral term, which also produces the broadened peak
 (which tends to a $\delta$-function at zero temperature). For such cases, the spectral representation
 above for finite size should be rearranged so there is no $\delta$-function in the traceless part.

The symmetry of the contact-term coefficient
\be
-i\langle[T_{\mu\nu},J_{\alpha\beta}]\rangle
\ee
under $\mu\nu\leftrightarrow\alpha\beta$ can be shown under some conditions or in some limits.
Using relation
(\ref{momflux}) again, the antisymmetric part is $-\half$ times the expectation value of
\be
[[H_0,J_{\mu\nu}],J_{\alpha\beta}]-[[H_0,J_{\alpha\beta}],J_{\mu\nu}]
=[H_0,[J_{\mu\nu},J_{\alpha\beta}]]
\ee
(this also gives the difference between the contact terms that would result in the $\mathbf{x}$ variables
from use of $T_{\mu\nu}$ versus $T_{\mu\nu}^{(\mathbf{X})}$ at the end of Sec.\ \ref{sec21}).
The commutator of the two $J$'s always contains only traceless parts, as the commutation relations can be
rewritten
\begin{align}
i\left[J_{\mu\nu},J_{\alpha\beta}\right]&=\delta_{\mu\beta}J_{\alpha\nu}-\delta_{\nu\alpha}J_{\mu\beta},
\nonumber\\
&=\delta_{\mu\beta}(J_{\alpha\nu}-\frac{1}{d}\delta_{\alpha\nu}J_{\gamma\gamma})
-\delta_{\nu\alpha}(J_{\mu\beta}-\frac{1}{d}\delta_{\mu\beta}J_{\gamma\gamma}).
\label{straincomm2}
\end{align}
Taking the expectation value, and using the relation (\ref{momflux}), we arrive at two expectations of
traceless parts of $T_{\mu\nu}$, which we have argued are small compared with $L^d$ in the thermodynamic
limit (or zero, in the magnetic field case in two dimensions). (In the rotationally-invariant case, we
look at parts symmetric under $\mu\leftrightarrow\nu$ and under $\alpha\leftrightarrow\beta$ only, and
then the commutator gives the angular momentum, and the expectation is zero even in finite size.)
Hence, if we are interested in the viscosity of the infinite (thermodynamic limit) system, then the
symmetry under $\mu\nu\leftrightarrow\alpha\beta$ does hold, and has the pleasing consequences mentioned
above.

One might wish for more in the finite-size case. Our basic definitions can, and perhaps should, be
modified to make the contact term in the stress-stress form symmetric, by taking its symmetric part. It is
not clear to us fundamentally why that would have to be done, but it would be useful anyway when performing
numerical calculations at finite size, to remove the antisymmetric part that should not be present in the
limit.
Alternatively, perhaps there is some physical meaning to the antisymmetric part of the coefficient of
$\delta(\omega)$ in $X''$, even though it does not contribute to energy absorption, or to the physical
(infinite-size) viscosity.

\section{Relation between viscosity and conductivity}\label{sec4}

In a Galilean-invariant system with particles that all have the same charge (which is one in our units)
and mass $m$, the momentum density is $m$ times the number current density.
We will now use the intensive form of the stress-strain response function to derive a general relation
between viscosity and conductivity for this case. We note that this does not require rotational
invariance, provided that the inverse mass tensor is the same for each particle.
In this section we will usually treat the zero and nonzero magnetic field case together, and so
specialize to $d=2$ dimensions; other cases are handled similarly.

We begin by using the translation-invariant system in a box of volume (area) $L^2$ with periodic boundary
conditions, with Hamiltonian $H_0$. The continuity equation for momentum density, eq.\ (\ref{fieldcont}),
can be written as
\be
\left(-\delta_{\nu\lambda}\frac{\partial}{\partial t}+\omega_c
\epsilon_{\nu\lambda}\right)g_\lambda(\mathbf{x},t) = \partial_\mu\tau_{\mu\nu}(\mathbf{x},t),
\ee
where $\omega_c=B/m$ is the cyclotron frequency. (For $B>0$, the following derivation still holds if there
is an anisotropic mass tensor, by making a suitable modification of $\omega_c\epsilon_{\nu\lambda}$ to a
different tensor.)
Due to the uniformity of the magnetic field, we can use this to derive additional modified Ward
identities, in particular starting from the stress-stress retarded response function:
\begin{widetext}
\begin{align}
\label{eqn:etatot_gg}
  q_\lambda q_\rho & \int_0^\infty dt\, e^{i \omega^+ t}
  \int d^2x\, e^{-i \mathbf{q} \cdot \mathbf{x}}
  \left\langle \left[ \tau_{\lambda\nu}(\mathbf{x},t), \tau_{\rho\beta}(\mathbf{0},0) \right]
  \right\rangle_0
  =\nonumber\\
 & \left(i \omega \delta_{\nu\lambda} + \omega_c \epsilon_{\nu\lambda} \right)
  \left( -i \omega \delta_{\beta\rho} + \omega_c \epsilon_{\beta\rho} \right)
  \int_0^\infty dt\, e^{i \omega^+ t}
  \int d^2x\, e^{-i \mathbf{q} \cdot \mathbf{x}}
  \left\langle \left[ g_{\lambda}(\mathbf{x},t), g_{\rho}(\mathbf{0},0) \right] \right\rangle_0
\nonumber \\
 &-i
  q_\gamma \int d^2x\, e^{-i\mathbf{q}\cdot\mathbf{x}}
  \left\langle \left[ \tau_{\gamma\nu}(\mathbf{x}), g_{\beta}(\mathbf{0}) \right] \right\rangle_0
 \nonumber \\
 &-\left( i \omega \delta_{\beta\rho} - \omega_c \epsilon_{\beta\rho} \right)
  \int d^2x\, e^{-i\mathbf{q}\cdot\mathbf{x}}
  \left\langle \left[ g_{\nu}(\mathbf{x}), g_{\rho}(\mathbf{0}) \right] \right\rangle_0.
\end{align}
\end{widetext}
Expressions similar to this have been used for zero magnetic field ($\omega_c=0$) by many authors,
in particular Taylor and Randeria \cite{Taylor2010}, however even in that case our approach differs in
some of the details.

Assuming Galilean invariance, the number current density $\mathbf{j}(\mathbf{x},t)$ is related to
the momentum density by $\mathbf{j}(\mathbf{x},t) = \mathbf{g}(\mathbf{x},t)/m$.
Hence, the retarded function on the right-hand side of Eq.~(\ref{eqn:etatot_gg}) is the same as the
one appearing in the Kubo formula for the electrical conductivity (current-current response function),
\begin{align}
  \lefteqn{\sigma_{\nu\beta}(\mathbf{q},\omega) =
  \frac{i \overline{n}}{m \omega^+}\delta_{\nu\beta}}\quad&\nonumber\\
 &{}+ \frac{1}{\omega^+}
  \int_0^\infty dt\, e^{i \omega^+ t}
  \int d^2x\, e^{-i\mathbf{q}\cdot\mathbf{x}}
  \left\langle \left[ j_{\nu}(\mathbf{x},t), j_{\beta}(\mathbf{0},0) \right] \right\rangle_0.
  \label{eqn:kubo_cond}
\end{align}
Meanwhile, the left-hand side of Eq.~(\ref{eqn:etatot_gg}) includes the time-integral part of the
intensive form of the response function for viscosity, with two factors of $\mathbf{q}$ contracted into
it, presently in finite size, as in eq.\ (\ref{chi1}). We will now account for the remaining terms on the
right-hand side, and aim to take the thermodynamic limit and only then expand in powers of $\mathbf{q}$ to
second order, to obtain the response function $\chi_{\mu\nu\alpha\beta}(\omega)$ from the conductivity. It
follows from the preceding derivation that, as the left-hand side of Eq.~(\ref{eqn:etatot_gg}) is of order
$\mathbf{q}^2$, all terms of order one and of order $\mathbf{q}$ on the right-hand side must cancel.

First, still in finite size, in the last term on the right-hand side of Eq.~(\ref{eqn:etatot_gg}), we can
use translation invariance to introduce integration over a variable $\mathbf{x}'$, divide by $L^2$, and
then evaluate, giving
\begin{align}
  \lefteqn{\frac{1}{L^2}\int d^2x\,d^2x'\, e^{-i\mathbf{q}\cdot(\mathbf{x}-\mathbf{x}')}
  \left\langle \left[ g_{\nu}(\mathbf{x}), g_{\rho}(\mathbf{\mathbf{x}'}) \right] \right\rangle_0
  }\quad& \nonumber\\
   &=\frac{1}{L^2}
  \sum_i \frac{1}{4}
  \left\langle \left[ \left\{ \pi^i_\nu , e^{-i \mathbf{q} \cdot \mathbf{x}^i} \right\}, \left\{
  \pi^i_\rho,
e^{i \mathbf{q} \cdot \mathbf{x}^i} \right\} \right] \right\rangle_0
  \nonumber\\
  &= \nonumber
  \frac{1}{L^2}
  \left(
  i  N B\epsilon_{\nu\rho}
  - q_\nu \left\langle G_\rho \right\rangle_0 - q_\rho \left\langle G_\nu \right\rangle_0
  \right),
  \label{eqn:gg}
\end{align}
where $\mathbf{G} = \sum_i \hbox{\boldmath$\pi$}^i$ is the total momentum. The thermodynamic limit of
this exists, and contains no term of order higher than $\mathbf{q}$ as $\mathbf{q}\to0$.

Next, the second term on the right-hand side of Eq.~(\ref{eqn:etatot_gg}) can be manipulated to produce
the contact term in $\chi$. First, we take the thermodynamic limit, and then the part quadratic
in $\mathbf{q}$, by expanding the exponential $e^{-i\mathbf{q}\cdot\mathbf{x}}$ to first order. We
also use translation invariance again, so that the expression
contains $[\tau_{\mu\nu}(\mathbf{0}),g_\beta(-\mathbf{x})]$. Then we recognize the occurrence of
the first term of the strain generator, if we write Eq.~(\ref{magjcompact}) in the form
\begin{eqnarray}
 J_{\mu\nu}
 &=& -\int d^2x\, x_\mu g_\nu(\mathbf{x})
 + \sum_i \half\mathcal{B}\epsilon_{\nu\alpha}x_\mu^i
 x_\alpha^i \vphantom{\frac{1}{2N}}\nonumber \\
 && {}+\frac{1}{2}\delta_{\mu\nu}\left\{\mathcal{B},\Xi(\{\mathbf{x}^i\},\mathcal{B})\right\}
 \label{magjlocal}
\end{eqnarray}
(or similarly without the terms containing $\mathcal{B}$, if $B=0$).
Then we can cast the contact term from Eq.~(\ref{chi2}) in the form,
\begin{equation} \label{eqn:contacttaug}
 \frac{1}{\omega^+}
\left\langle \left[\tau_{\mu\nu}(\mathbf{0}),J_{\alpha\beta}\right] \right\rangle_0
=
-\frac{1}{\omega^+}
\int d^2x\,
\left\langle\left[\tau_{\mu\nu}(\mathbf{0}),
x_\alpha g_\beta(\mathbf{x})
\right]\right\rangle_0,
\end{equation}
where for $B\neq 0$ the last two terms in Eq.~(\ref{magjlocal}) do not contribute, because they fail
to commute only with the kinetic part of the stress tensor, and the result of that commutator always
contains a product of a particle coordinate with a delta function of that coordinate at the origin.
This means that the term in question can be absorbed into the time-integral term on the left to produce
precisely $\chi$ contracted with two $\mathbf{q}$s.

Hence, Eq.~(\ref{eqn:etatot_gg}) turns into
a relation between the conductivity and a symmetrized part of the intensive strain-stress response
function; we define $\overline{\chi}_{\mu\nu\alpha\beta}(\omega)=
  \half[ \chi_{\mu\nu\alpha\beta}(\omega) + \chi_{\alpha\nu\mu\beta}(\omega)]$, then
\begin{widetext}
\begin{equation} \label{eqn:etatot_cond_B}
  \overline{\chi}_{\mu\nu\alpha\beta}(\omega)  =
  \half m^2
  \left( \omega \delta_{\nu\lambda} - i \omega_c \epsilon_{\nu\lambda} \right)
  \left. \frac{\partial^2 \sigma_{\lambda\rho}(\mathbf{q},\omega)}{\partial q_\mu \partial
  q_\alpha}  \right|_{\mathbf{q}=\mathbf{0}}
  \left( \omega \delta_{\rho\beta} - i \omega_c \epsilon_{\rho\beta} \right).
\end{equation}
This is the central result of this section.
When the magnetic field is zero, one can simply substitute $\omega_c=0$ in the above equations.
Let us note that a similar relation holds in three dimensions:
\begin{equation} \label{eqn:etatot_cond_B_3D}
  \overline{\chi}_{\mu\nu\alpha\beta}(\omega)  =
  \half m^2
  \left( \omega \delta_{\nu\lambda} - i \omega_c b_\gamma \epsilon_{\gamma\nu\lambda} \right)
  \left. \frac{\partial^2 \sigma_{\lambda\rho}(\mathbf{q},\omega)}{\partial q_\mu \partial q_\alpha}
  \right|_{\mathbf{q}=\mathbf{0}}
  \left( \omega \delta_{\rho\beta} - i \omega_c b_\delta \epsilon_{\delta\rho\beta} \right),
\end{equation}
\end{widetext}
where $\mathbf{b} = \mathbf{B}/|\mathbf{B}|$ is a unit vector in the direction of the magnetic field.

These results lend themselves to a simple interpretation.  For simplicity, we will concentrate on two
dimensions. If we expand the wavevector-dependent conductivity in powers of the wavevector $\mathbf{q}$,
\begin{equation}
   \sigma_{\lambda\rho}(\mathbf{q},\omega) = \sigma^{(0)}_{\lambda\rho}(\omega) +
   \sigma^{(2)}_{\lambda\rho}(\mathbf{q},\omega) + \cdots,
\end{equation}
then the zeroth order term (the response to a uniform electric field) is fixed by Galilean invariance to be
\begin{align} \label{eqn:sigma0}
  \sigma^{(0)}_{\lambda\rho} (\omega)
 & =
  - \frac{\overline{n}}{m} 
  \left(i \omega^+ \delta_{\lambda\rho} + \omega_c \epsilon_{\lambda\rho} \right)^{-1}\nonumber\\
  &=
  \frac{\overline{n}}{m \left( \omega^{+ 2} - \omega_c^2 \right)}
  (i\omega^+\delta_{\lambda\rho}-\omega_c\epsilon_{\lambda\rho})
\end{align}
By Eqs.~(\ref{eqn:etatot_cond_B}) and~(\ref{eqn:sigma0}), the second order in $\mathbf{q}$ term is
related to the viscosity through
\begin{equation} \label{eqn:cond_viscosity}
  \sigma^{(2)}_{\lambda\rho}(\mathbf{q},\omega)=
  - \sigma^{(0)}_{\lambda\nu} (\omega)
  \frac{1}{\overline{n}}
  q_\mu \chi_{\mu\nu\alpha\beta}(\omega)
  \frac{1}{\overline{n}}
  q_\alpha \sigma^{(0)}_{\beta\rho} (\omega).
\end{equation}
One can intuitively understand this expression as follows: in the presence of a nonuniform electric
field $\mathbf{E}(\mathbf{q},\omega)$, to the leading order the system responds with a nonuniform current,
$\langle j_\beta(\mathbf{q},\omega) \rangle = \sigma^{(0)}_{\beta\lambda} (\omega) E_\lambda
(\mathbf{q},\omega)$,
which implies a strain rate
$\dot{u}_{\alpha\beta} (\mathbf{q},\omega) = (i q_\alpha) \sigma^{(0)}_{\beta\lambda} (\omega)
E_\lambda(\mathbf{q},\omega)/\overline{n}$ (symmetrization over $\alpha$ and $\beta$ is not important,
as $\chi_{\mu\nu\alpha\beta}(\omega)$ is symmetric with respect to these two indices).
The strain rate results in an average stress $\langle \tau_{\mu\nu}(\mathbf{q},\omega) \rangle = -
\chi_{\mu\nu\alpha\beta}(\omega) \dot{u}_{\alpha\beta} (\mathbf{q},\omega) $, whose spatial derivative
gives a contribution to the effective electric field acting on the particles,
$E^\text{eff}_\nu (\mathbf{q},\omega) = -i q_\mu \langle \tau_{\mu\nu} (\mathbf{q},\omega) \rangle
/ \overline{n}$, which then affects the current (and the conductivity) via $\sigma^{(0)}_{\lambda\nu}
(\omega)$, resulting in Eq.~(\ref{eqn:cond_viscosity}).

\begin{widetext}
By Eq.~(\ref{viscnofield}) or (\ref{viscfield}),
to arrive at the viscosity one should subtract the inverse (internal) compressibility contribution
from $\overline{\chi}_{\mu\nu\alpha\beta}(\omega)$. Defining
$\overline{\eta}_{\mu\nu\alpha\beta}(\omega)
=\half[\eta_{\mu\nu\alpha\beta}(\omega)+\eta_{\alpha\nu\mu\beta}(\omega)]$,
we have
\begin{align} \label{eqn:eta_sig_kappa}
  \overline{\eta}_{\mu\nu\alpha\beta}(\omega)
  = &
  \half m^2
  \left( \omega \delta_{\nu\lambda} - i \omega_c \epsilon_{\nu\lambda} \right)
  \left. \frac{\partial^2 \sigma_{\lambda\rho}(\mathbf{q},\omega)}{\partial q_\mu \partial q_\alpha}
  \right|_{\mathbf{q}=\mathbf{0}}
  \left( \omega \delta_{\rho\beta} - i \omega_c \epsilon_{\rho\beta} \right) 
 -\frac{i\kappa^{-1}_\text{int}}{2  \omega^+} \left( \delta_{\mu\nu} \delta_{\alpha\beta} +
 \delta_{\mu\beta} \delta_{\nu\alpha} \right).
\end{align}

For rotationally-invariant two-dimensional systems, there are only three independent
(frequency-dependent) coefficients of viscosity: the bulk viscosity $\zeta$, shear viscosity
$\eta^\text{sh}$, and Hall viscosity $\eta^H$ [cf.\ Eqs.~(\ref{eqn:etas})--(\ref{eqn:etah1})].
The relation~(\ref{eqn:etatot_cond_B}) can be used to extract the viscosity coefficients at all
frequencies from the conductivity [taking $\mathbf{q}$ in the $x$- (i.e.\ $1$-) direction without
loss of generality],
\begin{align}
  \label{eqn:zeta_cond}
  \zeta(\omega) + \frac{i \kappa^{-1}_\text{int}}{\omega^+} = \overline{\chi}_{1111}(\omega) -
  \overline{\chi}_{1212}(\omega) = & \frac{m^2}{2}
  \left. \frac{\partial^2}{\partial q_x^2}
  \left\{
  \left( \omega^2 - \omega_c^2 \right) \left[ \sigma_{11}(\mathbf{q},\omega)
  - \sigma_{22}(\mathbf{q},\omega) \right]
  \right\} \right|_{\mathbf{q} = \mathbf{0}},
  \\
  \label{eqn:etas_cond}
  \eta^\text{sh}(\omega) = \overline{\chi}_{1212}(\omega) = & \frac{m^2}{2}
  \left. \frac{\partial^2}{\partial q_x^2}
  \left\{
  \omega^2 \sigma_{22}(\mathbf{q},\omega)
  + \omega_c^2 \sigma_{11}(\mathbf{q},\omega)
  + 2 i \omega \omega_c \sigma^H(\mathbf{q},\omega)
  \right\}
  \right|_{\mathbf{q} = \mathbf{0}},
  \\
  \label{eqn:etah_cond}
  \eta^H(\omega) = \frac{\overline{\chi}_{1112} (\omega) - \overline{\chi}_{1211} (\omega)}{2} = &
  \frac{m^2}{2} 
  \left.
  \frac{\partial^2}{\partial q_x^2}
  \left\{
  \left( \omega^2 + \omega_c^2 \right) \sigma^H(\mathbf{q},\omega)
  - i \omega \omega_c \left[ \sigma_{11}(\mathbf{q},\omega) + \sigma_{22}(\mathbf{q},\omega) \right]
  \right\}
  \right|_{\mathbf{q} = \mathbf{0}},
\end{align}
where $\sigma^H (\mathbf{q},\omega) = [\sigma_{12}(\mathbf{q},\omega)-\sigma_{21}(\mathbf{q},\omega)]/2$
is the Hall conductivity. Here we have separated $\zeta(\omega)$ from $i
\kappa_{\mathrm{int}}^{-1}/\omega$,
 according to our analysis of the trace part of the viscosity tensor in Sec.\ \ref{nondiv}. For zero
  magnetic field ($\omega_c=0$, and $\kappa^{-1}_{\mathrm{int}}=\kappa^{-1}$),
 relations similar to the first two are fairly well known, but often are written in terms of the
 transverse and longitudinal parts of $\sigma_{\lambda\rho}$; see for example Ref.\
  \onlinecite{Taylor2010}, in which however the $\kappa^{-1}$ term is absent. The Hall viscosity
  decouples in this case.

Alternatively, again for a rotationally-invariant system in two dimensions, we can invert
Eq.~(\ref{eqn:etatot_cond_B}) to obtain the $q^2$ part of the conductivity tensor, $\sigma^{(2)}_{\lambda
 \rho} (\mathbf{q},\omega)$.
For vanishing magnetic field we find:
\begin{equation} \label{eqn:sig2_gen_b0}
  \sigma^{(2)}_{\lambda \rho} (\mathbf{q},\omega)
  =
  \frac{1}{m^2 \omega^{+ 2}} 
  \left[
  \left(\zeta(\omega) + \frac{i \kappa^{-1}}{\omega^+} \right) q_\lambda q_\rho
  + \eta^\text{sh}(\omega) q^2 \delta_{\lambda \rho}
  + \eta^H(\omega) q^2 \epsilon_{\lambda \rho}
  \right].
\end{equation}
For zero magnetic field, we see that the Hall viscosity can be obtained just from the $q^2$ part of the
Hall conductivity, at all frequencies.

In the presence of a magnetic field, the expressions are more complicated, and for later use we retain
only terms that are non-vanishing at zero frequency, assuming that $\zeta$ and $\eta^H$ do not diverge
as $\omega\to0$, and that $\eta^\text{sh}$ does not diverge more rapidly than $1/\omega^+$. Then we have
as $\omega\to0$
\begin{equation} \label{eqn:sig2_gen_b}
  \sigma^{(2)}_{\lambda \rho} (\mathbf{q},\omega)
  \sim \frac{1}{m^2 \omega_c^2}
  \left[
  \left(\zeta(\omega=0) + \frac{i \kappa_\mathrm{int}^{-1}}{\omega^+} \right) q_\mu \epsilon_{\mu \lambda}
  q_\alpha \epsilon_{\alpha \rho}
  + \eta^\text{sh}(\omega) q^2 \delta_{\lambda \rho}
  + \left(\eta^H(\omega=0)-
  \frac{\kappa_\mathrm{int}^{-1}}{\omega_c}+\frac{2i\omega}{\omega_c}\eta^\text{sh}(\omega) \right) q^2
  \epsilon_{\lambda \rho}
  \right].
\end{equation}
This can be used to obtain $\eta^H(\omega=0)$ from the $q^2$ part of the conductivity at $\omega\to0$.
If $\eta^\text{sh}$ is non-diverging as $\omega\to0$, the antisymmetric part of the equation reduces to
the relation found by Hoyos and Son \cite{Hoyos2012} between the Hall viscosity, internal compressibility,
and $q^2$ part of the Hall conductivity at zero frequency in the presence of a magnetic field, which they
obtained for a gapped quantum Hall system at zero temperature (see eq.\ (\ref{eqn:kint_hoyosson}) for
the equivalence of $\kappa_{\mathrm{int}}^{-1}$ with their expression). That relation is now seen to hold
even when $\zeta(\omega=0)$ and $\eta^\text{sh}(\omega=0)$ are non-zero but finite, with a generalization
for the case of diverging $\eta^\text{sh}$. Our formulas above give the generalization to all frequencies
as well.

\end{widetext}


\section{Examples of the Kubo formulas and conductivity relation}\label{sec5}

In this section, we consider applications of the Kubo formulas to various simple model examples with
rotational invariance, at zero temperature except where otherwise stated.

\subsection{The free Fermi and Bose gases}\label{freefermi}

As a first application of our Kubo formula, let us compute the viscosity of the free Fermi gas in $d$
dimensions at zero temperature and zero magnetic field. The unperturbed Hamiltonian for this system is
\begin{equation}
H_0=\frac{1}{2m}\sum_ip_\mu^ip_\mu^i.
\end{equation}
For a system of $N$ particles enclosed in a very large box, the ground state energy $E_0$ has the form
\begin{equation}
E_0=\left<H_0\right>_0\propto (L^d)^{-\frac{2}{d}}.
\end{equation}
From Eq.~(\ref{momflux}), the stress tensor is given by
\begin{equation}
T _{\mu\nu}=-i\left[H_0,J_{\mu\nu}\right]=\frac{1}{m}\sum_ip^i_\mu p^i_\nu.
\end{equation}
Because $T _{\mu\nu}$ is independent of time, the time-integral term in the stress-stress form of the
Kubo formula~(\ref{stressstress}) vanishes, and we easily find
\begin{align}
X _{\mu\nu\alpha\beta}(\omega)&=\frac{1}{\omega^+}\left<\left[T
_{\mu\nu}(0),J_{\alpha\beta}(0)\right]\right> \nonumber \\
&=\frac{i}{\omega^+}\left(\delta_{\mu\beta}\left<T _{\nu\alpha}\right>_0+\delta_{\mu\alpha}\left<T
_{\nu\beta}\right>_0\right)\nonumber \\
&=\frac{2i}{d\omega^+}E_0\left(\delta_{\mu\beta}\delta_{\nu\alpha}+\delta_{\mu\alpha}\delta_{\nu\beta}\right),
\end{align}
where we have used rotational invariance to express $\left<T _{\mu\nu}\right>_0$ in terms of $E_0$:
$P=2E_0/(dL^d)$. Lastly, we have that
\begin{equation}
P-\kappa^{-1}=-\left(\frac{\partial E_0}{\partial (L^d)}\right)_N-L^d\left(\frac{\partial^2 E_0}{\partial
(L^d)^2}\right)_N=-\frac{4}{d^2L^d}E_0,
\end{equation}
and so from Eq.~(\ref{viscnofield}), we see that the zero-temperature viscosity tensor of the free Fermi
gas is given by
\begin{equation}
\frac{2i}{d\omega^+}\frac{E_0}{L^d}\left(\delta_{\mu\beta}\delta_{\nu\alpha}+\delta_{\mu\alpha}\delta_{\nu\beta}
-\frac{2}{d}\delta_{\mu\nu}\delta_{\alpha\beta}\right).
\end{equation}
This is purely a shear viscosity --- the bulk viscosity of the free Fermi gas is not only not infinite,
as anticipated, but identically zero at zero temperature. Furthermore, the dissipative shear viscosity
coefficient $\eta^{\mathrm{sh}}$, given by the real part of the scalar prefactor of
$\eta_{\mu\nu\alpha\beta}$, is
\begin{equation}
\eta^{\mathrm{sh}}(\omega)=\frac{2\pi E_0}{dL^d}\delta(\omega),
\end{equation}
which is zero for all $\omega\neq 0$, and infinite at $\omega=0$. Such delta-function divergences in
response functions are characteristic of noninteracting systems, and indicate that in response to shear
strains, the free Fermi gas accelerates without bound. In an interacting Fermi gas at non-zero temperature,
the $\delta$-function is broadened and the zero-frequency shear viscosity is finite. As the temperature
tends to zero, it approaches a $\delta$-function; the limit is continuous in the space of distributions.

We note that one can derive the same result by first finding the
$\mathbf{q}$-dependent conductivity through elementary Green function techniques,
\begin{multline}
  \sigma_{\nu\beta}(\mathbf{q},\omega) =
  \frac{i \overline{n}}{m \omega^+}\delta_{\nu\beta} 
  + \frac{2 i}{d m^2 \omega^{+ 3}} \frac{E_0}{L^d} 
  \left( \delta_{\nu \beta} \mathbf{q}^2
  + 2 q_\nu q_\beta
  \right) \\
  + O(\mathbf{q}^4),
\end{multline}
and substituting this expression into the zero magnetic field version of the viscosity-conductivity
relation, Eq.~(\ref{eqn:eta_sig_kappa}) [or, equivalently, comparing it with Eq.~(\ref{eqn:sig2_gen_b0})].

For the free Bose gas, the calculations are very similar, except of course that the Bose distribution
must be used in place of the Fermi distribution. In particular, at zero temperature, the ground state
is a Bose condensate with all particles in the $\bk=0$ state, instead of filling a Fermi sea. In this
case, the ground state energy $E_0$, pressure $P$, and inverse compressibility $\kappa^{-1}$, are all
zero. Then using similar arguments as above, the viscosity response tensor vanishes identically. At
positive temperature, the results take the same form as above, but $E_0$ is replaced by the average
energy $E=\langle H_0\rangle_0$. Then a $\delta$-function real shear viscosity, with coefficient
proportional to $E/L^d$, again appears; the bulk viscosity remains zero.

\subsection{The integer and fractional quantum Hall fluids}\label{sec6}

Let us now compute the viscosity for non-interacting electrons in an external magnetic field. The
Hamiltonian $H_0$ is
\begin{equation}
H_0=\frac{1}{2m}\sum_i{\pi_\mu^i\pi_\mu^i}.
\end{equation}
We can take our unperturbed state to have the lowest $\nu$ Landau levels occupied, in a region of area
$L^2$, and so the ground state energy is
\begin{equation}
E_0=\overline{n}\overline{s}\omega_cL^2=\frac{\nu^2}{2m\phi_0}B^2L^2,
\end{equation}
where as before $\overline{s}=\nu/2$ is minus the average orbital spin per particle. However, this is
unnecessarily restrictive here, and at any temperature and any average filling factor $\nu$ the average
energy is again $\langle H_0\rangle_0 =\overline{n}\overline{s}\omega_cL^2$. From Eq.~(\ref{momflux2}),
we have that
\begin{equation}
T _{\mu\nu}=-i\left[H_0,J_{\mu\nu}\right]=\frac{1}{2m}\sum_j\left\{\pi_\mu^j,\pi_\nu^j\right\}.
\end{equation}

We will calculate $X$ using both the strain-strain and stress-stress Kubo formulas. It will be convenient
to work in the symmetric gauge, where the dilation generator $K$ takes the simple form
\begin{equation}
K=-\frac{1}{2}\sum_i\left\{x_\mu^i,\pi_\mu^i\right\}+\left\{\mathcal{B},\mathcal{P}\right\}.
\end{equation}
We can diagonalize the Hamiltonian $H_0$ with two sets of creation annihilation operators. Writing
$z^j=x^j+iy^j$ and $\bar{z}^j=x^j-iy^j$, these are (see e.g.\ Ref.\ \onlinecite{Read2010})
\begin{align}
b^j&=\frac{1}{\sqrt{2\mathcal{B}}}\left(\pi^j_x+i\pi^j_y\right), \\
a^j&={b^j}^\dag-i\sqrt{\frac{\mathcal{B}}{2}}\bar{z}^j,
\end{align}
satisfying
\begin{align}
\left[b^i,{b^j}^\dag\right]&=\left[a^i,{a^j}^\dag\right]=\delta_{ij}, \\
\left[b^i,{a^j}^\dag\right]&=\left[b^i,a^j\right]=0.
\end{align}
In terms of these operators, the Hamiltonian takes the simple form
\begin{equation}
H_0=\frac{\mathcal{B}}{m}\sum_i\left({b^i}^\dag b^i+\frac{1}{2}\right).
\end{equation}
The stress tensor can be written as
\begin{equation}
T _{\mu\nu}=\delta_{\mu\nu}
H_0+\frac{\mathcal{B}}{2m}\sum_i{\left(({{b^i}^\dag}^2+{b^i}^2)\tau^{z}_{\mu\nu}+({{b^i}^\dag}^2-{b^i}^2)
\tau^{x}_{\mu\nu}\right)},
\end{equation}
where $\tau^x$ and $\tau^z$ are the standard $x$ and $z$ Pauli matrices. The shear generator
$J^{\mathrm{sh}}_{\mu\nu}$ from Eq.~(\ref{eqmagstraintless}) takes the form
\begin{align}
J^{\mathrm{sh}}_{\mu\nu}&=\sum_j\frac{i}{4}\left({{b^j}^\dag}^2-{b^j}^2-{a^j}^2+{{a^j}^\dag}^2\right)
\tau^z_{\mu\nu} \nonumber \\
&-\sum_j\frac{1}{4}\left({b^j}^2+{{b^j}^\dag}^2-{a^j}^2-{{a^j}^\dag}^2\right)\tau^x_{\mu\nu}\nonumber \\
&+\sum_j\frac{1}{2}\left({b^j}^\dag{b^j}-{a^j}^\dag{a^j}\right)\epsilon_{\mu\nu} \label{landaushear}
\end{align}
(note we could have written $\epsilon_{\mu\nu}=i\tau^{y}_{\mu\nu}$), and the dilation generator $K$ can
be written
\begin{equation}
K=\left\{\mathcal{B},\mathcal{P}\right\}+i\sum_j\left({a^j}^\dag{b^j}^\dag-{a^j}{b^j}\right).
\label{landaudilate}
\end{equation}
The shear generators agree with those presented by Read and Rezayi \cite{Read2010}. The last ingredient
we need is the commutation relations between $\mathcal{P}$ and the creation and annihilation operators.
A computation shows that these are
\begin{align}
\left[\mathcal{P},b^j\right]&=\frac{i}{2\mathcal{B}}{a^j}^\dag, \\
\left[\mathcal{P},a^j\right]&=\frac{i}{2\mathcal{B}}{b^j}^\dag.
\end{align}
It is worth noticing also that
\be
[K,a^i]=[K,b^i]=0.
\ee

Now, using the stress-stress Kubo formula~(\ref{stressstress}), we find rather directly that
\begin{align}
X_{\mu\nu\alpha\beta}(\omega)&=\frac{E_0}{{\omega^+}^2-4\omega_c^2}
\left[i\omega^+(\delta_{\mu\beta}\delta_{\nu\alpha}
-\epsilon_{\mu\beta}\epsilon_{\nu\alpha})\right.\nonumber \\
&\left.-2\omega_c(\delta_{\nu\alpha}\epsilon_{\mu\beta}-\delta_{\mu\beta}\epsilon_{\alpha\nu})\right]+\frac{iE_0}
{\omega^+}\delta_{\mu\nu}\delta_{\alpha\beta},
\end{align}
where in $\omega_c=B/m$, $B$ is again the value at the center of the narrow wavepacket over $\mathcal{B}$
values. Finally, applying
 Eqs.~(\ref{pint}) and (\ref{eqn:kint_hoyosson}), we have
\begin{align}
P_{\mathrm{int}}&=\overline{n}\overline{s}\omega_c=\frac{E_0}{L^2} \\
\kappa_{\mathrm{int}}^{-1}&=2\overline{n}\overline{s}\omega_c=2\frac{E_0}{L^2}
\end{align}
and hence
\begin{align}
\eta_{\mu\nu\alpha\beta}(\omega)&=\frac{E_0}{L^2({\omega^+}^2-4\omega_c^2)}\left[i\omega^+(\delta_{\mu\beta}
\delta_{\nu\alpha}-\epsilon_{\mu\beta}\epsilon_{\nu\alpha})\right.\nonumber \\
&\left.-2\omega_c(\delta_{\nu\alpha}\epsilon_{\mu\beta}-\delta_{\mu\beta}\epsilon_{\alpha\nu})\right].
\label{hallviscw}
\end{align}
At non-zero temperature, the result is the same, with the ground state energy density $E_0/L^2$ replaced
by $\langle H_0\rangle_0/L^2$. We notice that the trace on $\mu\nu$ or on $\alpha\beta$ vanishes at all
frequencies---the bulk viscosity is identically zero.
In the remainder, the spectral density consists solely of $\delta$-functions at the frequencies
$\pm 2\omega_c$, which represent transitions in which the Landau level index changes by $\pm 2$. This
is because of the quadrupolar nature of the stress, and the fact that only the operators $b^i$ appear
in $T_{\mu\nu}$. In the $\omega\rightarrow 0$ limit, the viscosity reduces to the Hall viscosity
\begin{equation}
\eta_{\mu\nu\alpha\beta}(\omega=0)=\half
\overline{n}\,\overline{s}\left(\delta_{\nu\alpha}\epsilon_{\mu\beta}
-\delta_{\mu\beta}\epsilon_{\alpha\nu}\right),
\end{equation}
in agreement with known results\cite{Avron1995,Levay1995,Read2009}. It is not surprising that the
zero-frequency bulk and shear viscosities vanish when the temperature is zero and there is a gap in the
spectrum; that they do so in other cases as well is due to the non-interacting nature of the system.

To make contact with the adiabatic calculation of the Hall
viscosity\cite{Avron1995,Levay1995,Read2009,Read2010}, we can also calculate the viscosity from
the strain-strain formula~(\ref{strainstrain}). Naturally, this must give the same result
as Eq.~(\ref{hallviscw}), however it is enlightening to see how this comes about. If we evaluate
just the equal-time contact term, we find
\begin{align}
-i\left<\left[J_{\mu\nu},J_{\alpha\beta}\right]\right>_0&=\half\left(\frac{E_0}{\omega_c}
-\sum_j\left<{a^j}^\dag{a^j}+\half\right>_0\right)\nonumber \\
&\times(\delta_{\nu\alpha}\epsilon_{\mu\beta}-\delta_{\mu\beta}\epsilon_{\alpha\nu}).
\end{align}
The intra-Landau level $\langle a^\dagger a \rangle$ term is larger than $O(N)$, but we know that it
must cancel. On calculating the time-integral contribution, we find as $\omega\to0$
\begin{align}
\omega^+\int_0^\infty&dt\,e^{i\omega^+t}\left<\left[J_{\mu\nu}(t),J_{\alpha\beta}(0)\right]\right>_0\sim
\nonumber \\
&\frac{iE_0}{\omega^+}\delta_{\mu\nu}\delta_{\alpha\beta}+\frac{\sum_j\left<a^{j\dag}{a^j}
+\frac{1}{2}\right>_0}{2}(\delta_{\nu\alpha}\epsilon_{\mu\beta}
-\delta_{\mu\beta}\epsilon_{\alpha\nu}).
\end{align}
Here the $i/\omega^+$ term arose because there is a part of the correlation function that is linear in
$t$ for large $t$. That is present because the trace $K=J_{\mu\mu}$ has time derivative
$-i[K,H_0]=-T_{\mu\mu}=-2H_0$, which is time independent.
Combining these, we see that the intra-Landau level contributions to $X$ exactly cancel in the
final result. The traceless part of this result was obtained previously by considering the transport
of degenerate subspaces in the disk geometry in the infinite plane in the adiabatic transport
formulation of viscosity
\cite{Read2010}.

We can also use the strain-strain form of the Kubo formula (\ref{strainstrain}) to calculate
the $\omega\rightarrow 0$ traceless viscosity of a fractional quantum Hall fluid in the lowest Landau
level. Although the presence of interactions makes manipulating the stress tensor rather cumbersome, we
can calculate the necessary matrix elements of the traceless strain generators (\ref{landaushear})
for certain trial states. Because of the non-standard time dependence of the dilation generator $K$,
we avoid computation of the diagonal response function.

Let us consider a fractional quantum Hall system in the lowest Landau level. We assume that the
interaction is rotationally invariant, and commutes with all the Landau level raising and lowering
operators ${b^i}^\dag $, $b^i$. Further, we assume it is one of the ``special'' Hamiltonians for which
exact zero-interaction-energy ground (and edge, and quasihole) states lying in the lowest Landau level
are known (for more details on these, see Ref.\ \onlinecite{Read2010}). Since
we are only concerned with the traceless strain generators, we are free to work with states
with a fixed magnetic field $B$.  We denote by $\left|0\right>$ the unique minimum angular momentum
ground state of the fluid.  Using the strain-strain form of the traceless response function
$X^{\mathrm{sh}}$ in Eq.~(\ref{strainstrain}), along with the commutation relations
Eq.~(\ref{straincomm}), we have
\begin{align}
X^{\mathrm{sh}}_{\mu\nu\alpha\beta}(\omega)&=\delta_{\nu\alpha}\left<0|J^{\mathrm{sh}}_{\mu\beta}|0\right>
-\delta_{\mu\beta}
\left<0|J^{\mathrm{sh}}_{\alpha\nu}|0\right>\nonumber \\
&+\omega^+\int_0^\infty dt\,e^{i\omega^+t}\left<0|\left[J^{\mathrm{sh}}_{\mu\nu}(t),J^{\mathrm{sh}}
_{\alpha\beta}
(0)\right]|0\right>.
\end{align}
Using the fact that the system is in the lowest Landau level, we can evaluate the first term above to get
\begin{align}
X^{\mathrm{sh}} _{\mu\nu\alpha\beta}(\omega)=&{}-\half\sum_j\left<0\right|
a^{j\dag} a^j\left|0\right>\left(\delta_{\nu\alpha}\epsilon_{\mu\beta}-\delta_{\mu\beta}
\epsilon_{\alpha\nu} \right) \nonumber \\
&+\omega^+\int_0^\infty dt\,e^{i\omega^+t}\left<0\right|\left[J^{\mathrm{sh}}_{\mu\nu}(t),J^{\mathrm{sh}}
_{\alpha\beta}
(0)\right]\left|0\right>\label{fqhkubo}.
\end{align}
In the limit $\omega\rightarrow 0$, we can evaluate the time-integral term. Note that in this limit,
the only nonzero contributions come from elements of the expectation value which are independent of
time (there is no other kind of non-oscillatory time-dependence for these traceless components).
Let $\{\left|D\right>:D=0,1,\ldots\}$ be an orthonormal basis for the subspace degenerate with the
ground state (including the ground state $\left|0\right>$), and $\{\left|e\right>:e=1,2\ldots\}$
an orthonormal basis for the space of all eigenstates with energy larger than that of the ground state
(thus we are assuming a discrete spectrum, as in a finite system).
Now we use the following fact:
Given a system with a discrete spectrum, and denoting by $P_0=\sum_{D}\left|D\right>\left<D\right|$
the projection operator onto the lowest-energy subspace, we have
\begin{align}
\lim_{\omega\rightarrow 0}\omega^+\int_0^\infty dt\,e^{i\omega^+t}\left<0\right|A(t)B(0)\left|0\right>
=i\left<0\right|A(0)P_0B(0)\left|0\right>
\end{align}
for any operators $A$ and $B$. (In fact, the identity continues to hold if $|0\rangle$ is replaced by any
state in the degenerate subspace.) Using this in $X^\text{sh}_{\mu\nu\alpha\beta}(\omega\to0)$, we obtain
exactly the expression that results from adiabatic transport of a degenerate subspace as in Ref.\
\onlinecite{Read2010}:
\be
X^\text{sh}_{\mu\nu\alpha\beta}(\omega=0)=-i\langle 0|J_{\mu\nu}P_\perp J_{\alpha\beta}|0\rangle
+i\langle 0|J_{\alpha\beta}P_\perp J_{\mu\nu}|0\rangle,
\ee
where $P_\perp=1-P_0$. Thus it is the expectation of the commutator of $J$'s, as in the contact term, but
in the intermediate-state sum, the states degenerate with the ground state are {\em omitted}. (The
noninteracting system considered above is a particular case, in which all intra-Landau level effects
cancel; this does not occur in fractional quantum Hall states in interacting systems.)

In the present case, using Eq.~(\ref{landaushear}) and noting that $\sum_ia^{i\dag 2}$ leaves the
ground state in the degenerate subspace while $\sum_ia^{i2}$ takes it out of that subspace\cite{Read2010},
we have
\begin{align}
\left<0\right|&J^{\mathrm{sh}}_{\mu\nu}P_0J^{\mathrm{sh}}_{\alpha\beta}\left|0\right>
-\left<0\right|J^{\mathrm{sh}}_{\alpha\beta} P_0J^{\mathrm{sh}}_{\mu\nu}\left|0\right>=\nonumber \\
&\frac{i}{8}\sum_{ij}\left<0\right|{a_i}^2{a_j^\dag}^2\left|0\right>\left(\delta_{\mu\beta}\epsilon_{\alpha\nu}
-\delta_{\alpha\nu}\epsilon_{\mu\beta}\right). \label{fraccorrection}
\end{align}
Inserting Eq.~(\ref{fraccorrection}) into the Kubo formula~(\ref{strainstrain}), we find
\begin{align}
\lefteqn{X^{\mathrm{sh}}_{\mu\nu\alpha\beta}(\omega\rightarrow
0)}\quad&\nonumber\\
&={}-\half\left(\delta_{\nu\alpha}\epsilon_{\mu\beta}-\delta_{\mu\beta}\epsilon_{\alpha\nu}\right)
\left<\sum_i {a^i}^\dag a^i-\frac{1}{4}\sum_{ij}{a^i}^2{a^j}^{\dag 2}\right>_0\nonumber \\
&=\half\overline{s}N\left(\delta_{\nu\alpha}\epsilon_{\mu\beta}-\delta_{\mu\beta}\epsilon_{\alpha\nu}
\right), \label{fqhchi}
\end{align}
(at leading order in the thermodynamic limit) where the second line follows from the matrix elements
computed in Ref.\ \onlinecite{Read2010}. Thus
again we obtain for the viscosity
\begin{equation}
\eta_{\mu\nu\alpha\beta}(\omega\rightarrow
0)=\half\overline{s}\overline{n}\left(\delta_{\nu\alpha}\epsilon_{\mu\beta}
-\delta_{\mu\beta}\epsilon_{\alpha\nu}\right).
\end{equation}

\subsection{The viscosity-conductivity relation for quantum Hall systems}
\label{subsec:etasigmahall}

In Sec.\ \ref{sec4}, we obtained a general relation between the viscosity and the $q^2$ part of
the conductivity for a system in a magnetic field.
We have already commented there that as $\omega\to0$ we obtain a relation
\be
\half B^2\frac{\partial^2}{\partial
q_x^2}\sigma^H(\mathbf{q},\omega=0)|_{\mathbf{q}=\mathbf{0}}=\eta^H(\omega=0)-
\frac{\kappa_\mathrm{int}^{-1}}{\omega_c}+\frac{2i\omega}{\omega_c}\eta^\text{sh}(\omega)
\ee
which is more general than that of Hoyos and Son \cite{Hoyos2012}, but which reduces to theirs [using
eq.\ (\ref{eqn:kint_hoyosson})] when the shear viscosity is non-diverging at $\omega\to0$.

It is interesting to compare the full frequency-dependent relation with an exact calculation for the
integer quantum Hall state at filling factor $\nu$, based on the results of Chen \text{et al.}
\cite{Chen1989} for $\sigma_{\mu\nu} (\mathbf{q},\omega)$. Extracting the quadratic in $\mathbf{q}$ terms
in the conductivity from their calculations we find:
\begin{align}
   \label{eqn:sig11}
   \left. \frac{\partial^2 \sigma_{11}(\mathbf{q},\omega)}{\partial q_x^2} \right|_{\mathbf{q}=\mathbf{0}}
   = &
   i \frac{\omega}{\omega_c} \frac{\nu^2 \ell^2}{2 \pi} 
   \left( \frac{\omega_c^2}{\omega^{+ 2} - 4 \omega_c^2} - \frac{\omega_c^2}{\omega^{+ 2} - \omega_c^2}
   \right), \\
   \label{eqn:sig22}
   \left. \frac{\partial^2 \sigma_{22}(\mathbf{q},\omega)}{\partial q_x^2} \right|_{\mathbf{q}=\mathbf{0}}
   = &
   -\frac{\omega_c}{i \omega^+} \frac{\nu^2 \ell^2}{2 \pi} 
   \left( \frac{4\omega_c^2}{\omega^{+ 2} - 4\omega_c^2} - \frac{3 \omega_c^2}{\omega^{+ 2} - \omega_c^2}
   \right), \\
   \label{eqn:sigh}
   \left. \frac{\partial^2 \sigma^H(\mathbf{q},\omega)}{\partial q_x^2} \right|_{\mathbf{q}=\mathbf{0}} = &
   -2 \frac{\nu^2 \ell^2}{2 \pi} 
   \left( \frac{\omega_c^2}{\omega^{+ 2} - 4 \omega_c^2} - \frac{\omega_c^2}{\omega^{+ 2} - \omega_c^2}
   \right),
\end{align}
where $\ell = 1/\sqrt{B}$ is the magnetic length.
Substituting these expressions into Eqs.~(\ref{eqn:zeta_cond})--(\ref{eqn:etah_cond}) we arrive at
(using again $\kappa^{-1}_\text{int} = \nu \overline{n} \omega_c$, $\overline{s}=\nu/2$):
\begin{align}
  \zeta (\omega) = & 0 ,\\
  \eta^\text{sh} (\omega) = & \overline{n} \overline{s} \frac{i \omega_c \omega}{4\omega_c^2-\omega^{+
  2}},\\
  \eta^H (\omega) = & \overline{n} \overline{s} \frac{2 \omega_c^2}{4\omega_c^2-\omega^{+ 2}},
\end{align}
in full agreement with Eq.~(\ref{hallviscw}).

\subsection{Complex $\ell$-wave paired superfluids in two dimensions}\label{sec7}

Lastly, we shall consider a complex $\ell$-wave paired superfluid of fermions in two dimensions. The
model mean-field Hamiltonian we shall consider takes the form
\begin{align}
H_0&=\int d^2x\,\psi^\dag(\mathbf{x})\left(-\frac{1}{2m}\nabla^2-\mu\right)\psi(\mathbf{x})\nonumber \\
&+\frac{1}{2}\int\int d^2x\,d^2x'\, \Delta(\mathbf{x}-\mathbf{x}')\psi^\dag(\mathbf{x})
\psi^\dag(\mathbf{x}')
+h.c., \label{pairedham}
\end{align}
where the pairing function $\Delta$ transforms as an $\ell$-wave under rotations. We note that we are
now working with a number-non-conserving system, in which the chemical potential appears as a parameter.
Previously we worked with systems at fixed number, however, generalizations to fixed chemical potential
(the grand canoncial ensemble), or even as here to systems in which particle number is not conserved,
should be reasonably self-evident. The terms in the Hamiltonian that violate number conservation
also violate angular momentum conservation; the system is not rotationally invariant. However,
the operator $\Theta=\epsilon_{\nu\mu}J_{\mu\nu}-\half\ell\hat{N}$ is still a conserved quantity, where
$\epsilon_{\nu\mu}J_{\mu\nu}$ is the angular momentum operator, and $\hat{N}$ is the number operator.

Using the continuity
equation~(\ref{contnofield}), we find that the second-quantized strain generator is given by
\begin{align}
J_{\mu\nu}&=-\int d^2x\,x_\mu g_\nu(\mathbf{x}) \nonumber \\
&=\frac{i}{2}\int d^2x\,x_\mu\left(\psi^\dag(\mathbf{x})\frac{\partial\psi(\mathbf{x})}{\partial
 x_\nu}-\frac{\partial\psi^\dag(\mathbf{x})}{\partial x_\nu}\psi(\mathbf{x})\right). \label{strain2q}
\end{align}
The general relations such as eq.\ (\ref{momflux}) between stress and strain generators still hold,
as do the Ward identities for the response function, and their consequences. We should view the system
as a non-rotationally-invariant case, and we commented on these as we went along.
To compute the viscosity, we shall use the strain-strain form of the Kubo formula, and specialize only to
the $\omega\rightarrow0$ limit. From Eqs.~(\ref{straincomm}) and~(\ref{strainstrain}), the response
function $X$ is given by
\begin{align}
X_{\mu\nu\alpha\beta}(\omega\rightarrow
0)&=\left(\delta_{\nu\alpha}\left<J_{\mu\beta}\right>_0-\delta_{\mu\beta}\left<J_{\alpha\nu}\right>_0
\right) \nonumber \\
&+\lim_{\omega\rightarrow
0}\omega^+\int_0^{\infty}dt\,e^{i\omega^+t}\left<\left[J_{\mu\nu}(t),J_{\alpha\beta}(0)\right]\right>_0,
\end{align}
with averages taken with respect to the ground state of the system in the plane geometry.

We now proceed to evaluate this expression. Introducing momentum space creation and annihilation operators
$c_\mathbf{k}$ and $c^\dag_{\mathbf{k}}$, Eq.~(\ref{strain2q}) for the strain generators becomes
\begin{equation}
J_{\mu\nu}=-\frac{1}{2L^2}\sum_{kk'}\int d^2x\,x_\mu(k_\nu+k'_{\nu})c_{k'}^\dag
c_{k}e^{i(\mathbf{k}-\mathbf{k}')\cdot\mathbf{x}},
\end{equation}
and in the limit of large system size $L\rightarrow\infty$
\begin{align}
J_{\mu\nu}=\frac{iL^2}{4(2\pi)^2}\int &d^2k\,d^2k'\,(k_\nu+k'_\nu)c_{k'}^\dag c_{k}\times\nonumber\\
&\times\left(\frac{\partial}{\partial k_\mu}-\frac{\partial}{\partial
k'_\mu}\right)\delta(\mathbf{k}-\mathbf{k'}).
\end{align}
(This clearly has the form of a strain generator in momentum space.)

To proceed further, we use the Bogoliubov transformation
\begin{align}
c_\mathbf{k}&=u_\mathbf{k}^*\alpha_\mathbf{k}-v_{-\mathbf{k}}\alpha^\dag_{-\mathbf{k}}, \nonumber \\
c^\dag_\mathbf{k}&=u_\mathbf{k}\alpha^\dag_\mathbf{k}-v_{-\mathbf{k}}^*\alpha_{-\mathbf{k}},
\end{align}
where
\be
\left\{\alpha_\mathbf{k},\alpha_\mathbf{k'}^\dagger\right\}=\delta_{\mathbf{k}\mathbf{k}'}
\ee
and other anticommutators vanish.
We have a gauge freedom in choosing the phases of $u_\mathbf{k}$ and $v_\mathbf{k}$, and for convenience
we shall work in the gauge where $u_\mathbf{k}$ is real. This implies that $v_\mathbf{k}$ transforms as
an $\ell$-wave under rotations. The explicit forms of $u_\mathbf{k}$ and $v_\mathbf{k}$ will not be needed.
The Hamiltonian then takes form (see e.g.\ Ref.\ \onlinecite{Read2000})
 \begin{equation}
 H_0=E_0+\sum_{\mathbf{k}}{\varepsilon_\mathbf{k}\alpha^\dag_\mathbf{k}\alpha_\mathbf{k}},
 \end{equation}
 where $E_0$ is the ground state energy, and
 \be
 \varepsilon_\mathbf{k}=\sqrt{\left(\frac{k^2}{2m}-\mu\right)^2+\left|\Delta_\mathbf{k}\right|^2}
 \ee
is the quasiparticle dispersion relation. We work at parameters for which $\varepsilon_\mathbf{k}>0$
(gapped) at all $\mathbf{k}$.

 We turn first to the contact term in Eq.~(\ref{strainstrain}). Using the properties of the
 $\alpha_{\mathbf{k}}$ operators, we find
 \begin{align}
 \left<J_{\mu\nu}\right>_0&=-\frac{iL^2}{4(2\pi)^2}\int d^2k\,k_\nu\left(v_{\mathbf{k}}\frac{\partial
 v^*_{\mathbf{k}}}{\partial k_\mu}-v^*_{\mathbf{k}}\frac{\partial v_{\mathbf{k}}}{\partial
 k_\mu}\right)\nonumber \\
 \end{align}
This is the continuum limit of the expression given in Ref.~\onlinecite{Read2010}. From rotational
covariance, it follows that this term is purely antisymmetric. Writing
\begin{equation}
v_\mathbf{k}=\left|v_k\right|e^{i\phi_{\mathbf{k}}},
\end{equation}
we have
\begin{align}
\left<J_{\mu\nu}\right>_0&=\frac{1}{2}\epsilon_{\mu\nu}\left<J_{xy}-J_{yx}\right>_0\nonumber \\
&=\frac{L^2}{4(2\pi^2)}\epsilon_{\mu\nu}\int{d^2k\,\left|v_k\right|^2\left(k_y\frac{\partial}
{\partial k_x}-k_x\frac{\partial}{\partial k_y}\right)\phi_{\mathbf{k}}} \nonumber\\
&=-\frac{L^2}{4(2\pi)^2}\epsilon_{\mu\nu}\int{kdk\,d\theta\,\left|v_k\right|^2\frac{\partial
\phi_{\mathbf{k}}}{\partial\theta}}\nonumber \\
&=-\frac{L^2}{4(2\pi)^2}\epsilon_{\mu\nu}\int{kdk\,\left|v_k\right|^2(\phi_\mathbf{k}(\theta=2\pi)
-\phi_\mathbf{k}(\theta=0))}.\nonumber
\end{align}
But since $v_\mathbf{k}$ transforms as an $\ell$-pole under rotations, we have
\begin{equation}
\phi_\mathbf{k}(\theta=2\pi)-\phi_\mathbf{k}(\theta=0)=2\pi\ell,\nonumber
\end{equation}
and hence
\begin{align}
\left<J_{\mu\nu}\right>_0&=-\frac{L^2}{4(2\pi)}\ell\epsilon_{\mu\nu}\int{kdk\,\left|v_{k}\right|^2} \\
&=-\frac{\left<N\right>_0\ell}{4}\epsilon_{\mu\nu}\\
&=\frac{1}{2}L^2\overline{n}\overline{s}\epsilon_{\mu\nu}.
\end{align}
Thus, the contact term gives the expected Hall viscosity $\eta^H=\frac{1}{2}\overline{s}\overline{n}$.

Next, we must consider the time integral term in the response function $X$. Because of the non-degenerate
ground state, and the gap in the
excitation spectrum, we expect that at low frequency, the only terms will be the pressure and
$\kappa^{-1}$ terms that go as $1/\omega^+$. But $\kappa^{-1}$ here will be the derivative of the
pressure with respect to size at fixed chemical potential $\mu$, not fixed number $N$, and will
vanish. The pressure itself is minus
the derivative of the ground state energy with volume, at fixed chemical potential, as usual
(this ``energy'' is really the grand thermodynamic potential), and is not expected to vanish (nor does
the usual inverse compressibility, which is defined using a derivative of pressure with volume at
fixed particle number, not fixed chemical potential). We have some difficulty with the formal
calculation, because
$\mathbf{k}$ space was most convenient to diagonalize the Hamiltonian,
but to make sense of the size dependence of the energy, we need the formalism of Appendix \ref{app},
with a confining potential (the use of periodic boundary conditions does not fit with the strain
generators, though we could use the approach of Appendix \ref{appC})). But we have seen in previous
sections how the pressure term emerges in the stress-stress
form, and thanks to the formalism, this is equivalent to the strain-strain form we wanted to use.
Hence we will not pursue this further.

Finally, let us apply the viscosity-conductivity relation~(\ref{eqn:etatot_cond_B}), which is still valid
in the present situation, even though number is not conserved. We consider only the $\ell=1$, or
$p_x + i p_y$, spinless superconductor.
The conductivity in that case was calculated by Lutchyn \textit{et al.} \cite{Lutchyn2008}.
The result for the Hall conductivity (the antisymmetric part) in the 2D limit to order $q^2$ reads
[cf.\ Eqs.\ (103) and (72) in that paper]:
\begin{equation}
  \sigma^H(\mathbf{q},\omega) = I(\omega) \frac{e^2}{4 \pi \hbar} \frac{v_F^2 q^2}{2 \omega^{+ 2}},
\end{equation}
where $v_F$ is the Fermi velocity and $I(\omega)$ is a dimensionless factor obeying $I(0)$=1.
Substituting this result into the viscosity-conductivity relation~(\ref{eqn:eta_sig_kappa}),
taking $\omega$ to zero, and remembering that the particle number density is related to the Fermi
wavevector $k_F$ by $\overline{n} = k_F^2 / (2 \pi)$ in 2D, we find that the
Hall viscosity is $\eta^H = \hbar \overline{n} \overline{s} / 2$, with $\overline{s}=1/2$,
in accordance with the discussion above, except for an apparent sign discrepancy: for $p+ip$,
$\overline{s}$ should be $-1/2$, not $+1/2$.

\section{Hall viscosity from electrodynamics of a fluid with orbital spin}
\label{electro}

In this section we present an alternative derivation of Eqs.~(\ref{eqn:etah1})-(\ref{eqn:etah2})
by examining the electrodynamics at small $\mathbf{q}$ and $\omega$ of a Galilean- and
rotationally-invariant fluid in the spirit of a low-energy effective description, in which we assume
there is an orbital spin $-\overline{s}$ per particle; we neglect bulk and
shear viscosity. We find the orbital spin contribution to the conductivity, and hence to the viscosity.

When one applies an electric field to such a fluid, to lowest order in $\mathbf{q}$ the electric
current response is
$\langle j_\alpha (\mathbf{q}, \omega) \rangle^{(0)} = \sigma^{(0)}_{\alpha \beta}(\omega)
E_\beta (\mathbf{q}, \omega)$;
by the continuity equation, the corresponding change in the particle number density is
$\delta \langle n(\mathbf{q}, \omega) \rangle^{(1)}  =
q_\alpha \langle j_\alpha(\mathbf{q}, \omega) \rangle^{(0)}/\omega^+$.
Since each particle carries orbital spin $-\overline{s}$,
this leads to a change in the magnetization density,
$\delta \mu (\mathbf{q},\omega) =
- \overline{s} \delta \langle n(\mathbf{q},\omega) \rangle^{(1)}/(2 m
)$.
Finally, this feeds into the electric current,
$\langle j_\nu (\mathbf{q},\omega) \rangle^{(2)}_M =
\epsilon_{\nu \mu} i q_\mu \delta \mu (\mathbf{q},\omega)$,
and gives the following order $\mathbf{q}^2$ contribution to the conductivity:
\begin{equation} \label{eqn:sig2m}
  \sigma^{(2),M}_{\nu \beta}(\mathbf{q},\omega) = -i \frac{\overline{s}}{2}
  \frac{q_\alpha q_\mu}{m \omega^+}
  \epsilon_{\nu \mu} \sigma^{(0)}_{\alpha \beta}(\omega).
\end{equation}
This contribution was previously discussed by Lutchyn et al.\ \cite{Lutchyn2008} in the zero magnetic
field case.

A similar contribution comes from the fact that by Maxwell's equations, a curl of the electric field
implies a time-dependent magnetic field,
$B(\mathbf{q},\omega) =
\epsilon_{\alpha \beta} q_\alpha E_\beta (\mathbf{q},\omega)/\omega$. The magnetic field couples
to the magnetization density, which is $-\overline{s}$ times the particle number density, divided by
$2m$; hence, it is equivalent to an electric potential
$V^B(\mathbf{q},\omega) = \overline{s} B(\mathbf{q},\omega)/(2 m
)$,
giving rise to an effective electric field
$E^B_\mu (\mathbf{q},\omega) = -i q_\mu V^B(\mathbf{q},\omega)$,
and thus to an electric current density
$\langle j_\nu(\mathbf{q},\omega) \rangle^{(2)}_B =
\sigma^{(0)}_{\nu \mu}(\omega) E^B_\mu(\mathbf{q},\omega)$.
The corresponding order $\mathbf{q}^2$ contribution to the conductivity is:
\begin{equation} \label{eqn:sig2b}
  \sigma^{(2),B}_{\nu \beta}(\mathbf{q},\omega) = -i \frac{\overline{s}}{2} \frac{q_\alpha q_\mu}
  {m \omega^+}
  \sigma^{(0)}_{\nu \mu}(\omega) \epsilon_{\alpha \beta}.
\end{equation}

Summing Eqs.~(\ref{eqn:sig2m})-(\ref{eqn:sig2b}) we find the total contribution of the orbital spin to
the conductivity:
\begin{equation} \label{eqn:sig2s}
  \sigma^{(2),\overline{s}}_{\nu \beta}(\mathbf{q},\omega) = -i \frac{\overline{s}}{2} \frac{q_\alpha
  q_\mu}{m \omega^+}
  \left[ \epsilon_{\nu \mu} \sigma^{(0)}_{\alpha \beta}(\omega) + \sigma^{(0)}_{\nu \mu}(\omega)
  \epsilon_{\alpha \beta} \right],
\end{equation}
where $\sigma^{(0)}_{\lambda \rho}(\omega)$ is given by Eq.~(\ref{eqn:sigma0}).

These terms, however, do not in general give the full conductivity tensor, even for gapped systems.
In that case, the only other possible contribution comes from a response with the form of the inverse
(internal) compressibility $\tilde{\kappa}^{-1}$, which is not necessarily the total one
$\kappa_\text{int}^{-1}$, as we will see.
As before, in the presence of a nonuniform electric field, to lowest order in $\mathbf{q}$ the
electric current response is
$\langle j_\alpha (\mathbf{q}, \omega) \rangle^{(0)} = \sigma^{(0)}_{\alpha \beta}(\omega)
E_\beta (\mathbf{q}, \omega)$.
Again, the corresponding change in the particle number density is
$\delta \langle n (\mathbf{q}, \omega) \rangle^{(1)} = q_\alpha \langle j_\alpha (\mathbf{q},
\omega) \rangle^{(0)}/\omega^+$.
Through the inverse compressibility, this leads to a contribution to the pressure
$\widetilde{\delta P} (\mathbf{q},\omega) = \tilde{\kappa}^{-1} \delta \langle n(\mathbf{q},\omega)
\rangle^{(1)}/\overline{n}$,
which is equivalent to an electric field
$E^P_\mu (\mathbf{q},\omega) = - i q_\mu \widetilde{\delta P} (\mathbf{q},\omega) / \overline{n}$.
Finally, this leads to a change in electric current,
$\langle j^{(2)} \rangle_\nu (\mathbf{q},\omega) = \sigma^{(0)}_{\nu \mu} (\omega) E^P_\mu
(\mathbf{q},\omega)$.
Combining all these expressions, we find an order $\mathbf{q}^2$ contribution to the conductivity,
\begin{equation} \label{eqn:sig2k}
  \sigma^{(2),\kappa}_{\nu \beta} =
  - i \tilde{\kappa}^{-1} \frac{q_\mu q_\alpha}{\overline{n}^2 \omega^+} \sigma^{(0)}_{\nu \mu} (\omega)
  \sigma^{(0)}_{\alpha \beta} (\omega).
\end{equation}

We then have the total $q^2$ conductivity
\be
\sigma^{(2)}_{\nu \beta}(\mathbf{q},\omega) =\sigma^{(2),\overline{s}}_{\nu \beta}(\mathbf{q},\omega)+
  \sigma^{(2),\kappa}_{\nu \beta}(\mathbf{q},\omega).
  \ee
Using Eq.~(\ref{eqn:sigma0}) we then obtain, in the absence of a magnetic field,
\be \label{eqn:sig2_b0}
  \sigma^{(2)}_{\nu \beta}(\mathbf{q},\omega) =
  i \tilde{\kappa}^{-1} \frac{q_\nu q_\beta}{m^2 \omega^{+ 3}} +
  \frac{\overline{n} \overline{s}}{2 m^2} \frac{q^2}{\omega^{+ 2}} \epsilon_{\nu \beta}.
\ee
The contributions of $\overline{s}$ and $\tilde{\kappa}^{-1}$ have different tensor structures, and the
result agrees with eq.\ (\ref{eqn:sig2_gen_b0}) if we identify $\tilde{\kappa}^{-1} = \kappa^{-1}$, and
$\eta^H=\half\overline{n}\,\overline{s}$.

On the other hand, for nonzero magnetic field we find,
\begin{align} \label{eqn:sig2sk_b}
  \sigma^{(2),\overline{s}}_{\nu \beta}(\mathbf{q},\omega) \sim &
  \frac{i\overline{n} \overline{s}}{m^2} \frac{q_\alpha  q_\mu}{ \omega^+ \omega_c} \epsilon_{\alpha
  \beta} \epsilon_{\mu \nu}
  -\frac{\overline{n} \overline{s}}{2 m^2} \frac{q^2}{\omega_c^2} \epsilon_{\nu \beta},
  \\
  \sigma^{(2),\kappa}_{\nu \beta}(\mathbf{q},\omega)  \sim &
   i\tilde{\kappa}^{-1}
  \frac{q_\alpha q_\mu}{m^2 \omega_c^2 \omega^+}
  \epsilon_{\mu \nu} \epsilon_{\alpha \beta} -
  \tilde{\kappa}^{-1} \frac{1}{m^2 \omega_c^3}
  q^2 \epsilon_{\nu \beta},
\end{align}
as $\omega\to0$, up to terms linear or higher order in $\omega$.
Adding these, we see that the total inverse internal compressibility is
$\kappa_\text{int}^{-1}=\tilde{\kappa}^{-1} +\overline{n} \overline{s} \omega_c$, and hence
\begin{multline}
  \sigma^{(2)}_{\nu \beta}(\mathbf{q},\omega)  \sim \\
   i\kappa_\text{int}^{-1}
  \frac{q_\alpha q_\mu}{i m^2 \omega_c^2 \omega^+}
  \epsilon_{\mu \nu} \epsilon_{\alpha \beta} +
  \left( \half \overline{n} \overline{s} \omega_c -\kappa_\text{int}^{-1}  \right) \frac{1}{m^2 \omega_c^3}
  q^2 \epsilon_{\nu \beta}.
\end{multline}
This agrees with the general form eq.\ (\ref{eqn:sig2_gen_b}), giving
again $\eta^H=\half\overline{n}\,\overline{s}$.

In this derivation of the Hall viscosity, the centerpiece was the use of the magnetization equal to
the orbital spin over twice the mass, for each particle. This is the standard result for zero
magnetic field, and can be obtained by taking minus the derivative of the kinetic energy with respect
to $B$ in the symmetric gauge, then setting $B$ to zero. But doing the same in a uniform non-zero
field gives $-({b^i}^\dagger b^i+\half)/m$ for the $i$th particle, which is {\em twice} the
orbital angular momentum of the cyclotron motion of a particle divided by twice the mass in the
non-interacting case. Of course, in the rigorous microscopic derivation, this was not the way that
the conductivity was calculated, so there is no contradiction, however, the underlying reason for
the assumed coupling seems less clear. Hence in the magnetic field case, the present argument should
perhaps be viewed as an interpretation of the correct result, rather than as an alternative derivation
from first principles.

\section{Conclusion}\label{sec8}

The motivation for and conclusions of this work have been described in detail in the Introduction.
Essentially, we have a fairly complete formalism for defining the stress response to an external strain
field, and obtaining from it the inverse compressibility and the viscosity tensor, in different
situations that include non-zero temperature or magnetic field, and finite systems with a
confining potential or periodic boundary conditions. The results from distinct formulations (such as
different boundary conditions) agree in the thermodynamic limit. The results illuminate the relation of
the Hall viscosity results, obtained from adiabatic transport, to more traditional Kubo formula
approaches; the Kubo formulas extend to more general situations.

A motivation for this work was to make it possible to study Hall viscosity experimentally. The generality
of the Kubo formulas should make it possible to analyze possible experimental set-ups to see when Hall
viscosity manifests itself. In particular, the use of optical techniques to obtain the current response
$\sigma(\mathbf{q},\omega)$ to an external perturbing electromagnetic field at small $\mathbf{q}$ and
$\omega$, in conjunction with the relation due to Hoyos and Son \cite{Hoyos2012}, which holds more broadly
as shown in Sec.\ \ref{sec4}, to obtain the Hall viscosity should be explored. The result would be of
interest in quantum Hall physics, because Hall viscosity is related to the so-called shift in the ground
state \cite{Read2009,Read2010}, which can distinguish between distinct possible ground states at a given
filling factor.

\acknowledgments

We are grateful to L.~Glazman, M.~Randeria, and D.T.~Son for helpful discussions. B.B. and N.R are
grateful for the hospitality of the Institut Henri Poincar\'{e}, Paris, where part of this work was
carried out. The work of B.B. and N.R. was supported by NSF grant no.\ DMR-1005895. M.G. was
supported by the Simons Foundation, the Fulbright Foundation, and the BIKURA (FIRST) program of the
Israel Science Foundation.

\appendix

\section{The stress tensor for pair interactions}\label{app_stress}

For completeness, we present a derivation of an expression for the local stress
$\tau^{(0)}_{\mu\nu}(\mathbf{x})$ and its integral $T_{\mu\nu}^{(0)}$ for the case of a system of
particles interacting through a two-particle interaction. In particular,
starting with the position-dependent generalization of eq.\ (\ref{stress_met_rel}) in a space with
coordinates $\mathbf{x}$ and a general metric $g_{\mu\nu}(\mathbf{x})$,
\begin{equation}
\tau^{(0)}_{\mu\nu}(\mathbf{x})=-\left.2\frac{\delta H_{\Lambda}}{\delta
g_{\mu\nu}(\mathbf{x})}\right|_{\Lambda=I}, \label{tau_var_deriv}
\end{equation}
and setting the metric to its standard flat space form after taking the functional derivative, we
shall recover the known Irving-Kirkwood form of the stress tensor\cite{Irving1950,Forster1975}, and
show that its integral over space agrees with eq.\ (\ref{momflux}). (We consider only the
rotationally-invariant case, as this was assumed in those references.) Note that any time dependence of
the metric does not
enter into this derivation (all variational derivatives are taken at equal time), and hence it has
been suppressed for brevity.

We begin with the Hamiltonian (in zero magnetic field, for simplicity)
\begin{align}
H_{\Lambda}&=\frac{1}{2m}\sum_i\int{d^dx\,
g^{\mu\nu}(\mathbf{x})\left\{p_\mu^i,\left\{p_\nu^i,\delta(\mathbf{x}-\mathbf{x}^i)\right\}\right\}}
\nonumber \\
&\quad{}+\half\sum_{i\neq j}V({\cal D}(\mathbf{x}^i,\mathbf{x}^j)).\label{pair_int_hamiltonian}
\end{align}
Here we have introduced the geodesic distance
\begin{equation}
{\cal D}(\mathbf{x},\mathbf{x}')\equiv\int_0^1d\xi\sqrt{g_{\mu\nu}(\mathbf{r}(\xi))\frac{d
r^\mu}{d\xi}\frac{dr^\nu}{\partial\xi}},\label{d_def}
\end{equation}
which is the length (computed using $g_{\mu\nu}(\mathbf{x})$) of the geodesic $\mathbf{r}(\xi)$,
the path from $\mathbf{x}$ to $\mathbf{x}'$ satisfying
\begin{align}
\mathbf{r}(0)&=\mathbf{x}, \\
\mathbf{r}(1)&=\mathbf{x}', \\
\frac{\delta}{\delta \mathbf{r}(\xi)}\left(\int_0^1d\xi\sqrt{g_{\mu\nu}(\mathbf{r}(\xi))
\frac{\partial r^\mu}{\partial\xi}\frac{\partial
r^\nu}{\partial\xi}}\right)&=0, \label{geodesicdef}
\end{align}
where the variational derivative is taken with fixed endpoints. For metrics close to the flat metric,
which is sufficient for our purposes, there is a unique path that is a local minimum (in the space of
paths) of the length, and this path is the geodesic. The
use of the geodesic distance allows us to use the same potential function in curved space as in flat,
and while in principle the choice of path entering the  distance function $\cal D$ is
somewhat arbitrary, we believe that the geodesic distance given above represents a natural choice.
Different choices could produce different stress tensors even in the flat-space limit, but they will
differ only by divergenceless tensors.

We now proceed to evaluate eq.\ (\ref{tau_var_deriv}). Looking first at the kinetic term, and using
the identity
\begin{equation}
\left.\frac{\delta g^{\alpha\beta}(\mathbf{x})}{\delta
g_{\mu\nu}(\mathbf{x}')}\right|_{g=I}=-\frac{1}{2}\delta(\mathbf{x}-\mathbf{x}')
\left(\delta_{\alpha\mu}\delta_{\beta\nu}+\delta_{\alpha\nu}\delta_{\beta\mu}\right),
\end{equation}
we easily find for the kinetic part of the stress tensor
\begin{equation}
\tau_{\mu\nu}^{(0),K}(\mathbf{x})=\frac{1}{4m}\sum_i\left\{p_\mu^i,\left\{p_\nu^i,
\delta(\mathbf{x}-\mathbf{x}^i)\right\}\right\},
\end{equation}
in agreement with standard results.

Moving on to the interaction term, we have
\begin{align}
-2\frac{\delta}{\delta g_{\mu\nu}(\mathbf{x})}&\left(\half\sum_{i\neq j}V({\cal D}(\mathbf{x}^i,\mathbf{x}^j))\right)=\nonumber \\
&=-\sum_{i\neq j}V'({\cal D}(\mathbf{x}^i,\mathbf{x}^j))\frac{\delta}{\delta g_{\mu\nu}(\mathbf{x})}{\cal D}(\mathbf{x}^i,\mathbf{x}^j).
\end{align}
Now comes the key observation. ${\cal D}(\mathbf{x}^i,\mathbf{x}^j)$ depends on the metric in two ways -
through its explicit dependence indicated in eq.\ (\ref{d_def}), and implicitly through the definition (\ref{geodesicdef}) of the geodesic $\mathbf{r}(\xi)$ for each pair $i$, $j$. However, because $\mathbf{r}(\xi)$ is defined such that the distance is stationary under variations of the path, the chain rule tells us that only variations of $\cal D$ with respect to the explicit metric dependence are nonvanishing. Thus, we have
\begin{equation}
\frac{\delta}{\delta g_{\mu\nu}(\mathbf{x})}{\cal D}(\mathbf{x}^i,\mathbf{x}^j)=\half\int_0^1 d\xi\,
\delta(\mathbf{x}-\mathbf{r}(\xi))\frac{1}{{\cal D}(\mathbf{x}^i,\mathbf{x}^j)}\frac{dr^\mu}{d\xi}
\frac{dr^\nu}{d\xi}.
\end{equation}
Now, at $g=I$, the geodesic $\mathbf{r}(\xi)$ is simply given by a straight-line path
\begin{equation}
\left.\mathbf{r}\right|_{g=I}=\mathbf{x}^i+\xi\left(\mathbf{x}^j-\mathbf{x}^i\right),
\end{equation}
and the distance $\cal D$ by the standard Euclidean norm
\begin{equation}
\left.{\cal D}(\mathbf{x}^i,\mathbf{x}^j)\right|_{g=I}=\left|\mathbf{x}^i-\mathbf{x}^j\right|.
\end{equation}
Thus, putting it all together, we find for the interaction contribution to the stress tensor
\begin{widetext}
\begin{equation}
\tau_{\mu\nu}^{(0),V}(\mathbf{x})=-\frac{1}{2}\sum_{i\neq
j}V'(\left|\mathbf{x}^i-\mathbf{x}^j\right|)\frac{(x^i-x^j)^\mu(x^i-x^j)^\nu}
{\left|\mathbf{x}^i-\mathbf{x}^j\right|}\int_0^1
d\xi\,\delta(\mathbf{x}-\mathbf{x}^i+\xi(\mathbf{x}^i-\mathbf{x}^j)),
\end{equation}
\end{widetext}
which is the well-known contribution due to two-particle interactions introduced by Irving and
Kirkwood\cite{Irving1950}. Physically, it means that the flow of momentum between particles $i$ and $j$
when they interact is treated as flowing along the straight line connecting them, and so there is a
momentum density at any $\mathbf{x}$ on that line, which necessitates the integral over $\xi$ from $0$
to $1$. The verification that this stress tensor satisfies the continuity equation for the momentum 
density, eq.\ (\ref{contnofield}), is straightforward or can be found in the literature.

Finally, integrating $\tau^{(0)}_{\mu\nu}=\tau^{(0),K}_{\mu\nu}+\tau^{(0),V}_{\mu\nu}$ over space yields
\begin{equation}
T_{\mu\nu}^{(0)}=\frac{1}{m}\sum_ip_\mu^ip_\nu^i-\frac{1}{2}\sum_{i\neq
j}V'(\left|\mathbf{x}^i-\mathbf{x}^j\right|)\frac{(x^i-x^j)^\mu(x^i-x^j)^\nu}{\left|\mathbf{x}^i
-\mathbf{x}^j\right|},
\end{equation}
which a simple calculation shows is equal to
\begin{equation}
T_{\mu\nu}^{(0)}=-i\left[H_0,J_{\mu\nu}\right]
\end{equation}
as expected.

\section{Strain generators for systems in a confining potential}\label{app}

Because the volume of a system in the infinite plane with zero magnetic field is poorly defined, we
would like to work in a finite-sized system when using the extensive forms of the stress response function
given in Section \ref{sec3}. Unfortunately, because the strain generators contain an explicit $\mathbf{x}$
operator, we cannot do this with periodic boundary conditions. Instead, let us add to the Hamiltonian
$H_\Lambda(t)$ a general confining potential (this differs slightly from what was mentioned in Sec.\
\ref{sec21}, for reasons that should become clear)
\begin{equation}
U^{(\mathbf{x})}=\sum_{i}u\left(Z \mathbf{x}^i\right),
\end{equation}
with an invertible matrix of shape parameters $Z_{\mu\nu}$, and we take the single-particle potential $u$
such that $u(\mathbf{x})\rightarrow\infty$ as $\mathbf{x}\to\infty$ in any direction. For example, the
most general harmonic confining potential can be written
\begin{equation}
U^{(\mathbf{x})}=\sum_i\frac{mC^2}{2}Z_{\mu\nu}Z_{\mu\alpha}x_\nu^ix_{\alpha}^i.
\end{equation}

Now, under an additional strain transformation we have $\mathbf{x}^i\rightarrow {\Lambda'}^T\mathbf{x}^i$,
and this $\Lambda'$ can be absorbed into $\Lambda$ by multiplication on the left.  According to the
definition just given, $U^{(\mathbf{x})}$ in the $\mathbf{x}$ variables should be unchanged, so we need
$Z\rightarrow Z{\Lambda'}^{T-1}$. To implement this transformation using operators, we are motivated by
our treatment of the magnetic field in Section (\ref{sec22}) to quantize the $Z_{\mu\nu}$. We introduce
operators $\mathcal{Z}_{\mu\nu}$ and their conjugate momenta $\mathcal{M}_{\mu\nu}$ such that
\begin{equation}
\left[\mathcal{M}_{\mu\nu},\mathcal{Z}_{\alpha\beta}\right]=-i\delta_{\mu\alpha}\delta_{\nu\beta},
\end{equation}
and take for the strain generators $J_{\mu\nu}$
\begin{equation}
J_{\mu\nu}=-\frac{1}{2}\left(\sum_i\left\{x_\mu^i,p_\nu^i\right\}
-\left\{\mathcal{M}_{\alpha\mu},\mathcal{Z}_{\alpha\nu}\right\}\right). \label{boxdilate}
\end{equation}
These satisfy $[J_{\mu\nu},(Z\mathbf{x}^i)_\alpha]=0$, and so also
\begin{equation}
\left[J_{\mu\nu},U^{(\mathbf{x})}\right]=0.
\end{equation}
For given eigenvalues $Z$ of $\mathcal{Z}$, the confining potential is fixed, independent of $\Lambda$,
and defines a box that is always fixed in the $\mathbf{x}$ variables.

To use this formalism, we first note that $\mathcal{M}$ does not appear in the Hamiltonian
$H_\Lambda+U^{(\mathbf{x})}$, and so with $\Lambda=I$ (for example), states with given eigenvalues $Z$ of
$\mathcal{Z}$ evolve with $Z$ fixed, however (as with $\mathcal{B}$), formal eigenstates of $\mathcal{Z}$
are not normalizable. One can, however, consider normalizable states that are eigenstates of
$H_0+U^{(\mathbf{x})}$ for each $Z$, with a narrow range of $Z$. These are almost as good as true
eigenstates for most purposes.

We want to show that both our basic relation~(\ref{momflux}) and our Kubo
formulas~(\ref{chifreq1}-\ref{strainstrain}) continue to hold unmodified in the presence of the confining
potential. First, we will verify that the stress tensor is given by minus the time derivative of
$J_{\mu\nu}$. This is simple, as $J_{\mu\nu}$ commutes with $U^{(\mathbf{x})}$, and the
$\{\mathcal{M},\mathcal{Z}\}$ terms in $J_{\mu\nu}$ commute with $H_0^{(\mathbf{x})}$. Hence
\begin{equation}
T_{\mu\nu}^{(0)}=-i[H_0^{(\mathbf{x})}+U^{(\mathbf{x})},J_{\mu\nu}].\label{ward}
\end{equation}

Next, we will verify that our Kubo formulas still hold with these strain generators. We have the
Hamiltonian $H_\Lambda(t)+U^{(\mathbf{x})}$ in the $\mathbf{x}$ variables (which include $\mathcal{Z}$ and
$\mathcal{M}$), and we make a time-dependent canonical transformation to $\mathbf{X}$ variables. The
analogs of $\mathcal{Z}$ and $\mathcal{M}$ in $\mathbf{X}$ variables will be denoted by
$\widehat{\mathcal{Z}}=S\mathcal{Z}S^{-1}=\mathcal{Z}\Lambda^{T-1}$ and
$\widehat{\mathcal{M}}=S\mathcal{M}S^{-1}$. Then the Hamiltonian in $\mathbf{X}$ variables is
$H=H_0^{(\mathbf{X})}+U^{(\mathbf{X})}+H_1$, where
\begin{align}
H_0^{(\mathbf{X})}+U^{(\mathbf{X})}&=\sum_i\frac{P_\mu^iP_\mu^i}{2m}+\frac{1}{2}\sum_{i\neq
j}V\left(\mathbf{X}^i-\mathbf{X}^j\right)\nonumber \\
&\quad{}+\sum_iu(\widehat{\mathcal{Z}}\mathbf{X}^i),\\
H_1&=-\frac{\partial\lambda_{\mu\nu}}{\partial t}J_{\mu\nu}
\end{align}
to order $\lambda$. In these variables, the $\{\widehat{\mathcal{M}},\widehat{\mathcal{Z}}\}$ terms in
$H_1$ cause $\widehat{\mathcal{Z}}$ to evolve in time, in such a way that if the $\mathbf{X}^i$s also
evolve by the $H_1$ term only, then they continue to lie inside the ``box'' (defined by $U$) if they
do so initially; this was the desired behavior. The stress in these variables,
$T_{\mu\nu}^{(\mathbf{X})}$, is the same expression as in the case with no potential $U$. Thus, the
stress-strain Kubo formula~(\ref{chifreq1}) obtained from standard linear response theory takes the same
form as before (though the strain generator is now different, and the Hamiltonian includes $U$).
Additionally, the Ward identity Eq.~(\ref{ward}) ensures that the stress-stress and strain-strain
Kubo formulas are still valid in the presence of a confining potential, as is the final formula
(\ref{viscfield}) for the viscosity, up to a choice of the volume $L^d$ to assign to the system.
(Technical justification of these statements is discussed further in Appendix \ref{appB}.) If the
potential $u$ is taken to have a hard-wall form, the volume can be taken as that within the walls.

As an example of this formalism, let us consider a system of noninteracting spinless fermions in a
harmonic potential in two dimensions, with Hamiltonian
\begin{equation}
H=\sum_i\left(\frac{p_\mu^ip_\mu^i}{2m}+\frac{mC^2}{2}\mathcal{Z}_{\mu\alpha}\mathcal{Z}_{\mu\beta}
x_\alpha^ix_\beta^i\right)
\end{equation}
 We consider the state $\left|0\right>$ in which the potential has angular frequency $C/L$, meaning
 that $\mathcal{Z_{\mu\nu}}\left|0\right>=L^{-1}\delta_{\mu\nu}\left|0\right>$ (strictly, this means
 a narrow normalizable wavepacket centered at this value, as we explained above), and the lowest $Q$
 levels filled. This state has $N=Q(Q+1)/2$ particles, and its energy is
\begin{equation}
E_Q=\frac{C}{L}\sum_{n=1}^Q{n^2}.
\end{equation}
The integrated stress tensor is $T_{\mu\nu}=\sum_ip^i_\mu p^i_\nu/m$.
We will use the stress-stress form of the Kubo formula, eq.\ (\ref{stressstress}).
Let us examine the contact term first. This gives
\begin{align}
\left<\left[T_{\mu\nu}(0),J_{\alpha\beta}(0)\right]\right>_0&=\frac{i}{m}\sum_i\left(\left<
p_\beta^ip_{\mu}^i \right>_0\delta_{\nu\alpha}+\left<p_\beta^ip_{\nu}^i\right>_0\delta_{\mu\alpha}
\right)\nonumber \\
&=\frac{iE_Q}{2}\left(\delta_{\mu\alpha}\delta_{\nu\beta}+\delta_{\nu\alpha}\delta_{\mu\beta}\right),
\end{align}
where the last line follows from an application of the virial theorem. To evaluate the time-integral
term, note that the stress tensor and the Hamiltonian are independent of $\mathcal{M}_{\mu\nu}$, and
hence $T_{\mu\nu}(t)$ leaves the ground state in the fixed $Z_{\mu\nu}=L^{-1}\delta_{\mu\nu}$ subspace.
Hence, we are free to evaluate the time-dependence in this subspace. Doing so, we find
\begin{align}
\int_0^{\infty}&{dte^{i\omega^+t}\left<\left[T _{\mu\nu}(t),T _{\alpha\beta}(0)\right]\right>_0}=\nonumber
\\
&=\frac{-iE_Q}{4}\frac{1}{1-\left(\frac{\omega^+L}{2C}\right)^2}\left(\delta_{\mu\alpha}\delta_{\nu\beta}
+\delta_{\nu\alpha}\delta_{\mu\beta}\right).
\end{align}
Hence we find that the response function for harmonically-trapped non-interacting fermions is
\be
X_{\mu\nu\alpha\beta}=\frac{iE_Q}{2\omega^+}\left(1-\frac{1}{2}\frac{1}{1-
\left(\frac{\omega^+L}{2C}\right)^2}\right)\left(\delta_{\mu\alpha}\delta_{\nu\beta}+\delta_{\nu\alpha}
\delta_{\mu\beta}\right).
\ee

The fluid in the harmonic potential is not homogeneous; its density is not uniform, and accordingly
its other properties, such as the expectation of $\tau_{\mu\nu}(\mathbf{x})$, are not uniform either,
and this effect does not disappear in the thermodynamic limit $L$, $N\to\infty$, if we take it with
the density at the center held fixed. A macroscopic fraction of particles experience a non-zero
potential (this is already apparent from the use of the virial theorem: the kinetic energy, which is
used to obtain the trace of $T_{\mu\nu}$, is only half the total energy). Therefore use of the results
involving $P$ and $\kappa^{-1}$ which relied on homogeneity is not justified. However, we find that if
we use the free Fermi gas results in Sec.\ \ref{freefermi} to obtain the pressure and viscosity at a given
density, and then average the results using the density profile of the harmonically-trapped gas, the
results agree in all details with the thermodynamic limit of the above. To obtain a fluid that is
homogeneous (up to a negligible boundary layer) in the thermodynamic limit, we need a confining potential
that is essentially zero in the interior, then rises rapidly very close to the edge.

\section{Stress response without use of strain generators}
\label{appC}

In this section we show how to obtain even more general forms of the stress response, which work for
either the infinite system with a confining potential, or for periodic boundary conditions, and are
analogous to the stress-strain and strain-strain forms, but do not require the use or existence of
the strain generators. The approach we use is based on that of Niu {\it et al.} for the conductivity
case \cite{Niu1985}, however we extend it to non-zero frequency as well as adapting it for the stress
response. For the case of infinite space, the expressions reduce to the forms in the text when written
in terms of strain generators.

We begin with the time-integral part of the extensive stress-stress form at zero temperature, which we
write in a spectral representation. For definiteness, one can assume periodic boundary conditions. Then
we have
\begin{eqnarray}
\lefteqn{\frac{1}{\omega^+}\int_0^{\infty}dt\,e^{i\omega^+t}\left<\left[T _{\mu\nu}(t),
T _{\alpha\beta}(0)\right]\right>_0=}&&\nonumber\\
&&\frac{1}{\omega^+}\int_0^{\infty}dt\,e^{i\omega^+t}\sum_{n}\left[e^{-i(E_n-E_0)t}\left<0|
T _{\mu\nu}(0)|n\rangle\langle n|T _{\alpha\beta}(0)|0\right>\right.\nonumber\\
&&\left.\qquad\qquad{}-e^{-i(E_0-E_n)t}\left<0|T _{\alpha\beta}(0)|n\rangle\langle n|
T _{\mu\nu}(0)|0\right>\right],
\end{eqnarray}
where $\{|n\rangle\}$ is an orthonormal set of energy eigenstates of $H_0$ (or $H_0+U$ if the confining
potential is used in place of periodic boundary conditions) with energies $E_n$, and we now write
$|0\rangle$ for the ground state, which we assume for simplicity is non-degenerate.
We now use the identity
\be
e^{-i(E_n-E_0)t}=\frac{\frac{d}{dt}e^{-i(E_n-E_0)t}}{-i(E_n-E_0)}
\ee
in the first term inside the integral, and the same with $n$ and $0$ switched in the second term; then
integrate by parts. Further, we recall that in the $\mathbf{x}$ variables, the Hamiltonian $H_\Lambda(t)$
to first order in $\lambda$ is
\be
H_\Lambda(t)=H_0-T_{\alpha\beta}\lambda_{\alpha\beta}(t)
\ee
(where $T_{\alpha\beta}=T_{\alpha\beta}(0)$ is at zero time in the Heisenberg picture), and so if we
define the ground state of $H_\Lambda$ for given $\lambda$ to be $|\varphi(\lambda)\rangle$, that is
\be
H_\Lambda|\varphi(\lambda)\rangle = E(\lambda)|\varphi(\lambda)\rangle
\ee
(with $|\varphi(0)\rangle=|0\rangle$, $E(0)=E_0$),
then perturbation theory to first order in $\lambda$ gives us
\be
\left|\frac{\partial \varphi}{\partial\lambda_{\alpha\beta}}\right\rangle
=\sum_n \frac{|n\rangle\langle n|T_{\alpha\beta}(0)|0\rangle}{E_n-E_0}.
\ee
(We leave it as understood that the partial derivative is evaluated at $\lambda=0$.)
Combining these results, and moving the term resulting from the lower limit in the integration by parts
to the other side, we arrive at two forms of the response function $X_{\mu\nu\alpha\beta}(\omega)$:
\begin{eqnarray}
\lefteqn{X_{\mu\nu\alpha\beta}(\omega)=}&&\nonumber\\
&&
\frac{i}{\omega^+}\left[\left\langle\varphi(0)\left|T_{\mu\nu}(0)\vphantom{\frac{\partial\varphi}{\partial
\lambda_{\alpha\beta}}}\right|\frac{\partial\varphi}{\partial
\lambda_{\alpha\beta}}\right\rangle+\left\langle\frac{\partial\varphi}
{\partial \lambda_{\alpha\beta}}\left|T_{\mu\nu}(0)
\vphantom{\frac{\partial\varphi}{\partial\lambda_{\alpha\beta}}}\right|\varphi(0)
\right\rangle\right]\nonumber\\
&&{}+\frac{1}{\omega^+}\int_0^\infty dt\,e^{i\omega^+t}\left\langle\varphi(0)|\left[T_{\mu\nu}(t),
T_{\alpha\beta}(0)\right]|\varphi(0)\right\rangle\\
&=&\int_0^\infty
dt\,e^{i\omega^+t}\left[\left\langle\varphi(0)\left|T_{\mu\nu}(t)\vphantom{\frac{\partial\varphi}{\partial
\lambda_{\alpha\beta}}}\right|\frac{\partial\varphi}{\partial
\lambda_{\alpha\beta}}\right\rangle\right.\nonumber\\
&&{}\qquad\qquad\qquad\qquad\left.+\left\langle\frac{\partial\varphi}
{\partial \lambda_{\alpha\beta}}\left|T_{\mu\nu}(t)
\vphantom{\frac{\partial\varphi}{\partial\lambda_{\alpha\beta}}}\right|\varphi(0)
\right\rangle\right].
\end{eqnarray}
The first of these is the stress-stress form, and the second is the stress-strain form.

In order to use the same identity and perform a second integration by parts, we generalize the
previous perturbation formula by using $T_{\mu\nu}(t)$ for given $t$ as the perturbation, and define
\begin{eqnarray}
\left|\frac{\partial \varphi(t)}{\partial\lambda_{\mu\nu}}\right\rangle
&=&\sum_n \frac{|n\rangle\langle n|T_{\mu\nu}(t)|0\rangle}{E_n-E_0}\\
&=&\sum_n \frac{|n\rangle\langle n|T_{\mu\nu}(0)|0\rangle}{E_n-E_0}e^{-i(E_0-E_n)t}.
\end{eqnarray}
Then the strain-strain form of $X$ is
\begin{eqnarray}
\lefteqn{X_{\mu\nu\alpha\beta}(\omega)=}&&\\
&&\qquad\qquad-i\left[\left\langle\frac{\partial\varphi(0)}{\partial
\lambda_{\mu\nu}}\left|\frac{\partial\varphi(0)}{\partial
\lambda_{\alpha\beta}}\right.\right\rangle-\left\langle\frac{\partial\varphi(0)}
{\partial \lambda_{\alpha\beta}}\left|\frac{\partial\varphi(0)}{\partial
\lambda_{\mu\nu}}\right.\right\rangle\right]\nonumber\\
&&{}+\omega^+\int_0^\infty dt\,e^{i\omega^+t}\left[\left\langle\frac{\partial\varphi(t)}{\partial
\lambda_{\mu\nu}}\left|\frac{\partial\varphi(0)}{\partial
\lambda_{\alpha\beta}}\right.\right\rangle-\left\langle\frac{\partial\varphi(0)}
{\partial \lambda_{\alpha\beta}}\left|\frac{\partial\varphi(t)}{\partial
\lambda_{\mu\nu}}\right.\right\rangle\right].\nonumber
\end{eqnarray}
The first part is the curvature of the Berry connection. This part reproduces the formulas for
the antisymmetric (Hall viscosity) part of $X$ at $\omega=0$ that result from adiabatic transport
in a gapped system (with the sign here corrected as explained earlier). The derivation of this
$\omega=0$ limit from the stress-strain form
is presumably equivalent to the one reviewed in Ref.\ \onlinecite{Read2010}.
The last term produces the $i\kappa^{-1}/\omega^+$ term in the trace part, as $\omega\to0$, similarly
as in Sec.\ \ref{sec6} (such parts, referred
to as ``persistent currents'' by analogy with the conductivity case, were suppressed in Ref.\
\onlinecite{Read2010} by subtracting the ground state energy $E(\lambda)$
from $H_\Lambda$). The strain-strain form given here can be viewed as generalizing the adiabatic curvature
(and the related Chern number for conductivity) to the full stress response tensor, to all frequencies,
and to systems that are gapless in the thermodynamic limit.
Because the formal derivation is general in form, similar ones can be given for conductivity at all
$\omega$, and for other transport properties also.

The similarity with the three forms of $X$ given in the text is evident, and can be made exact by use
of the formula
\be
|\varphi(\lambda)\rangle = e^{-i\lambda_{\alpha\beta}J_{\alpha\beta}}|\varphi(0)\rangle,
\ee
and expanding to first order. Then one can view
\be
\left|\frac{\partial \varphi(t)}{\partial\lambda_{\alpha\beta}}\right\rangle=-i
J_{\alpha\beta}(t)|\varphi(0)\rangle
=\sum_n \frac{|n\rangle\langle n|T_{\alpha\beta}(t)|0\rangle}{E_n-E_0}
\ee
as resulting from the basic relation, eq.\ (\ref{momflux}).
This is not merely an analogy; for the infinite-space geometry, the ground states of $H_\Lambda$ are
of exactly this form (with $t=0$), using the formalism for magnetic field and for a confining potential,
so they form a ``homogeneous bundle'' in the language of Ref.\ \onlinecite{Read2010} (here
for GL$(d,\mathbb{R})$, not just SL$(d,\mathbb{R})$ as mainly considered there). We note however that
strictly speaking for the infinite-space geometry, the formalism for a magnetic field or in
Appendix \ref{app} again produces issues regarding the existence of normalizable energy eigenstates, and
some arguments with wavepackets analogous to those in the text or in Appendix \ref{appB} are needed to
overcome these.

In any case, the forms given here for periodic boundary conditions are expected to yield the same results
for intensive quantities in the thermodynamic limit, justifying the use of the same notation
$X_{\mu\nu\alpha\beta}(\omega)$ in the formulas given here.

\section{Time-translation invariance and the Kubo formulas}
\label{appB}

In this Appendix, we give arguments that justify the use of time-translation invariance (TTI) in the
derivation of the Kubo formulas in Sec.\ \ref{sec31}, even though the states that must be used are not
true eigenstates of the Hamiltonian.
We say that a correlation or response function in the time domain has TTI
if shifting the time argument in each operator by the same constant has no effect on it.  For the zero
magnetic field case in the informal treatment without
a confining potential that we have mentioned in Sec.\ \ref{sec21}, one has to appeal to an assumed limit
in which
time-dependence of the state occurs only near the boundary, and the bulk of the system presumably
dominates the response over relevant time scales. This is intuitively appealing, but quite involved
to justify fully. Here we give more detailed arguments for the case with a
magnetic field, in which normalizable ground (energy eigen-) states in which a disk-shaped region
is occupied exist for given magnetic field. There is still an issue here, however, because our
strain generators involved the introduction of the magnetic field variable $\mathcal{B}$, and eigenstates
of $\mathcal{B}$ are not normalizable. Similar arguments are also given for the
case with a confining potential, in the formalism of Appendix \ref{app}.

The ``ground'' state that we described in Sec.\ \ref{sec22} is not strictly an energy eigenstate, because
though it is such for each value of $B$, the energy eigenvalue depends on $B$ (the Hamiltonian $H_0$
depends on $\mathcal{B}$ but not on $\mathcal{P}$, so parts of the state with different $B$ values do
not mix). Thus the contributions from different $B$ change phase with time at different rates. When
we calculate an expectation value of some operators, these time-dependent phase may cancel; in
particular, they will if those operators do not contain $\mathcal{P}$. Now the stress tensor
$T_{\mu\nu}$ does not contain $\mathcal{P}$, so the correlation function
\be
\langle[T_{\mu\nu}(t),T_{\alpha\beta}(t')]\rangle_0=f(t,t')
\ee
is TTI---it is unchanged if we replace $t$, $t'$ by $t+t_0$, $t'+t_0$. (In the following, the indices on
the operators will play no role, so they will not be recorded on $f$ and similar functions below.) Now
consider the correlation function
\be
\langle[T_{\mu\nu}(t),J_{\alpha\beta}(t')]\rangle_0=g(t,t').
\ee
From the identity (\ref{momflux2}),
\be
\frac{\partial}{\partial t'}g(t,t')=-f(t,t'),
\ee
and $f(t,t')=f(t-t')$ (say) by TTI. Integrating gives
\be
g(t,t')=\int_0^{t-t'}dt''\,f(t'') +g(t,t),
\ee
where
\be
g(t,t)=\langle e^{iH_0t}\,[T_{\mu\nu}(0),J_{\alpha\beta}(0)]\,e^{-iH_0t}\rangle_0.
\ee
The commutator $[T_{\mu\nu}(0),J_{\alpha\beta}(0)]$ does not contain $\mathcal{P}$ for any choice of
indices, and so $g(t,t)$ is again independent of $t$, or TTI. Hence $g(t,t')=g(t-t')$, say, is TTI. These
results justify the use of TTI to pass from the stress-strain to the stress-stress form of response.

For the strain-strain form, we proceed similarly with
\be
\langle[J_{\mu\nu}(t),J_{\alpha\beta}(t')]\rangle_0=h(t,t').
\ee
Then
\be
\frac{\partial}{\partial t}h(t,t')=-g(t-t'),
\ee
and so
\be
h(t,t')=-\int_0^{t-t'}dt''\,g(t'') +h(t,t).
\ee
The equal-time piece is
\be
h(t,t)=\langle [J_{\mu\nu}(t),J_{\alpha\beta}(t)]\rangle_0.
\ee
The commutators of two $J$'s are the $\mathfrak{gl}(d,\mathbb{R})$ Lie algebra relations, given
in (\ref{straincomm}). The generator corresponding to the trace (the generator of the gl(1) or
u($1$) subalgebra) never occurs on the right-hand side; see eq.\ (\ref{straincomm2}).
The traceless parts of $J$'s do not contain $\mathcal{P}$, and so $h(t,t)$ is independent of $t$, that
is, it is TTI. Hence $h(t,t')=h(t-t')$, say, is TTI.

In fact, TTI was not used to obtain the strain-strain form from the stress-strain form, but we have
learned that it can be applied to the final form. This enables us to reverse the argument, but with the
roles of the two pairs of indices interchanged. This leads to the same stress-stress form of response,
but with the two pairs of indices exchanged in the contact term. We showed in Sec.\ \ref{sumrules} that
this contact term is symmetric, which shows the arguments are consistent.

For the formalism of Appendix \ref{app}, with a confining potential, we can proceed similarly, using
a wavepacket in $Z$ space of energy eigenstates for each $Z$. The preceding line of argument goes through,
up to the point where $h(t,t)$ was studied. In the present case, all strain generators $J_{\mu\nu}$
contain $\mathcal{M}$, not only the trace. Hence $h(t,t)$ is not independent of $t$. If we take its
time derivative, we find
\be
\frac{\partial}{\partial t}h(t,t)=i\delta_{\alpha\nu}\langle
T_{\mu\beta}(t)\rangle_0-i\delta_{\mu\beta}\langle T_{\alpha\nu}(t)\rangle_0,
\ee
in which the trace of the expectation of the stress cancels. We have shown that the expectation of
the traceless part of $T_{\mu\nu}$ is time independent, so $h(t,t)$ is linear in $t$, and also its
time derivative is smaller than $O(L^d)$ as the thermodynamic limit is taken. Thus in the limit, for
our purposes $h(t,t')$ is again TTI. The strain-strain form in the main text is nonetheless correct
as written even in finite size.

\bibliography{kubo_visc}

\begin{thebibliography}{26}%
\makeatletter
\providecommand \@ifxundefined [1]{%
 \@ifx{#1\undefined}
}%
\providecommand \@ifnum [1]{%
 \ifnum #1\expandafter \@firstoftwo
 \else \expandafter \@secondoftwo
 \fi
}%
\providecommand \@ifx [1]{%
 \ifx #1\expandafter \@firstoftwo
 \else \expandafter \@secondoftwo
 \fi
}%
\providecommand \natexlab [1]{#1}%
\providecommand \enquote  [1]{``#1''}%
\providecommand \bibnamefont  [1]{#1}%
\providecommand \bibfnamefont [1]{#1}%
\providecommand \citenamefont [1]{#1}%
\providecommand \href@noop [0]{\@secondoftwo}%
\providecommand \href [0]{\begingroup \@sanitize@url \@href}%
\providecommand \@href[1]{\@@startlink{#1}\@@href}%
\providecommand \@@href[1]{\endgroup#1\@@endlink}%
\providecommand \@sanitize@url [0]{\catcode `\\12\catcode `\$12\catcode
  `\&12\catcode `\#12\catcode `\^12\catcode `\_12\catcode `\%12\relax}%
\providecommand \@@startlink[1]{}%
\providecommand \@@endlink[0]{}%
\providecommand \url  [0]{\begingroup\@sanitize@url \@url }%
\providecommand \@url [1]{\endgroup\@href {#1}{\urlprefix }}%
\providecommand \urlprefix  [0]{URL }%
\providecommand \Eprint [0]{\href }%
\providecommand \doibase [0]{http://dx.doi.org/}%
\providecommand \selectlanguage [0]{\@gobble}%
\providecommand \bibinfo  [0]{\@secondoftwo}%
\providecommand \bibfield  [0]{\@secondoftwo}%
\providecommand \translation [1]{[#1]}%
\providecommand \BibitemOpen [0]{}%
\providecommand \bibitemStop [0]{}%
\providecommand \bibitemNoStop [0]{.\EOS\space}%
\providecommand \EOS [0]{\spacefactor3000\relax}%
\providecommand \BibitemShut  [1]{\csname bibitem#1\endcsname}%
\let\auto@bib@innerbib\@empty
\bibitem [{\citenamefont {Kovtun}\ \emph {et~al.}(2005)\citenamefont {Kovtun},
  \citenamefont {Son},\ and\ \citenamefont {Starinets}}]{Kovtun2005}%
  \BibitemOpen
  \bibfield  {author} {\bibinfo {author} {\bibfnamefont {P.~K.}\ \bibnamefont
  {Kovtun}}, \bibinfo {author} {\bibfnamefont {D.~T.}\ \bibnamefont {Son}}, \
  and\ \bibinfo {author} {\bibfnamefont {A.~O.}\ \bibnamefont {Starinets}},\
  }\href {\doibase 10.1103/PhysRevLett.94.111601} {\bibfield  {journal}
  {\bibinfo  {journal} {Phys. Rev. Lett.}\ }\textbf {\bibinfo {volume} {94}},\
  \bibinfo {pages} {111601} (\bibinfo {year} {2005})}\BibitemShut {NoStop}%
\bibitem [{\citenamefont {Cao}\ \emph {et~al.}(2011)\citenamefont {Cao},
  \citenamefont {Elliott}, \citenamefont {Joseph}, \citenamefont {Wu},
  \citenamefont {Petricka}, \citenamefont {Schaefer},\ and\ \citenamefont
  {Thomas}}]{Cao2011}%
  \BibitemOpen
  \bibfield  {author} {\bibinfo {author} {\bibfnamefont {C.}~\bibnamefont
  {Cao}}, \bibinfo {author} {\bibfnamefont {E.}~\bibnamefont {Elliott}},
  \bibinfo {author} {\bibfnamefont {J.}~\bibnamefont {Joseph}}, \bibinfo
  {author} {\bibfnamefont {H.}~\bibnamefont {Wu}}, \bibinfo {author}
  {\bibfnamefont {J.}~\bibnamefont {Petricka}}, \bibinfo {author}
  {\bibfnamefont {T.}~\bibnamefont {Schaefer}}, \ and\ \bibinfo {author}
  {\bibfnamefont {J.~E.}\ \bibnamefont {Thomas}},\ }\href@noop {} {\bibfield
  {journal} {\bibinfo  {journal} {Science}\ }\textbf {\bibinfo {volume}
  {331}},\ \bibinfo {pages} {58} (\bibinfo {year} {2011})}\BibitemShut
  {NoStop}%
\bibitem [{\citenamefont {Avron}\ \emph {et~al.}(1995)\citenamefont {Avron},
  \citenamefont {Seiler},\ and\ \citenamefont {Zograf}}]{Avron1995}%
  \BibitemOpen
  \bibfield  {author} {\bibinfo {author} {\bibfnamefont {J.~E.}\ \bibnamefont
  {Avron}}, \bibinfo {author} {\bibfnamefont {R.}~\bibnamefont {Seiler}}, \
  and\ \bibinfo {author} {\bibfnamefont {P.~G.}\ \bibnamefont {Zograf}},\
  }\href {\doibase 10.1103/PhysRevLett.75.697} {\bibfield  {journal} {\bibinfo
  {journal} {Phys. Rev. Lett.}\ }\textbf {\bibinfo {volume} {75}},\ \bibinfo
  {pages} {697} (\bibinfo {year} {1995})}\BibitemShut {NoStop}%
\bibitem [{\citenamefont {Read}(2009)}]{Read2009}%
  \BibitemOpen
  \bibfield  {author} {\bibinfo {author} {\bibfnamefont {N.}~\bibnamefont
  {Read}},\ }\href {\doibase 10.1103/PhysRevB.79.045308} {\bibfield  {journal}
  {\bibinfo  {journal} {Phys. Rev. B}\ }\textbf {\bibinfo {volume} {79}},\
  \bibinfo {pages} {045308} (\bibinfo {year} {2009})}\BibitemShut {NoStop}%
\bibitem [{\citenamefont {Read}\ and\ \citenamefont {Rezayi}(2011)}]{Read2010}%
  \BibitemOpen
  \bibfield  {author} {\bibinfo {author} {\bibfnamefont {N.}~\bibnamefont
  {Read}}\ and\ \bibinfo {author} {\bibfnamefont {E.}~\bibnamefont {Rezayi}},\
  }\href {\doibase 10.1103/PhysRevB.84.085316} {\bibfield  {journal} {\bibinfo
  {journal} {Phys. Rev. B}\ }\textbf {\bibinfo {volume} {84}},\ \bibinfo
  {pages} {085316} (\bibinfo {year} {2011})}\BibitemShut {NoStop}%
\bibitem [{\citenamefont {Haldane}()}]{Haldane2009}%
  \BibitemOpen
  \bibfield  {author} {\bibinfo {author} {\bibfnamefont {F.~D.~M.}\
  \bibnamefont {Haldane}},\ }\href@noop {} {\ }\Eprint
  {http://arxiv.org/abs/0906.1854} {arXiv:0906.1854} \BibitemShut {NoStop}%
\bibitem [{\citenamefont {Nicolis}\ and\ \citenamefont {Son}()}]{Nicolis2011}%
  \BibitemOpen
  \bibfield  {author} {\bibinfo {author} {\bibfnamefont {A.}~\bibnamefont
  {Nicolis}}\ and\ \bibinfo {author} {\bibfnamefont {D.~T.}\ \bibnamefont
  {Son}},\ }\href@noop {} {\ }\Eprint {http://arxiv.org/abs/1103.2137}
  {arXiv:1103.2137} \BibitemShut {NoStop}%
\bibitem [{\citenamefont {Hoyos}\ and\ \citenamefont {Son}(2012)}]{Hoyos2012}%
  \BibitemOpen
  \bibfield  {author} {\bibinfo {author} {\bibfnamefont {C.}~\bibnamefont
  {Hoyos}}\ and\ \bibinfo {author} {\bibfnamefont {D.~T.}\ \bibnamefont
  {Son}},\ }\href {\doibase 10.1103/PhysRevLett.108.066805} {\bibfield
  {journal} {\bibinfo  {journal} {Phys. Rev. Lett.}\ }\textbf {\bibinfo
  {volume} {108}},\ \bibinfo {pages} {066805} (\bibinfo {year}
  {2012})}\BibitemShut {NoStop}%
\bibitem [{\citenamefont {Green}(1952)}]{Green1952}%
  \BibitemOpen
  \bibfield  {author} {\bibinfo {author} {\bibfnamefont {M.~S.}\ \bibnamefont
  {Green}},\ }\href@noop {} {\bibfield  {journal} {\bibinfo  {journal} {J.
  Chem. Phys.}\ }\textbf {\bibinfo {volume} {20}},\ \bibinfo {pages} {1281}
  (\bibinfo {year} {1952})}\BibitemShut {NoStop}%
\bibitem [{\citenamefont {Kubo}(1957)}]{Kubo1957}%
  \BibitemOpen
  \bibfield  {author} {\bibinfo {author} {\bibfnamefont {R.}~\bibnamefont
  {Kubo}},\ }\href@noop {} {\bibfield  {journal} {\bibinfo  {journal} {J. Phys.
  Soc. Japan}\ }\textbf {\bibinfo {volume} {12}},\ \bibinfo {pages} {570}
  (\bibinfo {year} {1957})}\BibitemShut {NoStop}%
\bibitem [{\citenamefont {Mori}(1958)}]{Mori1958}%
  \BibitemOpen
  \bibfield  {author} {\bibinfo {author} {\bibfnamefont {H.}~\bibnamefont
  {Mori}},\ }\href {\doibase 10.1103/PhysRev.112.1829} {\bibfield  {journal}
  {\bibinfo  {journal} {Phys. Rev.}\ }\textbf {\bibinfo {volume} {112}},\
  \bibinfo {pages} {1829} (\bibinfo {year} {1958})}\BibitemShut {NoStop}%
\bibitem [{\citenamefont {Kadanoff}\ and\ \citenamefont
  {Martin}(1963)}]{Kadanoff1963}%
  \BibitemOpen
  \bibfield  {author} {\bibinfo {author} {\bibfnamefont {L.~P.}\ \bibnamefont
  {Kadanoff}}\ and\ \bibinfo {author} {\bibfnamefont {P.~C.}\ \bibnamefont
  {Martin}},\ }\href@noop {} {\bibfield  {journal} {\bibinfo  {journal} {Ann.
  Phys.}\ }\textbf {\bibinfo {volume} {24}},\ \bibinfo {pages} {419} (\bibinfo
  {year} {1963})}\BibitemShut {NoStop}%
\bibitem [{\citenamefont {Luttinger}(1964)}]{Luttinger1964}%
  \BibitemOpen
  \bibfield  {author} {\bibinfo {author} {\bibfnamefont {J.~M.}\ \bibnamefont
  {Luttinger}},\ }\href@noop {} {\bibfield  {journal} {\bibinfo  {journal}
  {Phys. Rev.}\ }\textbf {\bibinfo {volume} {135}},\ \bibinfo {pages} {A1505}
  (\bibinfo {year} {1964})}\BibitemShut {NoStop}%
\bibitem [{\citenamefont {Taylor}\ and\ \citenamefont
  {Randeria}(2010)}]{Taylor2010}%
  \BibitemOpen
  \bibfield  {author} {\bibinfo {author} {\bibfnamefont {E.}~\bibnamefont
  {Taylor}}\ and\ \bibinfo {author} {\bibfnamefont {M.}~\bibnamefont
  {Randeria}},\ }\href {\doibase 10.1103/PhysRevA.81.053610} {\bibfield
  {journal} {\bibinfo  {journal} {Phys. Rev. A}\ }\textbf {\bibinfo {volume}
  {81}},\ \bibinfo {pages} {053610} (\bibinfo {year} {2010})}\BibitemShut
  {NoStop}%
\bibitem [{\citenamefont {Tan}(2008{\natexlab{a}})}]{Tan2008_1}%
  \BibitemOpen
  \bibfield  {author} {\bibinfo {author} {\bibfnamefont {S.}~\bibnamefont
  {Tan}},\ }\href@noop {} {\bibfield  {journal} {\bibinfo  {journal} {Ann.
  Phys.}\ }\textbf {\bibinfo {volume} {323}},\ \bibinfo {pages} {2952}
  (\bibinfo {year} {2008}{\natexlab{a}})}\BibitemShut {NoStop}%
\bibitem [{\citenamefont {Tan}(2008{\natexlab{b}})}]{Tan2008_2}%
  \BibitemOpen
  \bibfield  {author} {\bibinfo {author} {\bibfnamefont {S.}~\bibnamefont
  {Tan}},\ }\href@noop {} {\bibfield  {journal} {\bibinfo  {journal} {Ann.
  Phys.}\ }\textbf {\bibinfo {volume} {323}},\ \bibinfo {pages} {2971}
  (\bibinfo {year} {2008}{\natexlab{b}})}\BibitemShut {NoStop}%
\bibitem [{\citenamefont {Tan}(2008{\natexlab{c}})}]{Tan2008_3}%
  \BibitemOpen
  \bibfield  {author} {\bibinfo {author} {\bibfnamefont {S.}~\bibnamefont
  {Tan}},\ }\href@noop {} {\bibfield  {journal} {\bibinfo  {journal} {Ann.
  Phys.}\ }\textbf {\bibinfo {volume} {323}},\ \bibinfo {pages} {2987}
  (\bibinfo {year} {2008}{\natexlab{c}})}\BibitemShut {NoStop}%
\bibitem [{\citenamefont {Irving}\ and\ \citenamefont
  {Kirkwood}(1950)}]{Irving1950}%
  \BibitemOpen
  \bibfield  {author} {\bibinfo {author} {\bibfnamefont {J.~H.}\ \bibnamefont
  {Irving}}\ and\ \bibinfo {author} {\bibfnamefont {J.~G.}\ \bibnamefont
  {Kirkwood}},\ }\href@noop {} {\bibfield  {journal} {\bibinfo  {journal} {J.
  Chem. Phys.}\ }\textbf {\bibinfo {volume} {18}},\ \bibinfo {pages} {817}
  (\bibinfo {year} {1950})}\BibitemShut {NoStop}%
\bibitem [{\citenamefont {Forster}(1975)}]{Forster1975}%
  \BibitemOpen
  \bibfield  {author} {\bibinfo {author} {\bibfnamefont {D.}~\bibnamefont
  {Forster}},\ }\href@noop {} {\emph {\bibinfo {title} {{Hydrodynamic
  Fluctuations, Broken Symmetry, and Correlation Functions}}}}\ (\bibinfo
  {publisher} {W. A. Benjamin Inc.},\ \bibinfo {address} {Reading},\ \bibinfo
  {year} {1975})\BibitemShut {NoStop}%
\bibitem [{\citenamefont {Cooper}\ \emph {et~al.}(1997)\citenamefont {Cooper},
  \citenamefont {Halperin},\ and\ \citenamefont {Ruzin}}]{Cooper1997}%
  \BibitemOpen
  \bibfield  {author} {\bibinfo {author} {\bibfnamefont {N.~R.}\ \bibnamefont
  {Cooper}}, \bibinfo {author} {\bibfnamefont {B.~I.}\ \bibnamefont
  {Halperin}}, \ and\ \bibinfo {author} {\bibfnamefont {I.~M.}\ \bibnamefont
  {Ruzin}},\ }\href {\doibase 10.1103/PhysRevB.55.2344} {\bibfield  {journal}
  {\bibinfo  {journal} {Phys. Rev. B}\ }\textbf {\bibinfo {volume} {55}},\
  \bibinfo {pages} {2344} (\bibinfo {year} {1997})}\BibitemShut {NoStop}%
\bibitem [{\citenamefont {Abrikosov}\ and\ \citenamefont
  {Khalatnikov}(1959)}]{Abrikosov1959}%
  \BibitemOpen
  \bibfield  {author} {\bibinfo {author} {\bibfnamefont {A.~A.}\ \bibnamefont
  {Abrikosov}}\ and\ \bibinfo {author} {\bibfnamefont {I.~M.}\ \bibnamefont
  {Khalatnikov}},\ }\href@noop {} {\bibfield  {journal} {\bibinfo  {journal}
  {Rep. Prog. Phys.}\ }\textbf {\bibinfo {volume} {22}},\ \bibinfo {pages}
  {329} (\bibinfo {year} {1959})}\BibitemShut {NoStop}%
\bibitem [{\citenamefont {Lévay}(1995)}]{Levay1995}%
  \BibitemOpen
  \bibfield  {author} {\bibinfo {author} {\bibfnamefont {P.}~\bibnamefont
  {Lévay}},\ }\href {\doibase 10.1063/1.531066} {\bibfield  {journal}
  {\bibinfo  {journal} {J. Math. Phys.}\ }\textbf {\bibinfo {volume} {36}},\
  \bibinfo {pages} {2792} (\bibinfo {year} {1995})}\BibitemShut {NoStop}%
\bibitem [{\citenamefont {Chen}\ \emph {et~al.}(1989)\citenamefont {Chen},
  \citenamefont {Wilczek}, \citenamefont {Witten},\ and\ \citenamefont
  {Halperin}}]{Chen1989}%
  \BibitemOpen
  \bibfield  {author} {\bibinfo {author} {\bibfnamefont {Y.~H.}\ \bibnamefont
  {Chen}}, \bibinfo {author} {\bibfnamefont {F.}~\bibnamefont {Wilczek}},
  \bibinfo {author} {\bibfnamefont {E.}~\bibnamefont {Witten}}, \ and\ \bibinfo
  {author} {\bibfnamefont {B.~I.}\ \bibnamefont {Halperin}},\ }\href@noop {}
  {\bibfield  {journal} {\bibinfo  {journal} {Int. J. Mod. Phys. B}\ }\textbf
  {\bibinfo {volume} {3}},\ \bibinfo {pages} {1001} (\bibinfo {year}
  {1989})}\BibitemShut {NoStop}%
\bibitem [{\citenamefont {Read}\ and\ \citenamefont {Green}(2000)}]{Read2000}%
  \BibitemOpen
  \bibfield  {author} {\bibinfo {author} {\bibfnamefont {N.}~\bibnamefont
  {Read}}\ and\ \bibinfo {author} {\bibfnamefont {D.}~\bibnamefont {Green}},\
  }\href {\doibase 10.1103/PhysRevB.61.10267} {\bibfield  {journal} {\bibinfo
  {journal} {Phys. Rev. B}\ }\textbf {\bibinfo {volume} {61}},\ \bibinfo
  {pages} {10267} (\bibinfo {year} {2000})}\BibitemShut {NoStop}%
\bibitem [{\citenamefont {Lutchyn}\ \emph {et~al.}(2008)\citenamefont
  {Lutchyn}, \citenamefont {Nagornykh},\ and\ \citenamefont
  {Yakovenko}}]{Lutchyn2008}%
  \BibitemOpen
  \bibfield  {author} {\bibinfo {author} {\bibfnamefont {R.~M.}\ \bibnamefont
  {Lutchyn}}, \bibinfo {author} {\bibfnamefont {P.}~\bibnamefont {Nagornykh}},
  \ and\ \bibinfo {author} {\bibfnamefont {V.~M.}\ \bibnamefont {Yakovenko}},\
  }\href {\doibase 10.1103/PhysRevB.77.144516} {\bibfield  {journal} {\bibinfo
  {journal} {Phys. Rev. B}\ }\textbf {\bibinfo {volume} {77}},\ \bibinfo
  {pages} {144516} (\bibinfo {year} {2008})}\BibitemShut {NoStop}%
\bibitem [{\citenamefont {Niu}\ \emph {et~al.}(1985)\citenamefont {Niu},
  \citenamefont {Thouless},\ and\ \citenamefont {Wu}}]{Niu1985}%
  \BibitemOpen
  \bibfield  {author} {\bibinfo {author} {\bibfnamefont {Q.}~\bibnamefont
  {Niu}}, \bibinfo {author} {\bibfnamefont {D.~J.}\ \bibnamefont {Thouless}}, \
  and\ \bibinfo {author} {\bibfnamefont {Y.-S.}\ \bibnamefont {Wu}},\ }\href
  {\doibase 10.1103/PhysRevB.31.3372} {\bibfield  {journal} {\bibinfo
  {journal} {Phys. Rev. B}\ }\textbf {\bibinfo {volume} {31}},\ \bibinfo
  {pages} {3372} (\bibinfo {year} {1985})}\BibitemShut {NoStop}%
\end{thebibliography}%
\end{document}